\documentclass[onecollarge,natbib]{mysvjour2}
\bibpunct{[}{]}{,}{n}{}{,} % to get "[numbered]" references from natbib
\smartqed  % flush right qed marks, e.g. at end of proof
\usepackage{graphicx}
\usepackage{subfig}
 
\usepackage{txfonts}
\usepackage{multirow}
\usepackage{float}
\newcommand{\beq}{\begin{eqnarray}}
\newcommand{\eeq}{\end{eqnarray}}
\newcommand{\be}{\begin{eqnarray*}}
\newcommand{\ee}{\end{eqnarray*}}

\newcommand{\ie}{{\it i.e.}}

\newcommand{\cf}[1]{{Fig.~\ref{#1}}}

\def\lsim{\raise0.3ex\hbox{$<$\kern-0.75em\raise-1.1ex\hbox{$\sim$}}}
\def\gsim{\raise0.3ex\hbox{$>$\kern-0.75em\raise-1.1ex\hbox{$\sim$}}}

\usepackage{lineno}
\def\pPb  {$p$Pb}

\def\RpPb    {\mbox{$R_{p\rm Pb}$}}

\def\beq     {\begin{equation}}
\def\eeq     {\end{equation}}

\usepackage{ifpdf}
\ifpdf
\usepackage[pdftex]{hyperref}
\else
\usepackage[hypertex]{hyperref}
\fi

\hypersetup{
  pdftitle={},%
  pdfauthor={},%
  pdfsubject={},%
  pdfkeywords={},%
  pdfstartview={},%
  bookmarksopen=true, breaklinks=true, debug=true, %
  colorlinks=true, linkcolor=red, citecolor=blue, urlcolor=blue
}

\journalname{Few Body Systems}
\begin{document}

\title{Towards an automated tool to evaluate the impact of the nuclear modification 
of the gluon density on quarkonium, $D$ and $B$ meson production in proton-nucleus collisions}

\author{Jean-Philippe Lansberg \and Hua-Sheng Shao}%

\institute{Jean-Philippe Lansberg 
           \at IPNO, Universit\'e Paris-Saclay, Univ. Paris-Sud, CNRS/IN2P3, F-91406, Orsay, France
           \and  Hua-Sheng Shao \at Theoretical Physics Department, CERN, CH-1211 Geneva 23, Switzerland 
}

\date{}

\maketitle

\begin{abstract}
We propose a simple and model-independent procedure to account for the impact 
of the nuclear modification of the gluon density as encoded in nuclear collinear PDF sets
on two-to-two partonic hard 
processes in proton-nucleus collisions. This applies to a good approximation to 
quarkonium, $D$ and $B$ meson production, generically referred to $\cal H$. 
Our procedure consists in parametrising the square of the parton scattering 
amplitude, ${\cal A}_{gg \to {\cal H} X}$ and constraining it from the 
proton-proton data. Doing so, we have been able to compute the 
corresponding nuclear modification factors for $J/\psi$, $\Upsilon$ and $D^0$ as 
a function of $y$ and $P_T$ at $\sqrt{s_{\rm NN}}=5$ and $8$ TeV in the kinematics of the various
LHC experiments in a model independent way. It is of course justified since
the most important ingredient in such evaluations is the probability of each kinematical 
configuration. Our computations for $D$ mesons can also be extended
to $B$ meson production.
To further illustrate the potentiality of the tool, we provide --for the first time-- predictions
for the nuclear modification factor for $\eta_c$ production in \pPb\ collisions at the LHC.
\keywords{$J/\psi$, $\Upsilon$, $D$, $B$, $\eta_c$ production \and  heavy-ion collisions \and nuclear PDFs \and LHC}
\end{abstract}

\section{Introduction}
\label{intro}

For many years, open and closed heavy-flavour production in hadron-hadron, hadron-nucleus 
and nucleus-nucleus collisions has been a 
major subject of investigations, on both experimental and theoretical sides (see 
\cite{Andronic:2015wma} for a review in the context of the first LHC results and
\cite{Lansberg:2015uxa,Brambilla:2010cs,Rapp:2008tf,Frawley:2008kk,Lansberg:2006dh} for earlier reviews).
In addition of helping us to understand the interface between the perturbative and non-perturbative
regimes of QCD in hadron-hadron collisions, these reactions are also sensitive to --and thus probe-- 
the properties of the possible deconfined state of matter (QGP) resulting from nucleus-nucleus ($AA$) 
collisions at ultra-relativistic energies. 

Yet, heavy-flavour production can also be affected by other nuclear 
effects\footnote{In what follows, we will called them "cold nuclear matter effects".} which are not
related to a phase transition; they should in principle be subtracted in a way or another
to study the QGP. These are typically believed to be the only ones acting in 
proton/deuteron-nucleus ($pA$) collisions at fixed-target, RHIC and LHC energies. Experimental results 
from RHIC and the LHC in $pA$ collisions~\cite{Andronic:2015wma} have shown that the yields and the spectra
of $J/\psi$, $\Upsilon$, $D$ and $B$ are indeed modified in a magnitude which cannot simply be
ignored in QGP studies. Many effects can be at play:  break up within the nucleus~\cite{Gerschel:1988wn,Vogt:1999cu} 
or with comovers for the quarkonia~\cite{Ferreiro:2014bia,Capella:2005cn,Capella:2000zp,Gavin:1990gm}, coherent or incoherent energy loss~\cite{Arleo:2012hn,Sharma:2012dy,Arleo:2010rb,Brodsky:1992nq,Gavin:1991qk},  colour filtering~\cite{Brodsky:1989ex},
saturation/small-$x$/coherence effects~\cite{Ducloue:2015gfa,Ma:2015sia,Fujii:2013gxa,Qiu:2013qka,Kopeliovich:2001ee}, and
the modification of the parton fluxes, as encoded in nuclear Parton Distribution Functions (nPDFs)~\cite{Kovarik:2015cma,Owens:2012bv, deFlorian:2011fp,Eskola:2009uj,Hirai:2007sx}. 

In what follows, we will focus on the latter effects as a baseline for comparisons with experimental data. 
Our aim here is not to argue that it is indeed the dominant effect at RHIC and the LHC. Yet, a couple of recent 
comparisons~\cite{Albacete:2016veq,Andronic:2015wma,Ferreiro:2013pua}
have shown that the magnitude of the gluon modification in usual nPDF fits is in reasonnable agreement with quarkonium, $D$ and 
$B$ meson data in $p$Pb collisions at the LHC. 

nPDF fits are constantly updated with new data, recently from the LHC, and we have found it useful 
to propose a simple and model-independent 
procedure to account for the nPDF impact, in the particular case of gluon-induced  $2 \to 2$ reactions\footnote{We stress
that a similar procedure could devised for Drell-Yan pair, $W$ and $Z$ production.}. Such 
a procedure, to be encoded in a user friendly forthcoming tool, would then allow anybody to 
make up one's mind about the typical expected magnitude of the gluon nuclear modificatons on a given probe.

In the past, shortcut procedures using simplified kinematics (like the one Drell-Yan at LO, that 
is $2\to 1$) have widely been used~\cite{Vogt:2004dh,Arleo:2006qk,Arleo:2008zc,Lourenco:2008sk,Vogt:2010aa}. However, it has been 
shown~\cite{Ferreiro:2008qj,Ferreiro:2008wc,Ferreiro:2009ur} that it can yield to systematics 
differences and, in principle, it cannot account for the $P_T$ dependence of the yield. In general, 
it is just better to rely on a more proper $2 \to 2$ kinematics, altough some higher QCD corrections could involve more than 2 hard particles
in the final state at large $P_T$. For this purpose, a probabilistic Glauber Monte Carlo code, 
{\sc Jin}~\cite{Ferreiro:2008qj,Ferreiro:2008wc,Ferreiro:2009ur}, dedicated to the quarkonium case, has been
developped to account for the geometry of the nuclear collisions and the impact parameter dependence of the nuclear effects at play
along with the nPDF effect with an exact kinematics. However, as for now, the code deals with a limited number of processes (including
though $b$ production~\cite{delValle:2014wha}) and of
nPDFs; a simpler tool focusing on a $2 \to 2$ kinematics as the one we propose here is therefore  very complementary. 
Eventually, both tools could interfaced or merged.

As will be explained below, the tool which we propose is based on {\sc\small HELAC-Onia}~\cite{Shao:2012iz,Shao:2015vga}
(but is not restricted to quarkonia) can use any nPDF set included in the 
library LHAPDF5~\cite{Whalley:2005nh,Bourilkov:2006cj} and LHAPDF6~\cite{Buckley:2014ana}
 and does not rely on any model
for the hard-probe production, but on $pp$ measurements which are used to tune
the partonic scattering elements.

\section{Our approach}
\label{sec:1}

As announced, our approach is based on a data-driven modelling of the scattering
at the partonic level. Once folded with proton PDFs, they yield $pp$
cross sections and, when folded with one proton PDF and one nuclear PDF, they
yield $pA$ cross sections. Such a choice is essentially motivated by 
the case of inclusive quarkonium hadroproduction. Firstly, it makes the computations faster
with a limited loss of generality. Secondly, we have to acknowledge that we do not have at 
present time a global and consistent theoretical description of inclusive 
quarkonium production in the whole transverse momentum domain at hadron colliders.
Thirdly, most of available models on the market show uncertainties larger 
than those of the data which they are meant to describe (and which sometimes they do not).
Some of these observation also apply to $D$ and $B$ production.

This translates into the following advantages:
\begin{enumerate}
\item one can describe  single quarkonium, $D$ and $B$ production in $pp$ collisions in a very satisfactory way with only $2-3$ tuned parameters for each meson;
\item the uncertainty within our approach is well controlled by the available $pp$ data which, as just said, is much smaller than the theoretical uncertainties of the state-of-the-art calculations;
\item the method is much more efficient to generate events, with significantly reduced Monte-Carlo uncertainty, owing to the simplicity of the computation.
\end{enumerate}

\subsection{$pp$ cross section and partonic amplitude}

As for the partonic scattering, we use a functional form for  $\overline{|\mathcal{A}(k_1 k_2\rightarrow {\cal H}+k_3)|^2}$ 
initially proposed in~\cite{Kom:2011bd} and then 
successfully used in~\cite{Lansberg:2014swa,Lansberg:2015lva,Shao:2016wor,Borschensky:2016nkv} to model single quarkonium production
at Tevatron and LHC energies in the context of double-parton scattering (DPS) studies. It reads
\beq
\overline{|\mathcal{A}(k_1 k_2\rightarrow {\cal H}+k_3)|^2}=\frac{\lambda^2\kappa s x_1x_2}{M_{\cal H}^2}\exp\Big(-\kappa\frac{{\rm min}(P_T^2,\langle P_T \rangle^2)}{M_{\cal H}^2}\Big)\left(1+\theta(P_T^2-\langle P_T \rangle^2)\frac{\kappa}{n}\frac{P_T^2-\langle P_T \rangle^2}{M_{Q}^2}\right)^{-n},
\label{eq:crystalball}
\eeq
where $k_i$ denote the partons involved in the hard scattering, $x_{1,2}$ are the momentum fractions carried by $k_{1,2}$, 
$s$ is the square of the center-of-mass energy of the hadron collision, $P_T$ ($M_{\cal H}$) is the transverse momentum (mass) of the produced particle, ${\cal H}$, and $\theta(x)$ is the Heaviside step function. $\overline{|\mathcal{A}|^2}$ is meant to account for the the squared amplitude averaged (summed) over the initial (final) helicity/colour factors. It contains 4 parameters $\lambda,\kappa, \langle P_T\rangle,n$, to be determined from the $pp$ experimental data via a fit after the usual convolution with the PDFs: 
\beq
\frac{d\sigma(pp\rightarrow {\cal H}+X)}{d\Phi_2}=\frac{1}{2s}\int{dx_1dx_2x_1f^p(x_1)x_2f^p(x_2)\overline{|\mathcal{A}(k_1k_2\rightarrow Q+k_3)|^2}},
\eeq
where $f^p$ denotes the proton PDF and  $\Phi_2$ is the relativistic two-body phase space measure for the $2\to 2$ scattering.

In what follows, we will only consider processes which are dominated by gluon fusion at LHC energies. All the procedure can readily be generalised
to other partonic initial states. 

\subsection{Accounting for the nuclear PDF impact}

As announced, we will also only consider the nuclear modification of 
the PDF among the possible effects acting on quarkonia, $D$ and $B$ mesons. 
Such a restriction would probably not yield a good description of the quarkonium 
excited states~\cite{Lansberg:2015uxa}, which we therefore do not discuss. 
Along the same lines, we will focus on the LHC regime where the nuclear absorption
is likely negligible. It may not be so at RHIC and even less at fixed-target energies.

Whereas one could think that the proposed procedure can be used to evaluate the sole impact of the nPDF on 
the excited states or the ground states at lower energies, one may want to be careful that in 
presence of other significant effects, the impact of the nPDFs may be affected. A clear
example is a $b$-dependent antishadowing, which would tend to generate more $J/\psi$
in the center of the overlap zone, which then may have more chance to be broken up
by the nuclear absorption than those produced in the periphery of the overlap zone. Yet, the procedure should give a right order of magnitude
of the nPDF impact even if other effects are at play.

As it is costumary, the yield of a particle $\cal H$  in $pA$ collisions
is obtained from that corresponding to the simple superposition of the 
equivalent number of $pp$ collisions corrected by a factor encoding
the nuclear modification of the parton flux. This is absolutely equivalent to directly using
nuclear PDFs (normalised to the nucleus atomic number $A$) instead of proton PDFs. As aforementionned, 
our procedure does not currently rely on a Glauber code and we will thus restrict
our studies to mininum bias collisions, \ie\ integrated on all possible impact parameters $b$.

As such, the correction factor can be expressed in terms of the ratios $R_i^A$ of the
nuclear PDF (nPDF) in a nucleon belonging to a nucleus~$A$ to the
PDF in the free nucleon:
\begin{equation}
\label{eq:shadow-corr-factor}
R^A_i (x,Q^2) = \frac{f^A_i (x,Q^2)}{ A f^{p}_i (x,Q^2)}\ , \ \
i = q, \bar{q}, g \ .
\end{equation}

To illustrate the potentiality of our procedure, we will only use two of 
the most up-to-date nPDF parameterisations resulting from global analyses with uncertainties. 
The first is EPS09~\cite{Eskola:2009uj}, which provides the fit uncertainties at
 both leading order (LO, dubbed EPS09LO) and next-to-leading order (NLO, dubbed 
EPS09NLO)  and is available in the library LHAPDF5~\cite{Whalley:2005nh}. 
The nPDF effects are given in terms of $R^A_i (x,Q^2)$ for all the flavours.

A new set, nCTEQ15~\cite{Kovarik:2015cma}, has recently been released. It is 
available in the library LHAPDF6~\cite{Buckley:2014ana} and provides NLO nuclear 
PDFs. As such, it is important to use the very same proton PDF as the one used for 
the fit. We have thus used CT14NLO~\cite{Dulat:2015mca}. In the case of EPS09, 
which provides ratios, the proton PDF to be used is less critical. In principle, we should 
have used CTEQ6(L1 or M) by consistency with EPS09, or CT14NLO for a good comparison 
of the yields with nCTEQ. Since CT14NLO is not available in LHAPDF5 and the code 
cannot load two PDF libraries at a time, we have preferred to use CT10NLO~\cite{Lai:2010vv}
which anyhow yields very similar gluon PDFs.

\section{Fitting the LHC $pp$ cross sections}

\begin{figure}[H]
\begin{center} 
\subfloat[$J/\psi$~\cite{Aaij:2011jh}]{\includegraphics[width=0.33\textwidth,keepaspectratio]{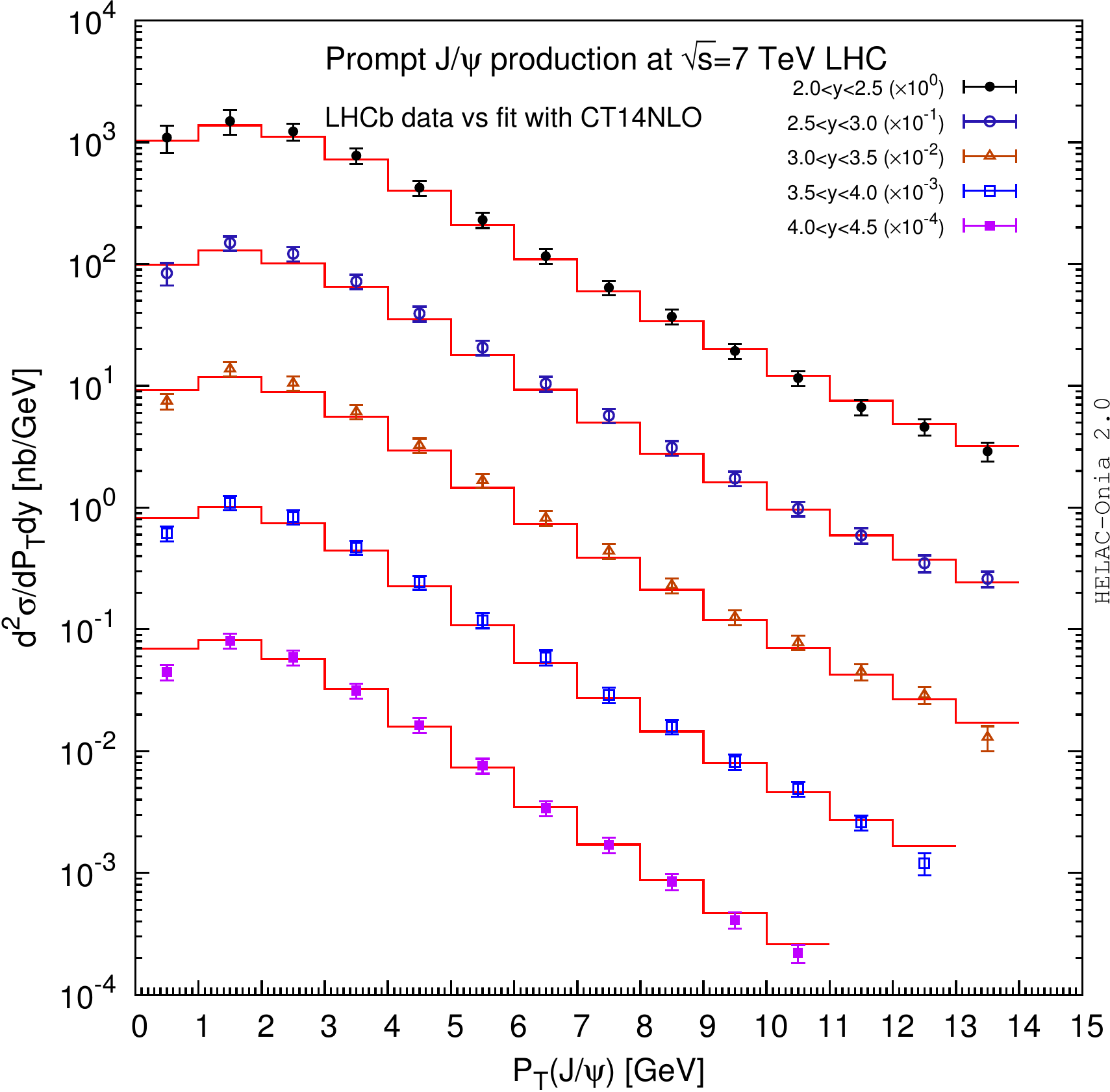}}
\subfloat[$J/\psi$~\cite{Aaij:2013yaa}]{\includegraphics[width=0.33\textwidth,keepaspectratio]{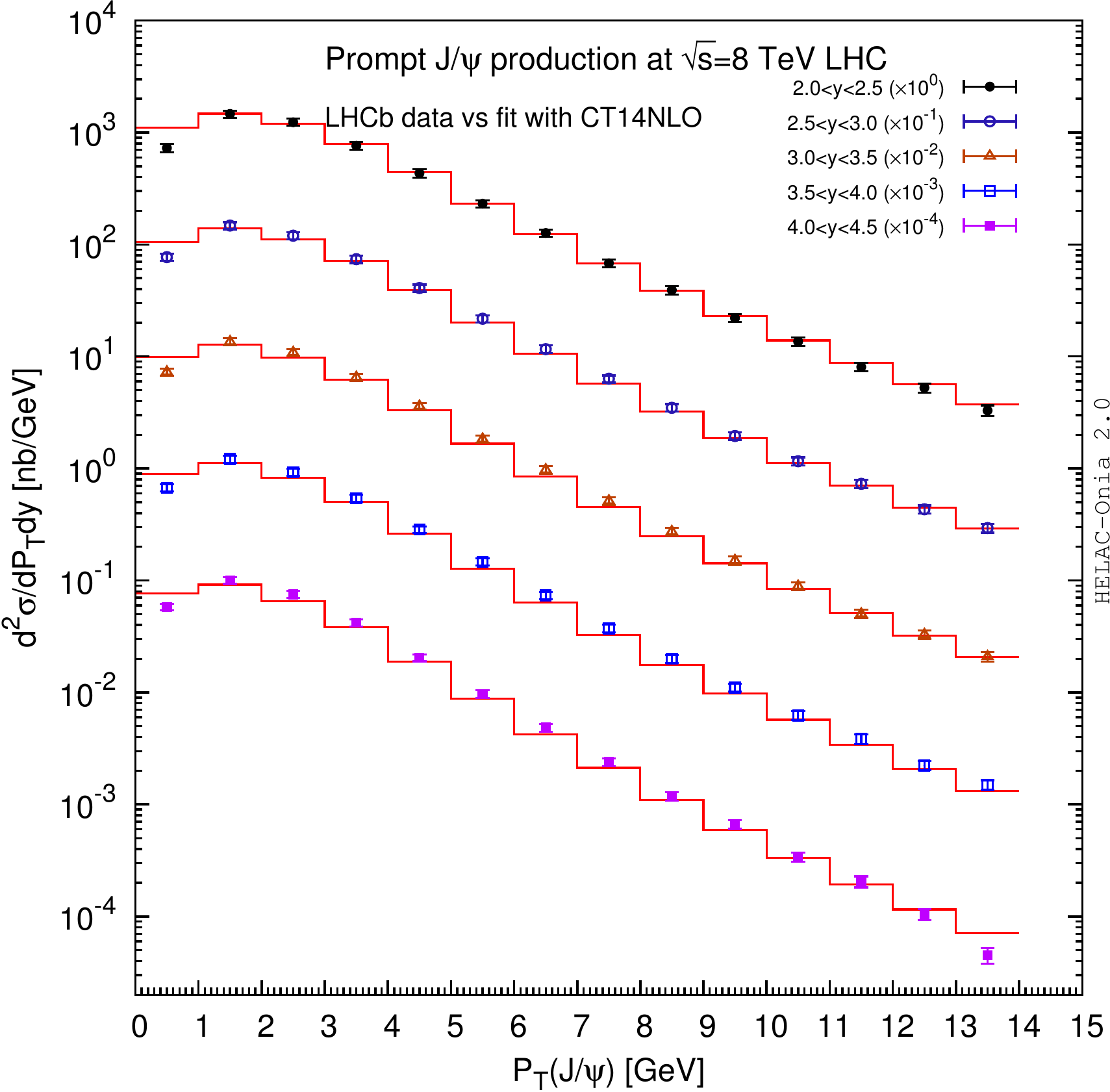}}
\subfloat[$J/\psi$~\cite{Aad:2015duc}]{\includegraphics[width=0.33\textwidth,keepaspectratio]{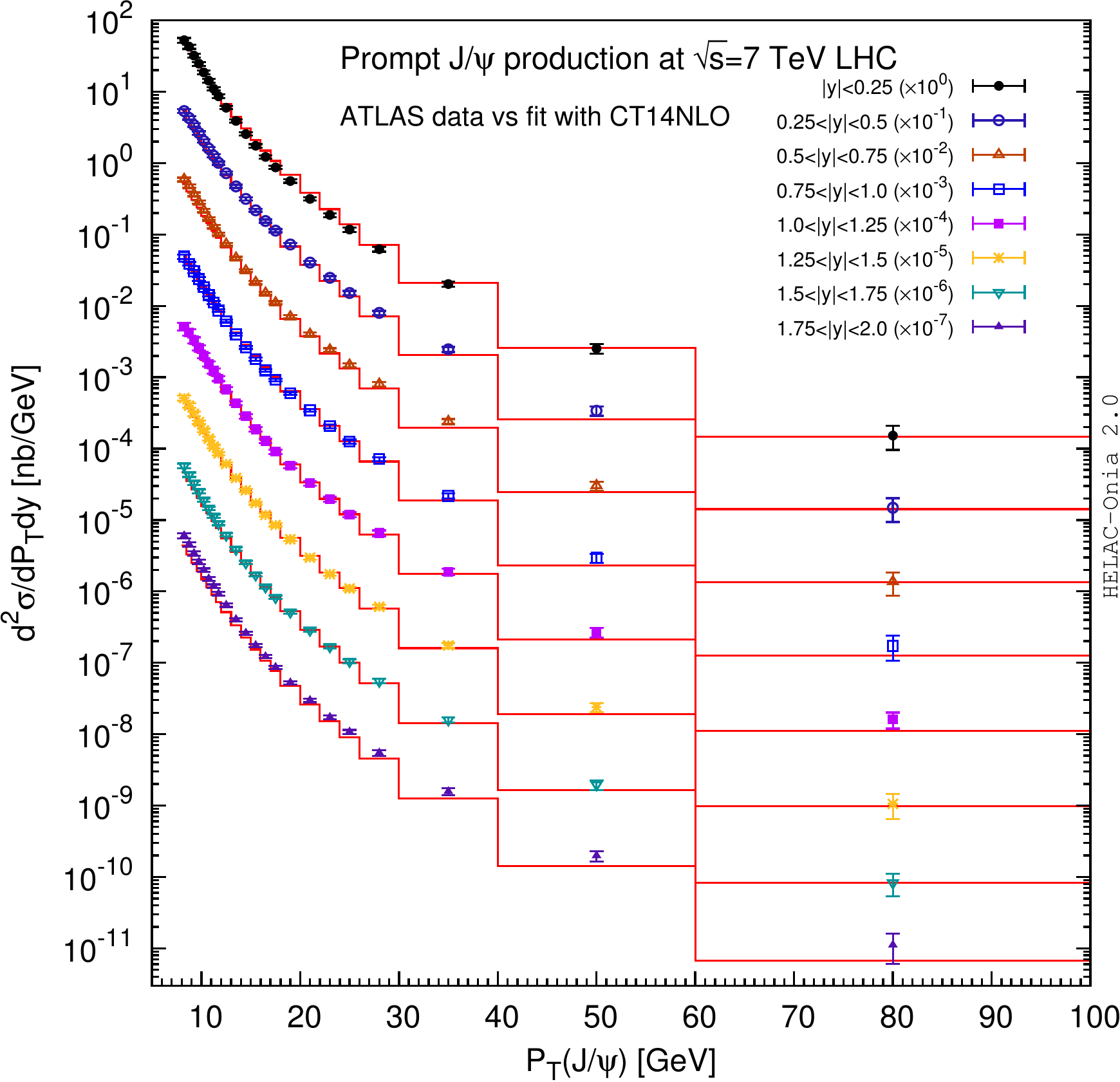}}\\
\subfloat[$J/\psi$~\cite{Khachatryan:2015rra}]{\includegraphics[width=0.33\textwidth,keepaspectratio]{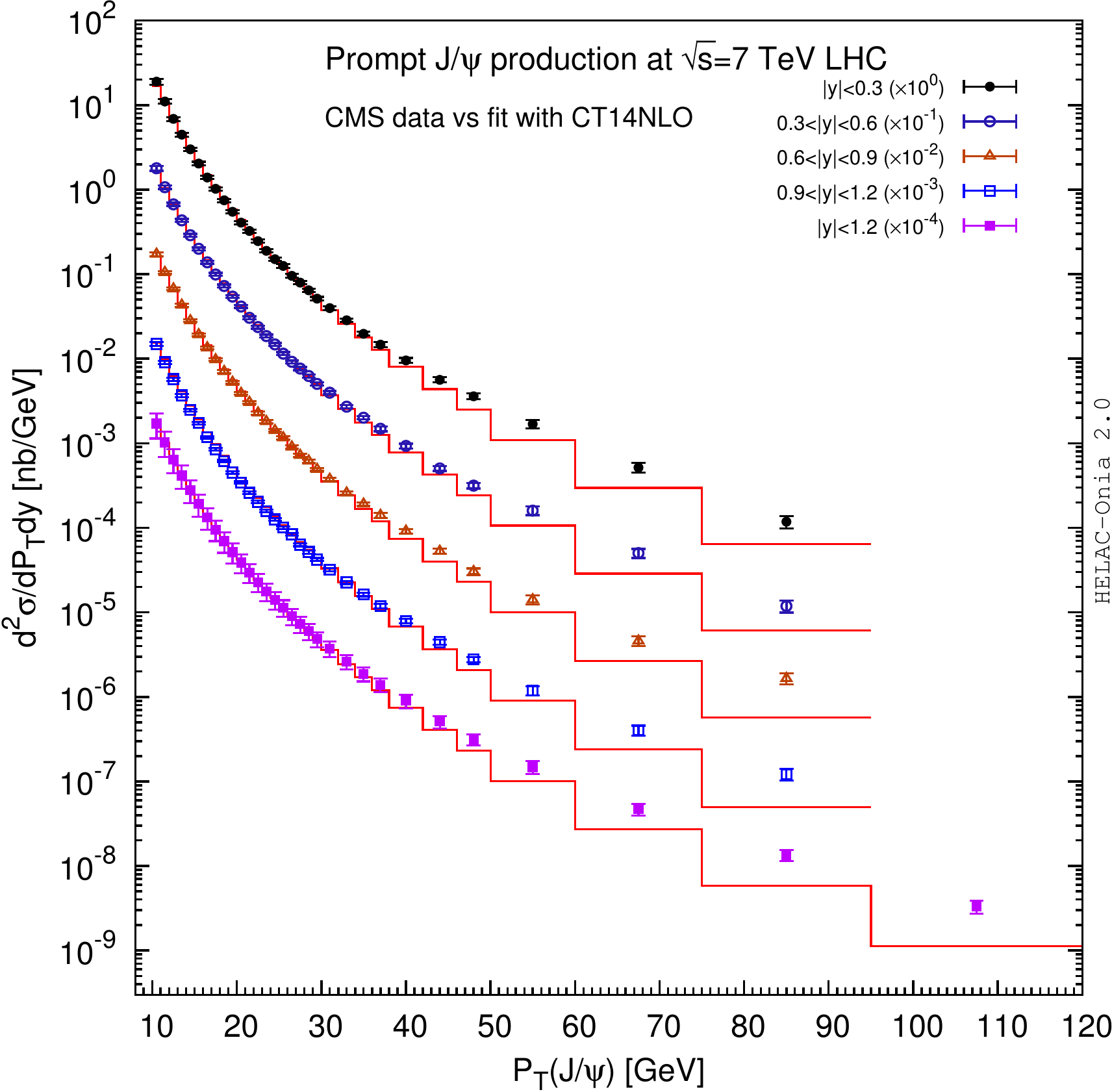}}
\subfloat[$\Upsilon$~\cite{Abelev:2014qha}]{\includegraphics[width=0.33\textwidth,keepaspectratio]{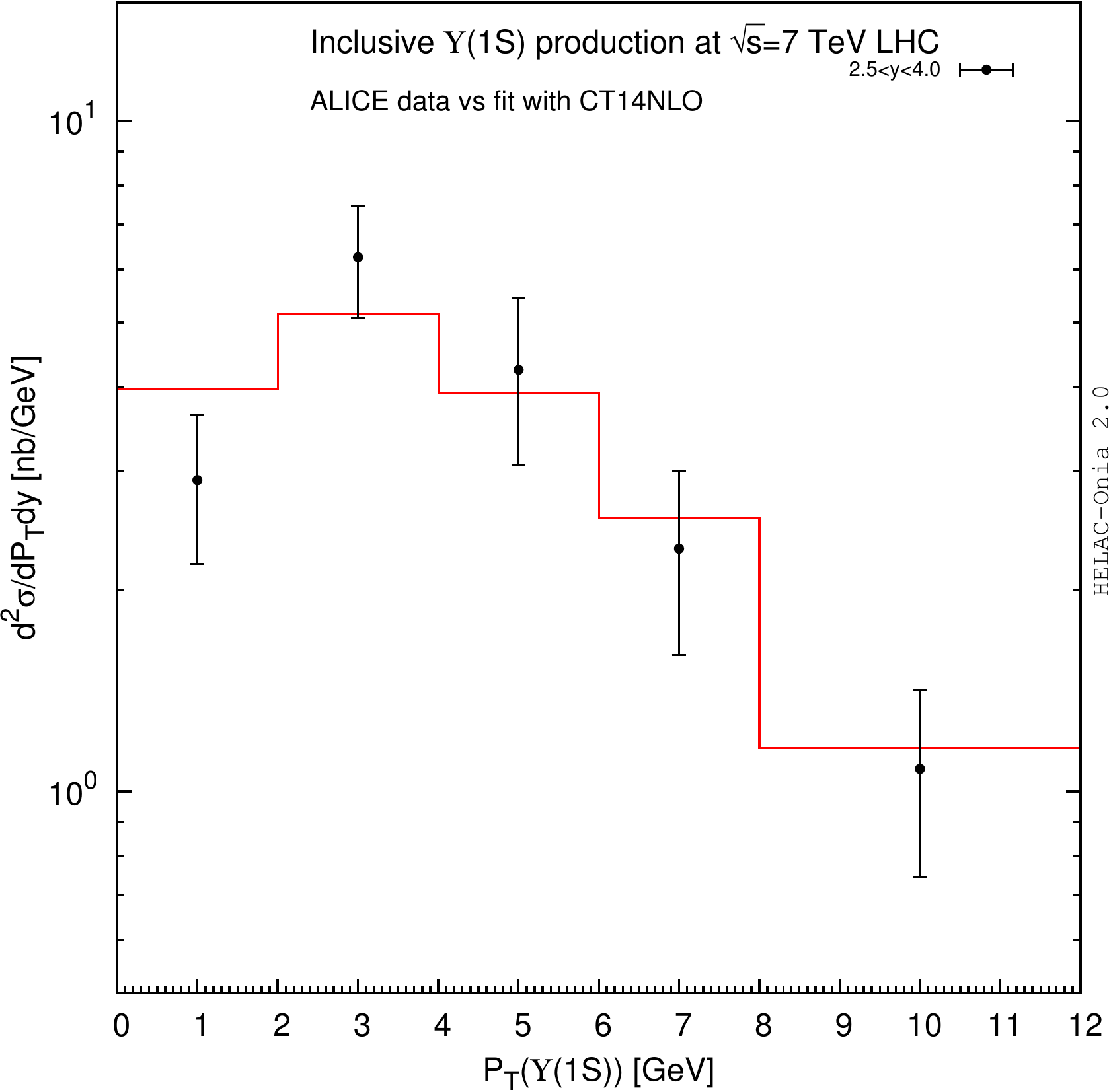}}
\subfloat[$\Upsilon$~\cite{Chatrchyan:2013yna}]{\includegraphics[width=0.33\textwidth,keepaspectratio]{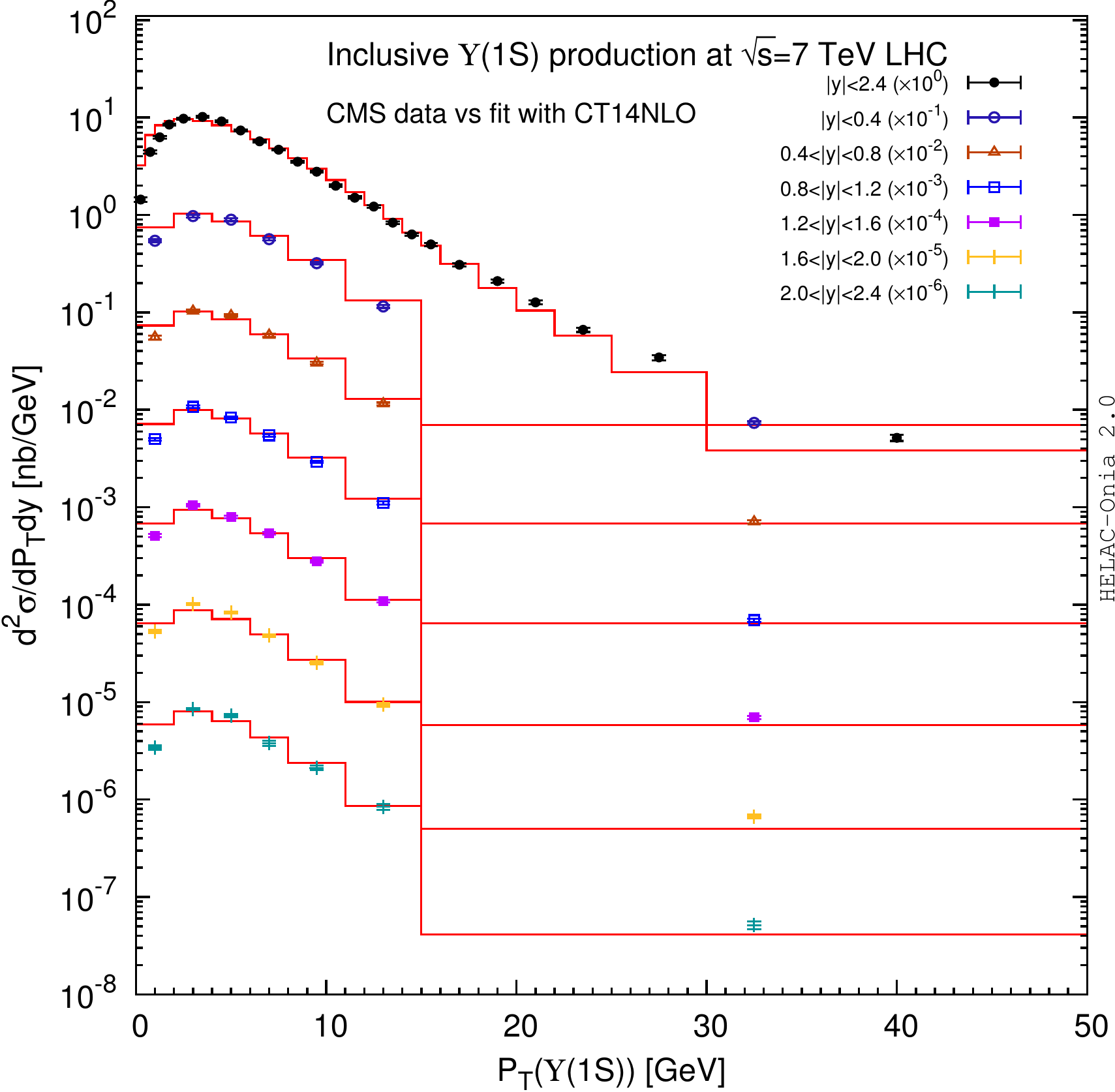}}\\
\subfloat[$\Upsilon$~\cite{LHCb:2012aa}]{\includegraphics[width=0.33\textwidth,keepaspectratio]{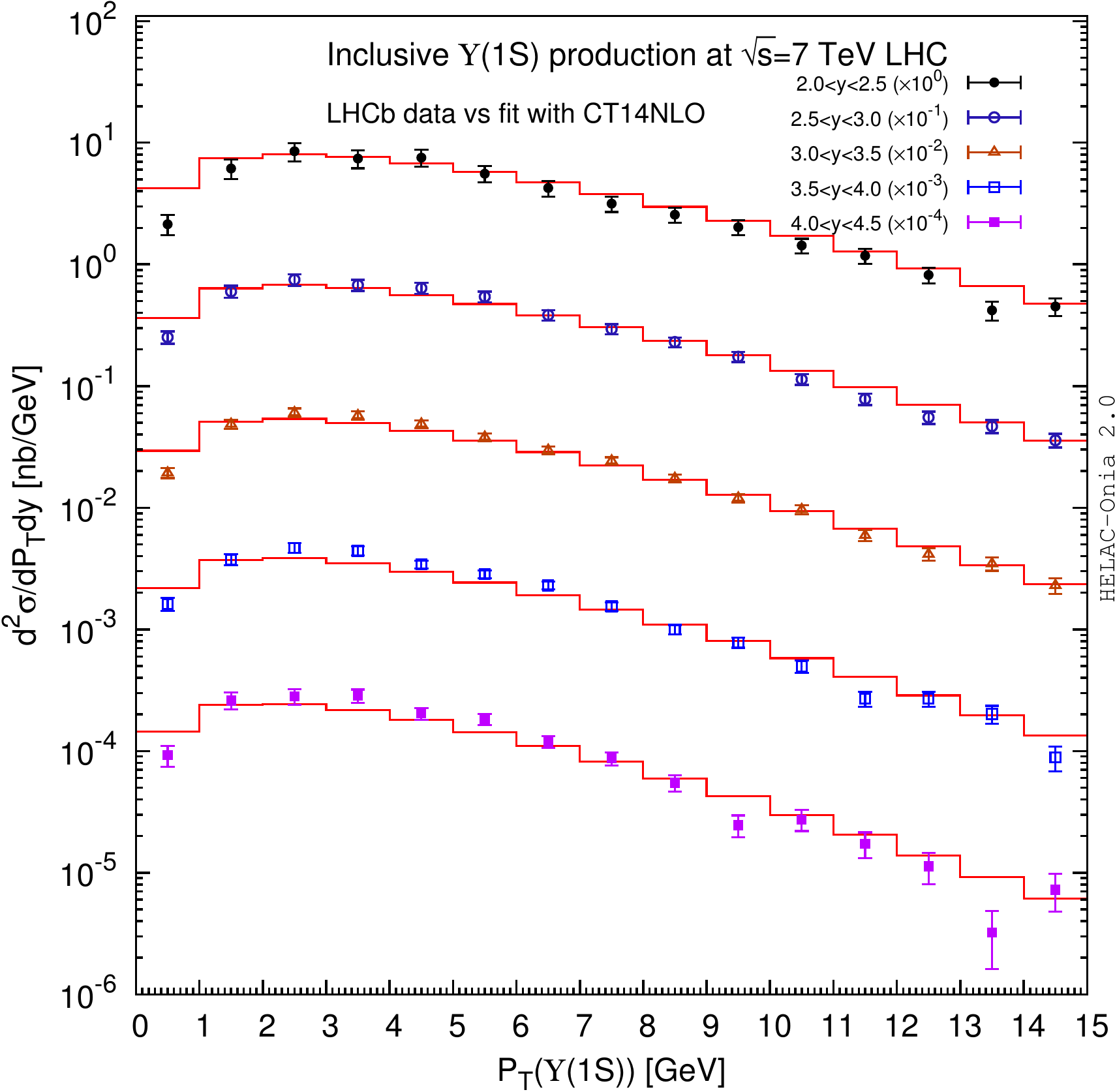}}
\subfloat[$\eta_c$~\cite{Aaij:2014bga}]{\includegraphics[width=0.33\textwidth,keepaspectratio]{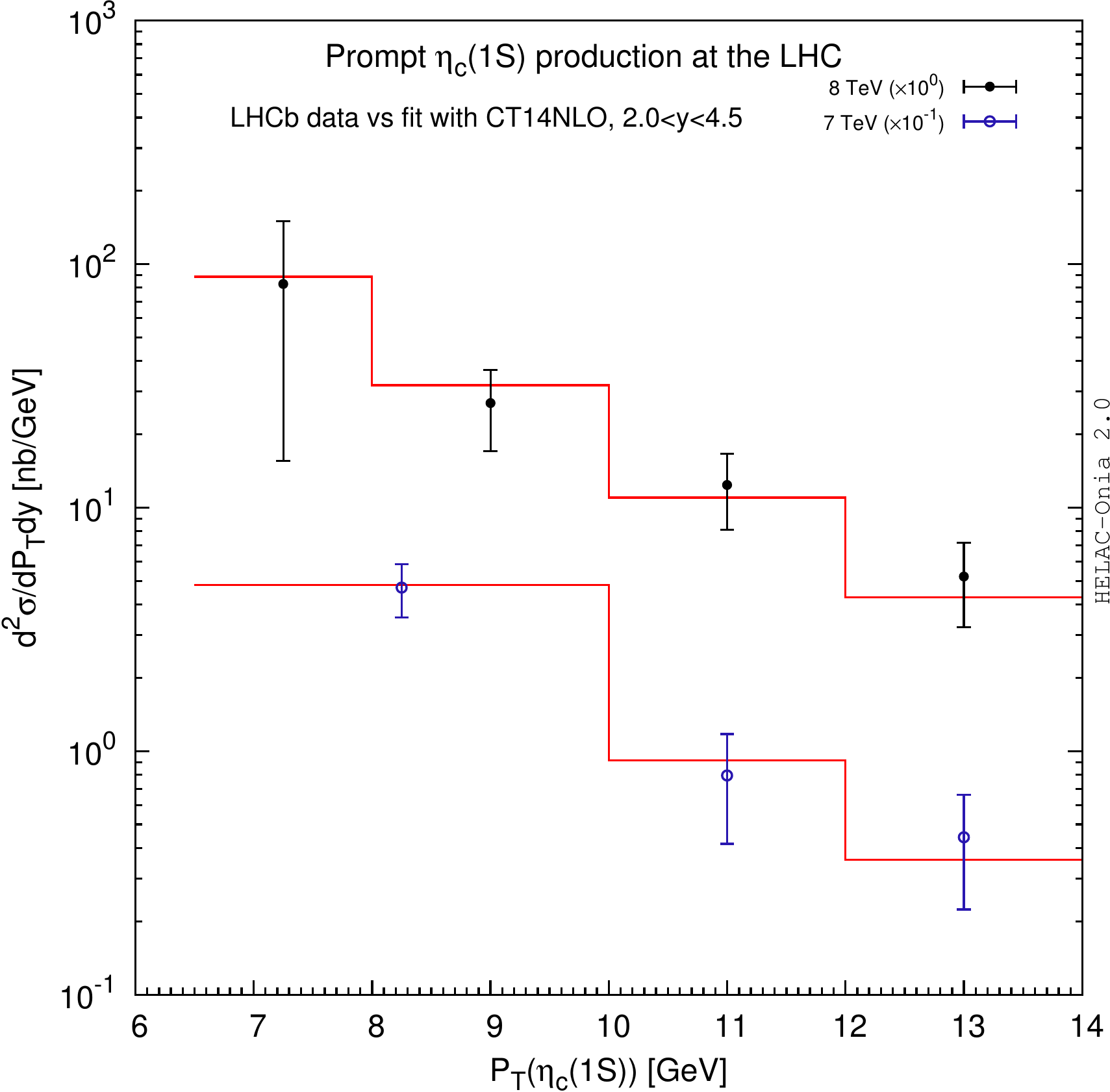}
\label{fig:etac-pp}}
\subfloat[$D^0$~\cite{Aaij:2013mga}]{\includegraphics[width=0.33
\textwidth,keepaspectratio]{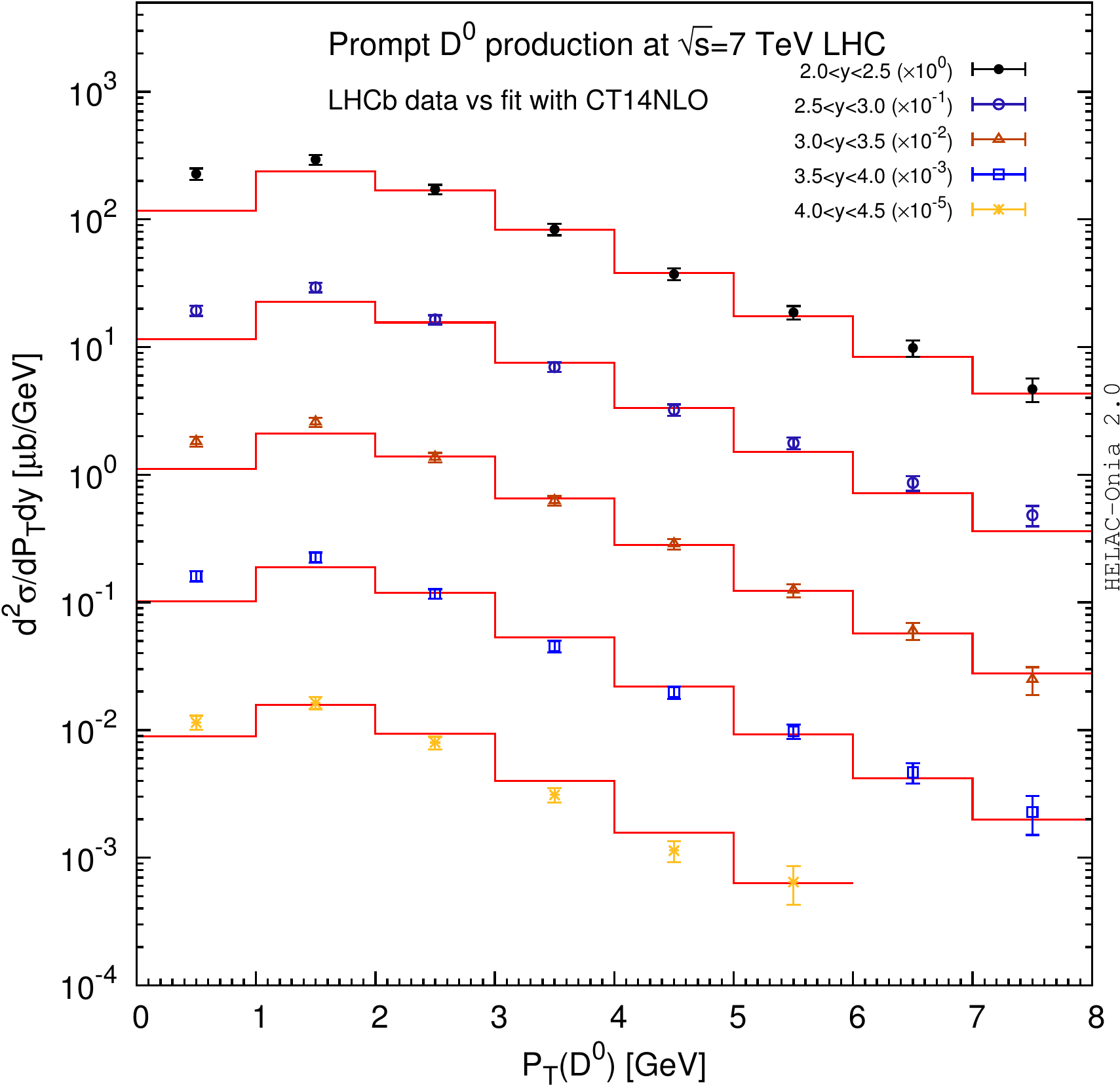}
\label{fig:D0-pp}}
\caption{Comparison of our fit results with the prompt $J/\psi$ (a-d), inclusive $\Upsilon$ (e-g), prompt $\eta_c$ (h) and prompt $D^0$  (i) production data in $pp$ collisions 
at the LHC with CT14NLO as our proton PDF.\label{fig:jpsi-upsi-pp}}
\end{center}
\end{figure}

At the LHC, we can essentially divide the inclusive (prompt) $J/\psi$ production cross-sections measurements into 2 classes: the slightly forward and low $P_T$ (from 0 up to roughly 20 GeV) data of LHCb and ALICE and those from ATLAS and CMS at "high"  $P_T$ (from 6--8 up to roughly 100 GeV)\footnote{ALICE has also measured low $P_T$ central $J/\psi$  but with a limited statistical precision and a $b$ feed-down contamination. The forward
ALICE data are also prone to such a $b$ feed-down contimanation. As such, we will focus
on the LHCb data for our fits in the forward and low $P_T$ region. We also note that CMS has the capacity to cover $P_T$ down to 3 GeV (even below in specific cases) in its most forward/backward acceptance.}.
We have performed 2 times 2 $\chi^2$ fits of $d^2\sigma/dP_Tdy$ of prompt $J/\psi$ production in $pp$ collisions with 2 PDF sets (CT14NLO and CT10LO) using, on the one hand, LHCb data~\cite{Aaij:2011jh,Aaij:2013yaa} and, on the other, ATLAS~\cite{Aad:2015duc} and CMS~\cite{Khachatryan:2015rra} data. The fit parameters ($\lambda,\kappa,\langle P_T\rangle$ and $n$ of Eq.~\ref{eq:crystalball}) are shown in Table.~\ref{tab:fitjpsi}.  A comparison of our fit results with the experimental data is shown in Fig.~\ref{fig:jpsi-upsi-pp} (a) -- (d). The procedure is
particularly successful, but for a few marginal bins\footnote{Let us in particular note the slight discrepancy with the CMS very high $P_T$ data. The very same fits are however consistent with the ATLAS data in the same $P_T$ regime, which
possibly indicates an underestimation of the systematical experimental uncertainties in that regime.}. These will nevertheless do
 not have a visible impact on the $pA$ observables to be discussed later.

\begin{table}[H]
\centering\renewcommand{\arraystretch}{1.2}
\begin{tabular}{c|c|cccc} 
PDF & data &  $\lambda$ & $\kappa$ & $\langle P_T\rangle$ & $n$\\\hline\hline
\multirow{2}{*}{CT14NLO} & LHCb~\cite{Aaij:2011jh,Aaij:2013yaa} & $0.296\pm0.118$ & $0.558$ & $4.5$ (fixed) & $2$ (fixed)\\
& ATLAS~\cite{Aad:2015duc} \& CMS~\cite{Khachatryan:2015rra} & $0.378$ & $0.743\pm0.0395$ & $4.5$ (fixed) & $2$ (fixed)\\\hline
\multirow{2}{*}{CT10NLO} & LHCb~\cite{Aaij:2011jh,Aaij:2013yaa} & $0.297$ & $0.532$ & $4.5$ (fixed) & $2$ (fixed)\\
& ATLAS~\cite{Aad:2015duc} \& CMS~\cite{Khachatryan:2015rra} & $0.383$ & $0.750\pm0.0364$ & $4.5$ (fixed) & $2$ (fixed)\\\hline
\end{tabular}
\caption{\label{tab:fitjpsi} Results of a fit of $d^2\sigma/dP_Tdy$ of prompt $J/\psi$ in  $pp$ collisions using CT10NLO and CT14NLO, where we have fixed the values of $n$ and $\langle P_T \rangle$. [The uncertainties from the $\chi^2$ fit below the per cent level are not shown.]}
\begin{tabular}{c|cccc}
PDF &  $\lambda$ & $\kappa$ & $\langle P_T\rangle$ & $n$\\\hline\hline
CT14NLO &  $0.768$ & $0.0841\pm 0.0271$ & $13.5$ (fixed) & $2$ (fixed)\\\hline
CT10NLO & $0.687\pm0.367$ & $0.0864$ & $13.5$ (fixed) & $2$ (fixed)\\\hline
\end{tabular}
\caption{\label{tab:fity1s} Results of a fit of $d^2\sigma/dP_Tdy$ of inclusive $\Upsilon(1S)$ in $pp$ collisions using CT10NLO and CT14NLO, where we have fixed the values of $n$ and $\langle P_T \rangle$. The experimental data used in the fit are from ALICE~\cite{Abelev:2014qha}, LHCb~\cite{LHCb:2012aa,Aaij:2015awa}, ATLAS~\cite{Aad:2012dlq} and CMS~\cite{Chatrchyan:2013yna}. [The uncertainties from the $\chi^2$ fit below the per cent level are not shown.]}
\begin{tabular}{c|cccc}
PDF &  $\lambda$ & $\kappa$ & $\langle P_T\rangle$ & $n$\\\hline\hline
CT14NLO &  $0.558$ & $0.398$ & $4.5$ (fixed) & $2$ (fixed)\\\hline
CT10NLO & $0.337$ & $0.291$ & $4.5$ (fixed) & $2$ (fixed)\\\hline
\end{tabular}
\caption{\label{tab:fitetac} Results of a fit of $d^2\sigma/dP_Tdy$ of prompt $\eta_c(1S)$ in $pp$ collisions using CT10NLO and CT14NLO, where we have fixed the values of $n$ and $\langle P_T \rangle$. The experimental data used in the fit are from LHCb~\cite{Aaij:2014bga}. [The uncertainties from the $\chi^2$ fit below the per cent level are not shown.]}
\begin{tabular}{c|cccc}
PDF &  $\lambda$ & $\kappa$ & $\langle P_T\rangle$ & $n$\\\hline\hline
CT14NLO &  $2.29$ & $1.11$ & $0.88$ & $2$ (fixed)\\\hline
CT10NLO & $2.38$ & $1.62$ & $0.521$ & $2$ (fixed)\\\hline
\end{tabular}
\caption{\label{tab:fitD0} Results of a fit of $d^2\sigma/dP_Tdy$ of prompt $D^0$ in $pp$ collisions using CT10NLO and CT14NLO, where the value of $n$ was fixed. The experimental data used in the fit are from LHCb~\cite{Aaij:2013mga}. [The uncertainties from the $\chi^2$ fit below the per cent level are not shown.]}
\end{table}

For the $\Upsilon(1S)$ case, all the experiments have access to low $P_T$ data and 
there is no $b$ feed-down contamination. We have performed 2 fits (with CT14NLO and CT10LO) 
using data from ALICE~\cite{Abelev:2014qha}, LHCb~\cite{LHCb:2012aa,Aaij:2015awa}, ATLAS~\cite{Aad:2012dlq} and CMS~\cite{Chatrchyan:2013yna} altogether. See Table.~\ref{tab:fity1s} for the fit results and Fig.~\ref{fig:jpsi-upsi-pp}e--h for comparison with 
the fit spectra.

For the prompt $\eta_c$ case, we have performed 2 fits (with CT14NLO and CT10LO) 
 from the sole LHCb~\cite{Aaij:2014bga} data. See Table.~\ref{tab:fitetac} 
for the fit results and \cf{fig:etac-pp} for comparison with 
the fit spectra.

As for the $D^0$, we have performed 2 fits (with CT14NLO and CT10LO) 
 from the LHCb~\cite{Aaij:2013mga} data. See Table.~\ref{tab:fitD0} 
for the fit results and \cf{fig:D0-pp} for comparison with 
the fit spectra.

Just as for the $J/\psi$ fits, the procedure works very well 
for $\Upsilon(1S)$, $\eta_c$ and $D^0$ and gives us confidence
that using the corresponding parametrised squared amplitudes 
will provide us with a reliable mapping of the $x_{1,2}$, $y$ and $P_T$ space.

\section{Results}

\subsection{Rapidity and transverse-momentum dependence of the production cross-section in $p$Pb collisions at $\sqrt{s_{NN}}=8$~TeV}

Now that we have described our approach, we can present our results for the cross-section for quarkonium and $D^0$ production in proton-lead ($p$Pb) collisions at the LHC. 
In the following, we show comparisons with all the existing data. As announced, we have employed three different nPDF EPS09LO, EPS09NLO and nCTEQ15. The sole nPDF uncertainties  are displayed. In particular, we have not varied the factorisation scale despite the fact that it can indeed alter our results. Our histograms are calculated under the same cuts as the experimental data.

\begin{figure}[H]
\begin{center} 
\subfloat[]{\includegraphics[width=0.33\textwidth,keepaspectratio]{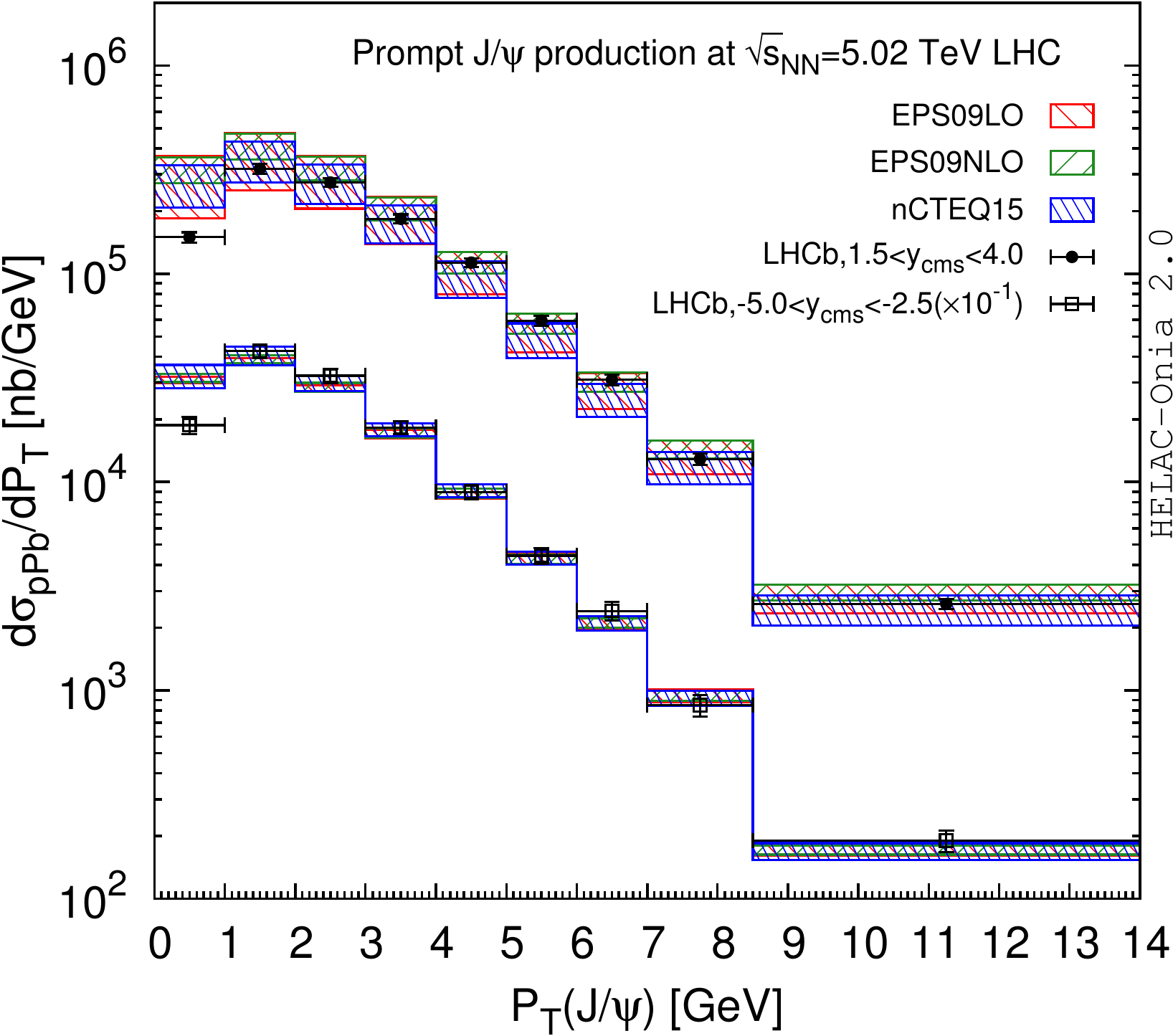}\label{fig:ptjpsisigmapPba}}
\subfloat[]{\includegraphics[width=0.33\textwidth,keepaspectratio]{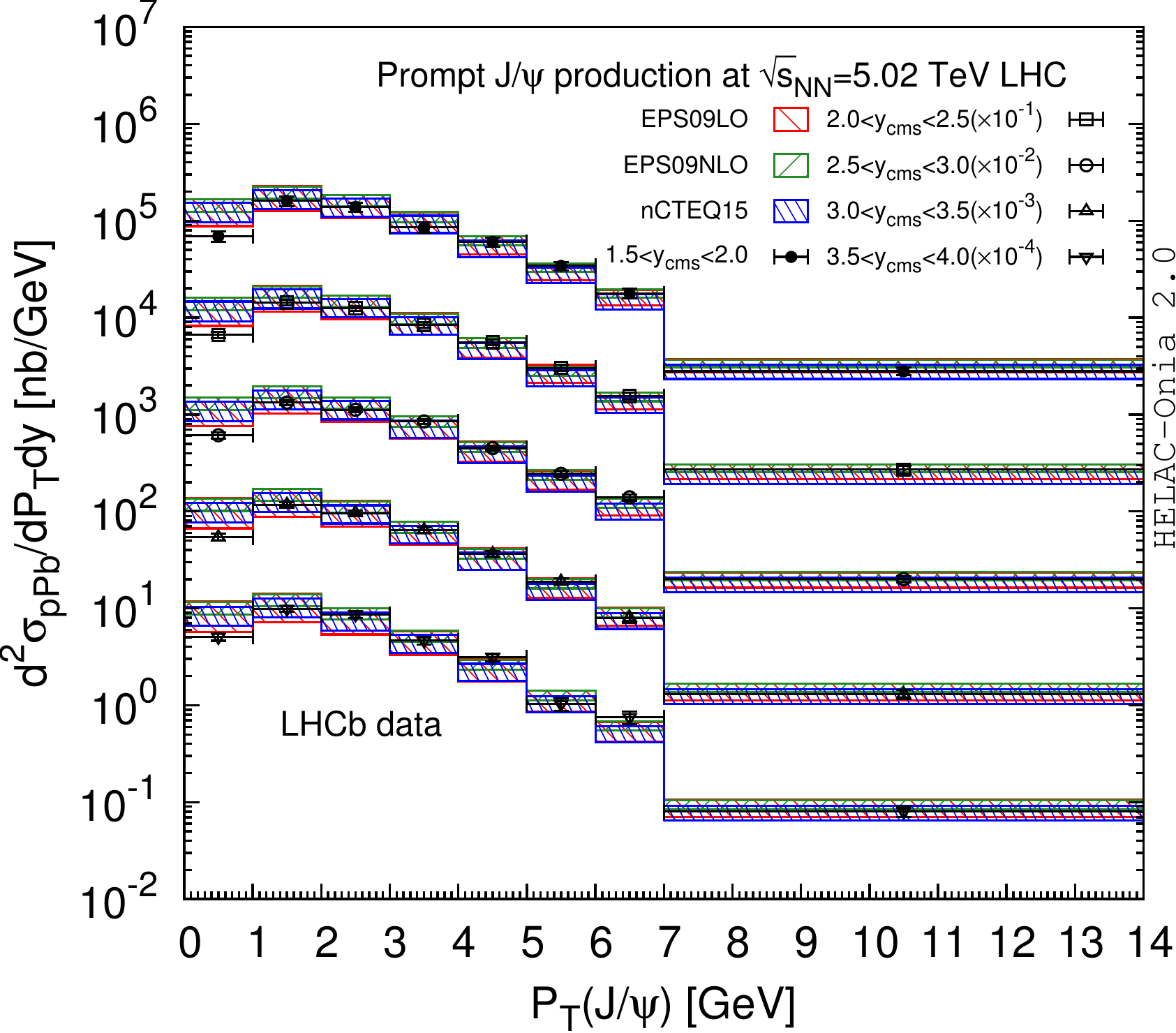}\label{fig:ptjpsisigmapPbb}}\\
\subfloat[]{\includegraphics[width=0.33\textwidth,keepaspectratio]{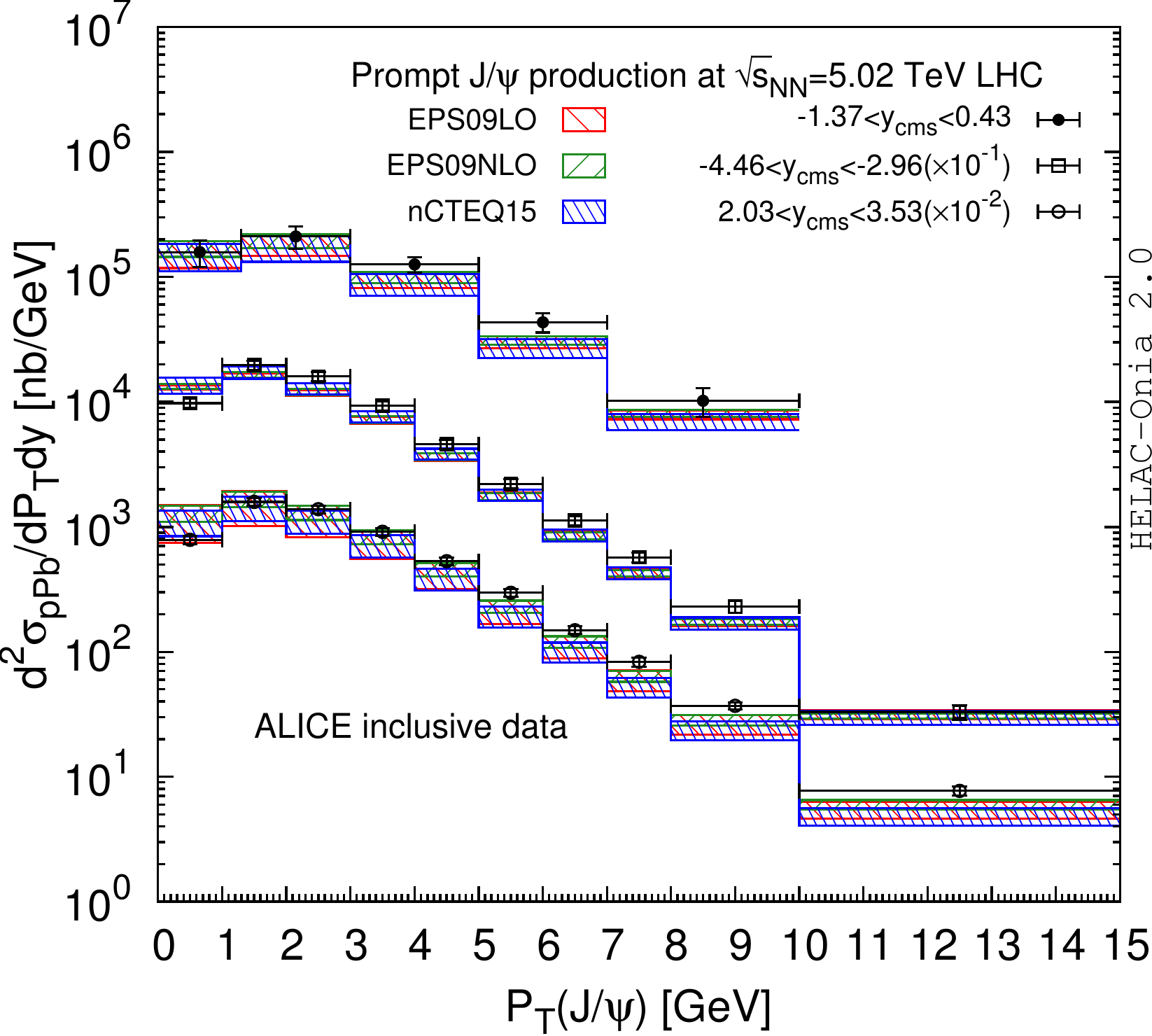}\label{fig:ptjpsisigmapPbc}}
\subfloat[]{\includegraphics[width=0.33\textwidth,keepaspectratio]{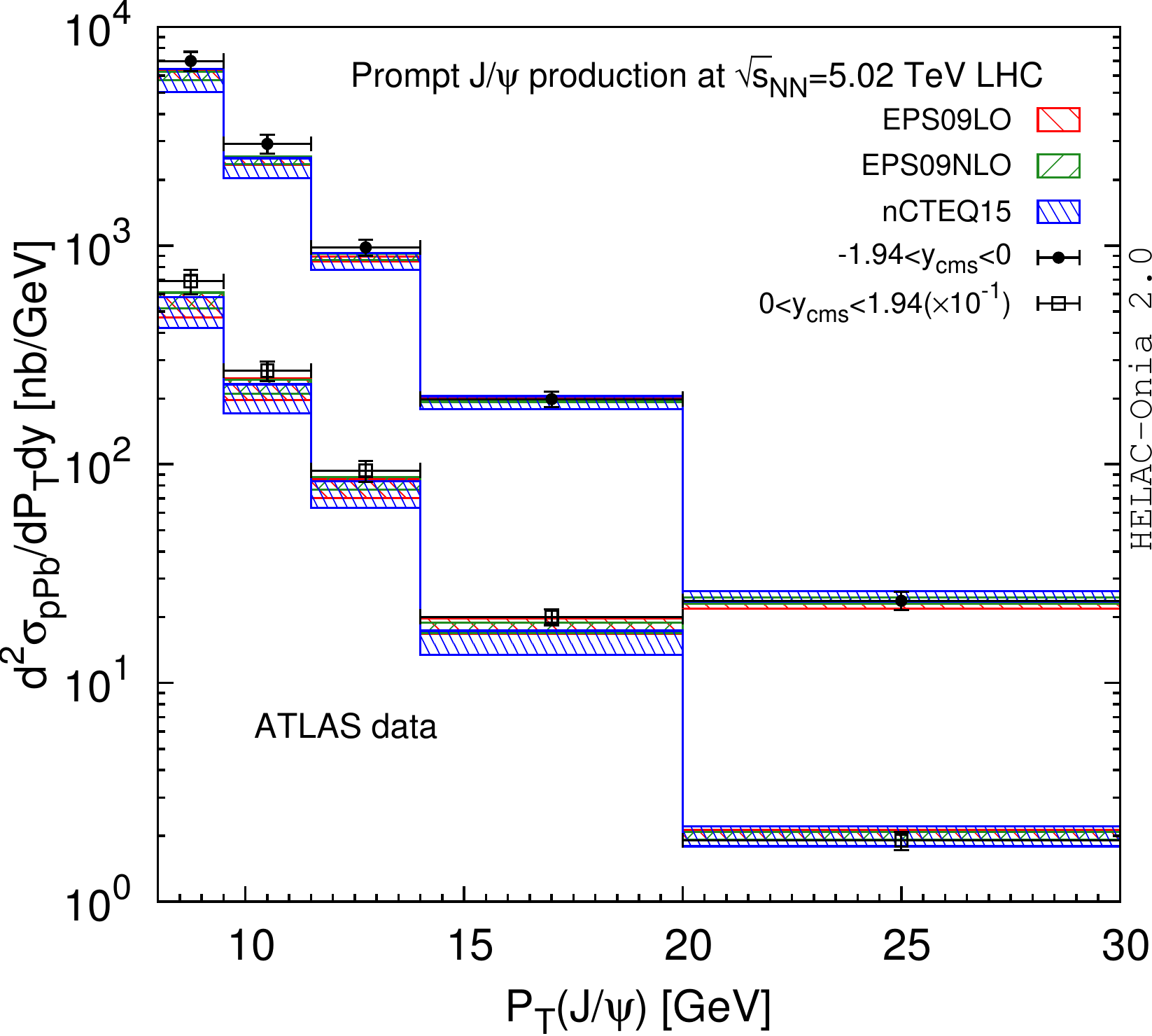}\label{fig:ptjpsisigmapPbd}}
\caption{Transverse-momentum dependence of the cross-section for prompt $J/\psi$ production in $p$Pb collisions at $\sqrt{s_{NN}}=5.02$ TeV: comparison between our results and the measurements of LHCb~\cite{Aaij:2013zxa}, ALICE~\cite{Adam:2015iga} and ATLAS~\cite{Aad:2015ddl}.  [The uncertainty bands represent the nuclear PDF uncertainty only].\label{fig:ptjpsisigmapPb}}
\end{center}
\end{figure}

The transverse-momentum $P_T$ spectra ($d\sigma_{p\rm Pb}/dP_T$) of promptly produced 
$J/\psi$ 
 in $p$Pb collisions at $\sqrt{s_{NN}}=5.02$ TeV are shown in Fig.~\ref{fig:ptjpsisigmapPb}. Comparisons are made with the LHCb prompt $J/\psi$ production data~\cite{Aaij:2013zxa} in both the forward ($1.5<y^{J/\psi}_{\rm c.m.s.}<4.0$)~\footnote{Unless indicated, all rapidity $y$ (or $y_{\rm c.m.s.}$) mean the rapidity in the center-of-mass frame of  nucleon-nucleon collision. In particular, rapidities in the laboratory frame would read $y_{\rm lab}$.} and backward ($-5.0<y^{J/\psi}_{\rm c.m.s.}<-2.5$) rapidity regions in Fig.~\ref{fig:ptjpsisigmapPba}. Fig.~\ref{fig:ptjpsisigmapPbb} shows a comparison with the double differential cross sections $d^2\sigma_{p\rm Pb}/dP_Tdy$ of $J/\psi$ production of LHCb. Similarly, comparisons with the ALICE data~\cite{Adam:2015iga} and ATLAS data~\cite{Aad:2015ddl} are given in Fig.~\ref{fig:ptjpsisigmapPbc} and Fig.~\ref{fig:ptjpsisigmapPbd} respectively. We note that ALICE data do not exclude the contribution from $b$-hadron decays. In general, the agreement with the yields differential in $P_T^{J/\psi}$ is satisfactory both at low $P_T$ and high $P_T$.

\begin{figure}[H]
\begin{center} 
\subfloat[]{\includegraphics[width=0.33\textwidth,keepaspectratio]{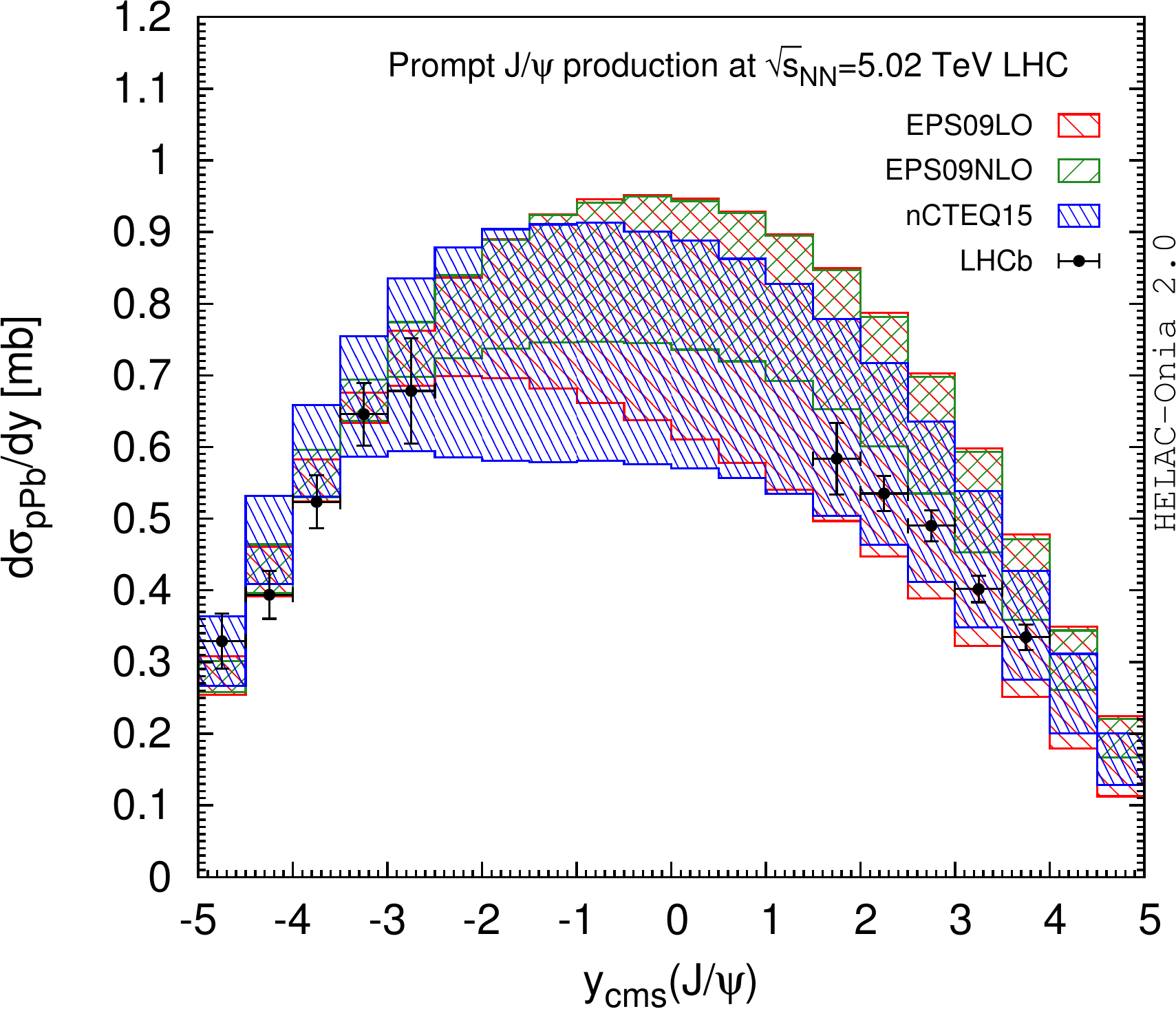}\label{fig:yjpsisigmapPba}}
\subfloat[]{\includegraphics[width=0.33\textwidth,keepaspectratio]{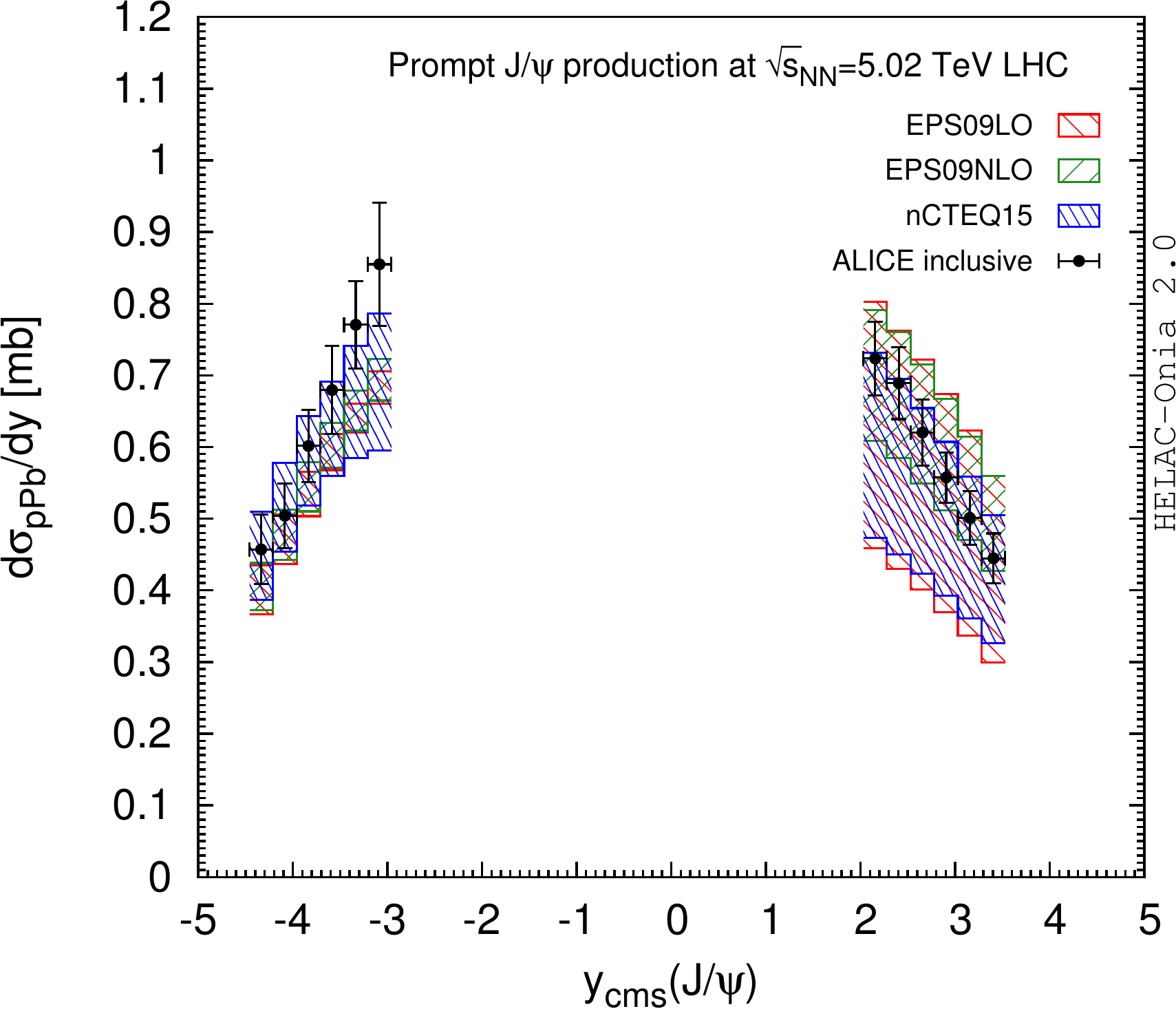}\label{fig:yjpsisigmapPbb}}
\subfloat[]{\includegraphics[width=0.33\textwidth,keepaspectratio]{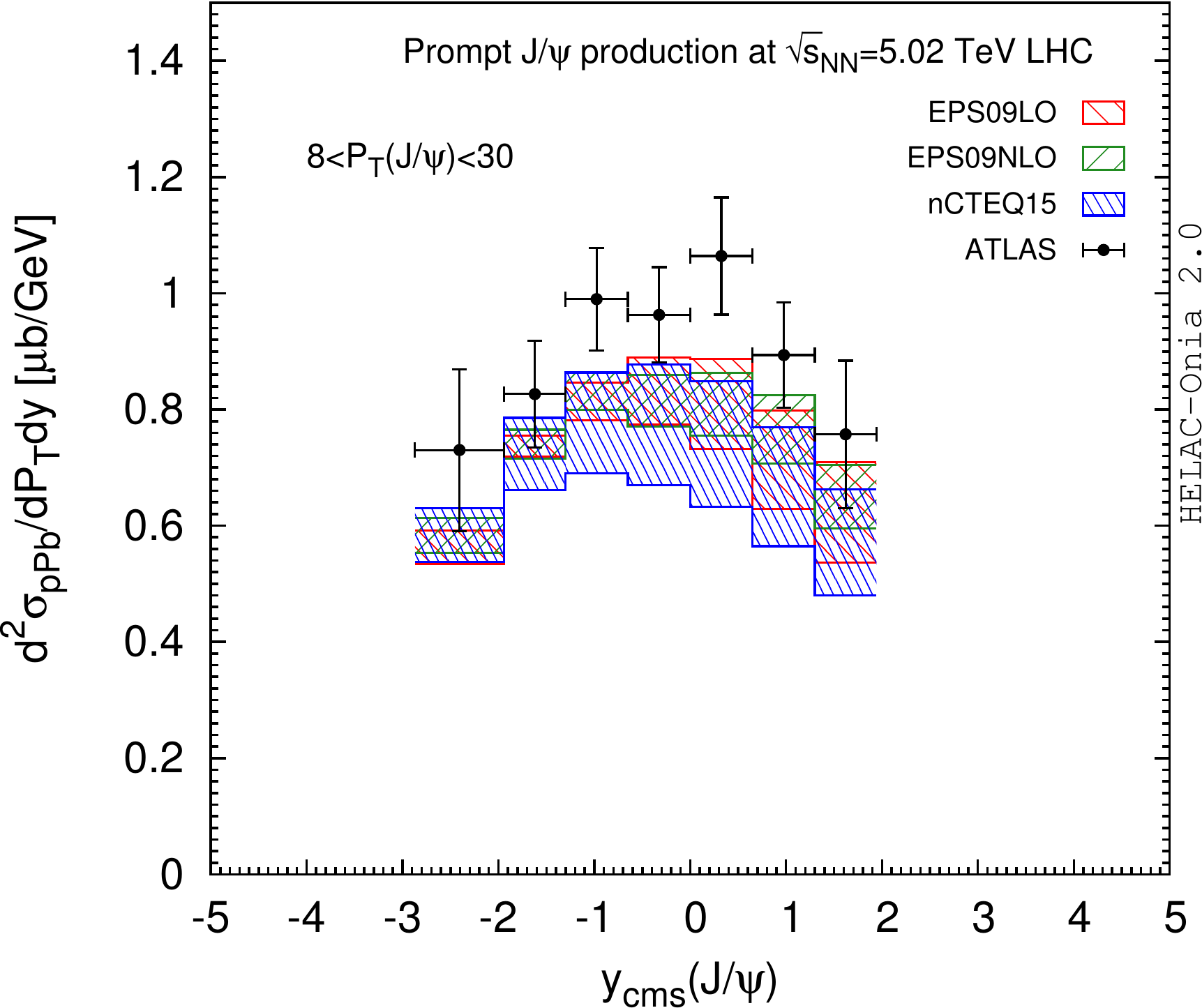}\label{fig:yjpsisigmapPbc}}
\caption{Rapidity dependence of the cross-section for prompt $J/\psi$ production in $p$Pb collisions at $\sqrt{s_{NN}}=5.02$ TeV: comparison between our results and the measurements of LHCb~\cite{Aaij:2013zxa}, ALICE~\cite{Abelev:2013yxa} and ATLAS~\cite{Aad:2015ddl}.  %[The uncertainty bands represent the nuclear PDF uncertainty only].
\label{fig:yjpsisigmapPb}}
\end{center}
\end{figure}

\begin{figure}[H]
\begin{center} 
\subfloat[]{\includegraphics[width=0.33\textwidth,keepaspectratio]{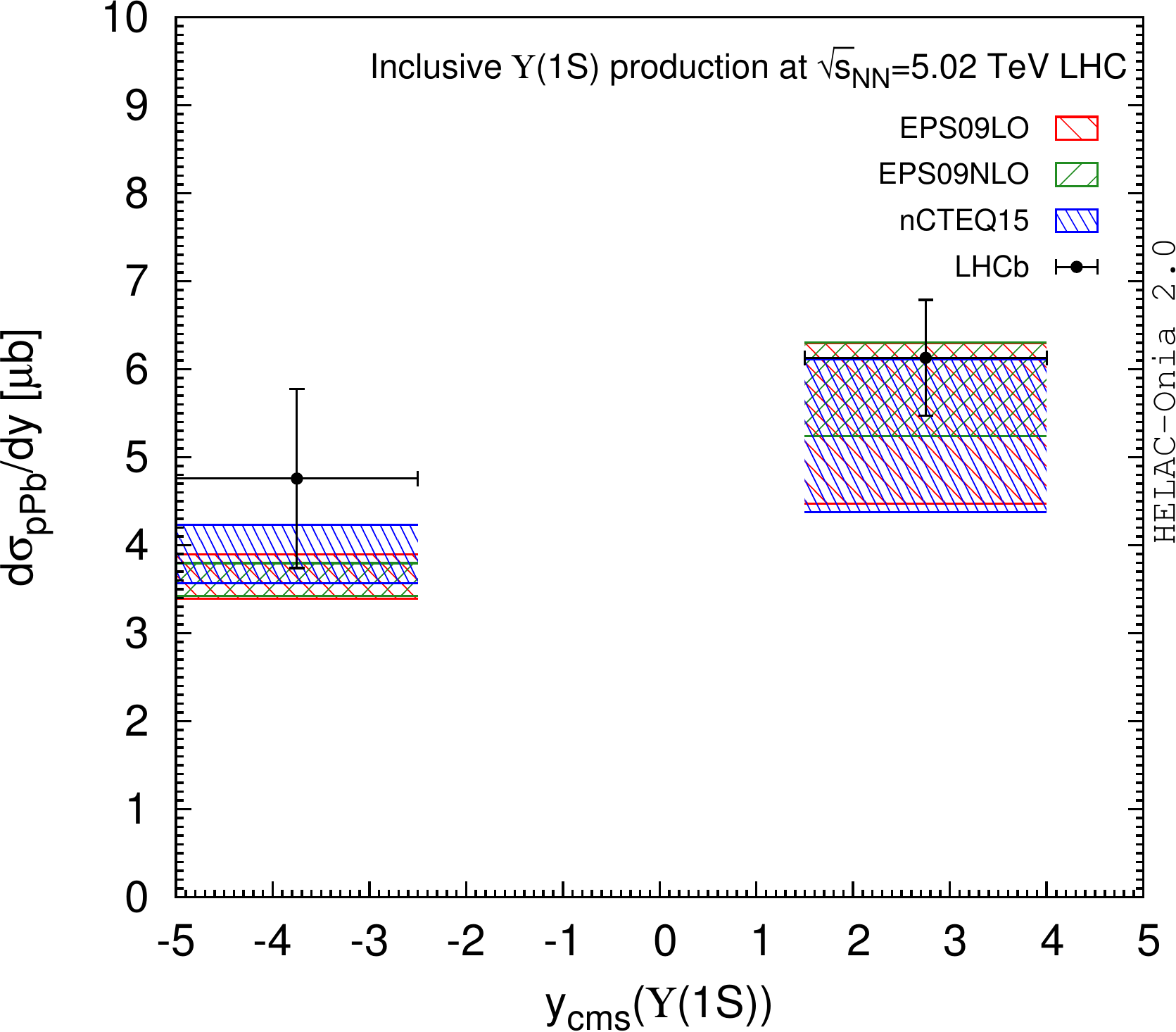}\label{fig:yy1ssigmapPba}}
\subfloat[]{\includegraphics[width=0.33\textwidth,keepaspectratio]{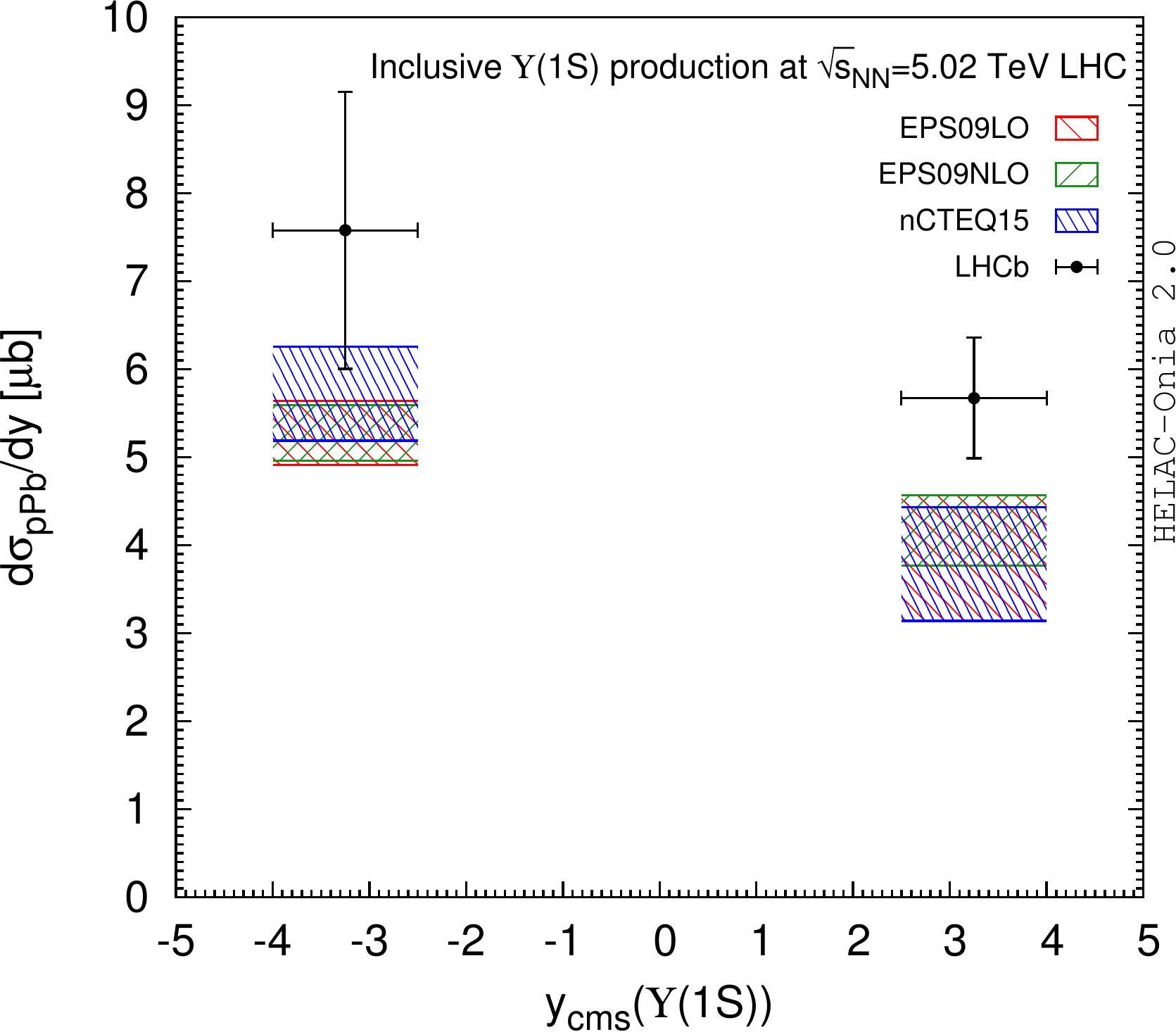}\label{fig:yy1ssigmapPbb}}
\subfloat[]{\includegraphics[width=0.33\textwidth,keepaspectratio]{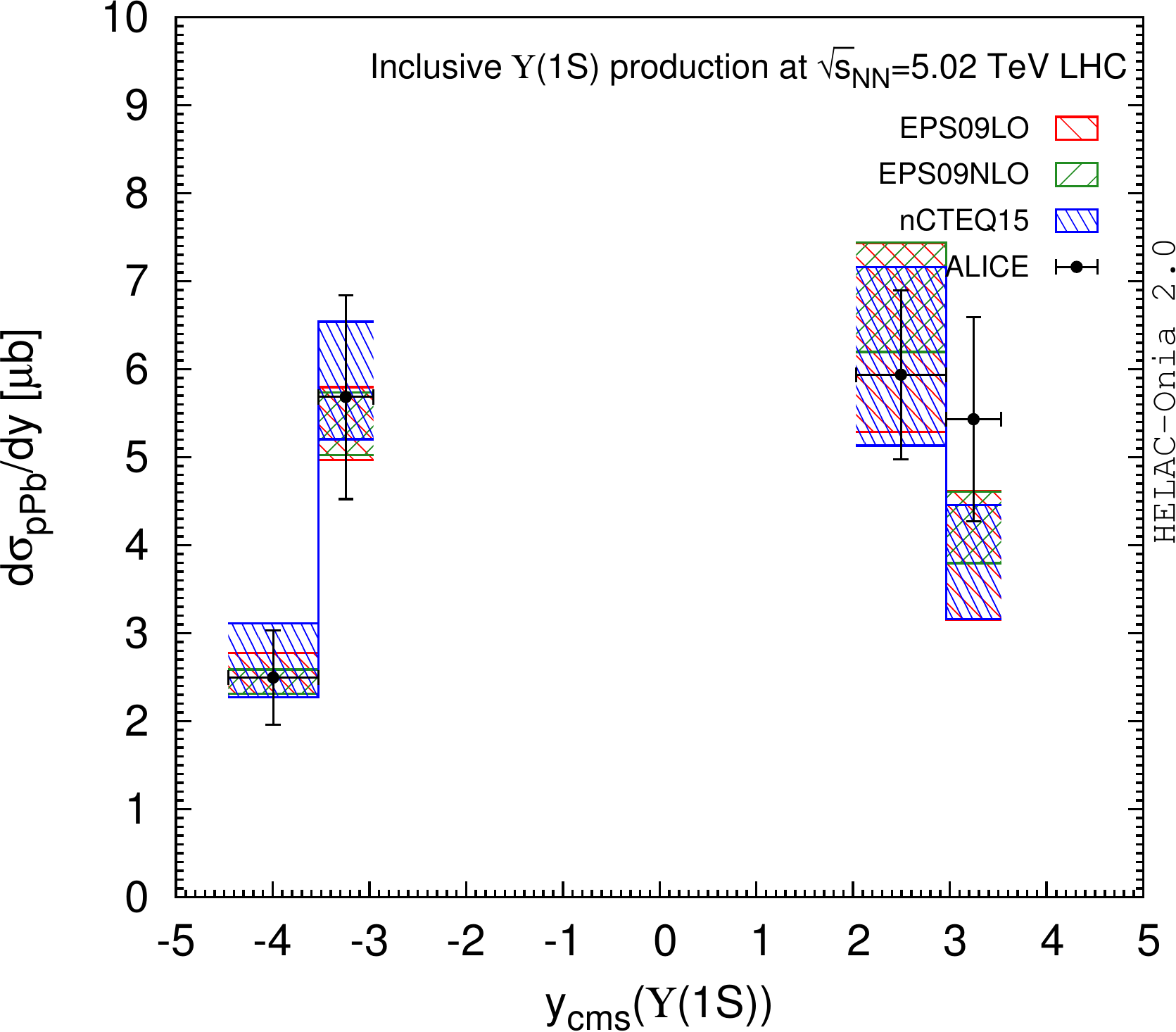}\label{fig:yy1ssigmapPbc}}\\
\subfloat[]{\includegraphics[width=0.33\textwidth,keepaspectratio]{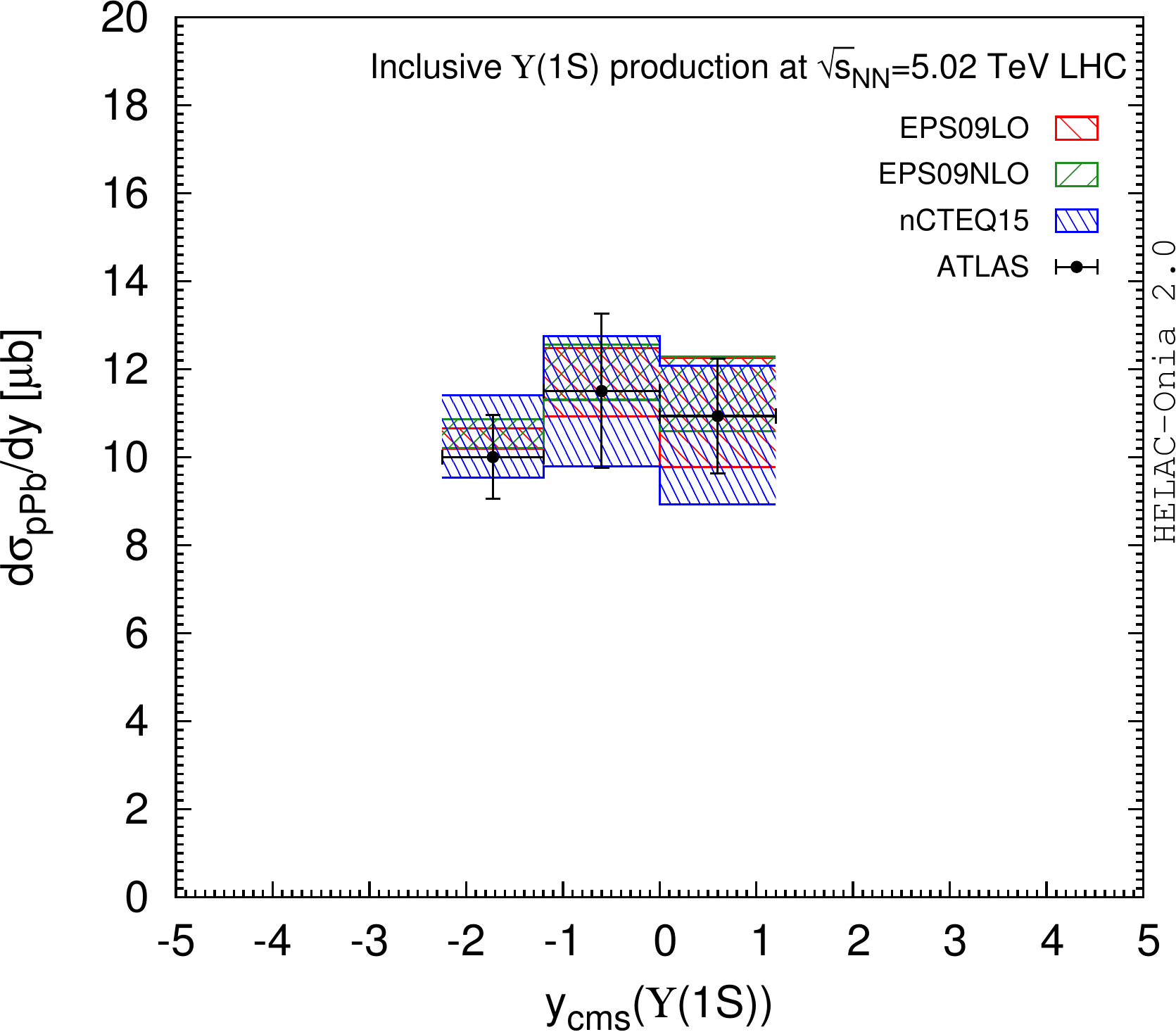}\label{fig:yy1ssigmapPbd}}
\subfloat[]{\includegraphics[width=0.33\textwidth,keepaspectratio]{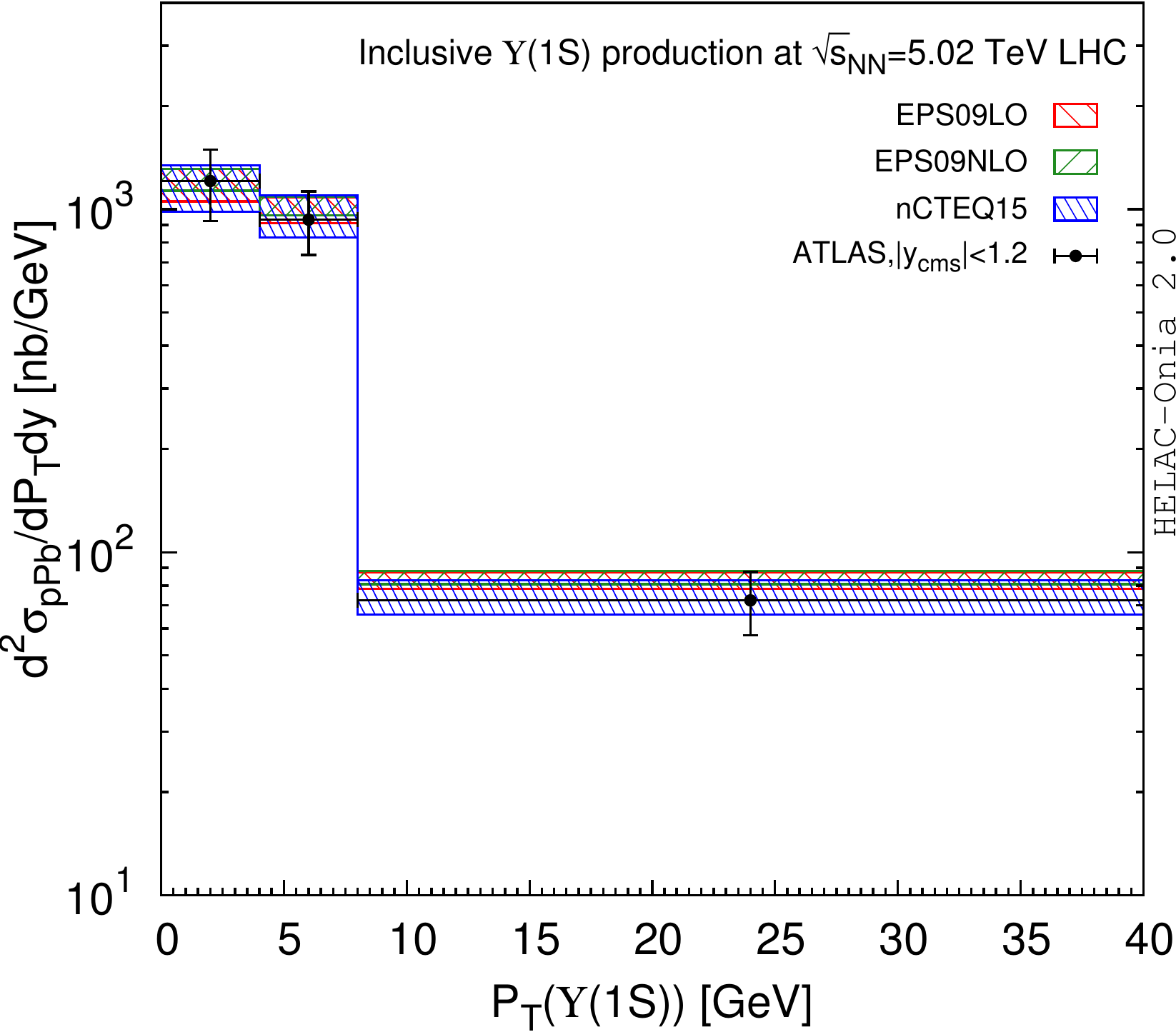}\label{fig:pty1ssigmapPb}}
\caption{Differential cross-section for inclusive $\Upsilon(1S)$ production in $p$Pb collisions at $\sqrt{s_{NN}}=5.02$ TeV: comparison of (a-d) the rapidity dependence obtained with our procedure with the measurements by LHCb~\cite{Aaij:2014mza}, ALICE~\cite{Abelev:2014oea} and ATLAS~\cite{ATLAS-CONF-2015-050} and (e) the transverse-momentum dependence as measured by ATLAS~\cite{ATLAS-CONF-2015-050}.
%[The uncertainty bands represent the nuclear PDF uncertainty only.]
\label{fig:y1ssigmapPb}}
\end{center}
\end{figure}

In Fig.~\ref{fig:yjpsisigmapPb}, we have compared the LHCb~\cite{Aaij:2013zxa}, ALICE~\cite{Abelev:2013yxa} and ATLAS data~\cite{Aad:2015ddl} with the $J/\psi$ cross-section differential in $y$. It is interesting to notice that the results with the three nPDF show different uncertainties. In the forward region (low $x_2$), the result with EPS09NLO has the smallest uncertainty and tend to overshoot the LHCb data~\cite{Aaij:2013zxa} (see Fig.~\ref{fig:yjpsisigmapPba}). Such a discrepancy does not appear in Fig.~\ref{fig:yjpsisigmapPbb}. One can also note that the  EPS09LO uncertainty can be considered as the combination of both EPS09NLO and nCTEQ15 uncertainties in the forward region. In the backward region, owing to both the significant experimental and nPDF uncertainties, the three nPDFs are compatible with the data. 
At high $P_T$ (in Fig.~\ref{fig:yjpsisigmapPbc} for the ATLAS data~\cite{Aad:2015ddl}), although the central values of the experimental data are systematically higher than our theoretical bands, they remain compatible within one standard deviation. There could indeed be an overestimation of the nPDF suppression in this region or an offset in the ATLAS data.

As for the $\Upsilon(1S)$, \cf{fig:yy1ssigmapPba} and \ref{fig:yy1ssigmapPbb} 
show comparisons with the LHCb data. The agreement is better when the full LHCb
 range is considered as opposed to that when the LHCb acceptance is restricted 
to a range where equal positive and negative $y$ can be accessed 
(\cf{fig:yy1ssigmapPbb}). A good agreement is also obtained with the ALICE data
(\cf{fig:yy1ssigmapPbc}) in a similar rapidity domain. In the ATLAS acceptance,
all three nPDF magnitudes correctly account for the yield differential in $y$
and $P_T$ (\cf{fig:yy1ssigmapPbd} and~\ref{fig:pty1ssigmapPb}).

\cf{fig:D0sigmapPb-1} and \cf{fig:D0sigmapPb-2} show comparisons for the $D^0$
 case between our results for the 3 nPDFs and the LHCb and ALICE measurements. The 
agreement is overall good. The yields tend to lie on the upper half of the 
uncertainty band. The nPDF uncertainties are however larger than for the 
quarkonia owing to the smaller value of the factorisation scale. As for the 
discrepancy in the first $P_T$ bin of~\cf{fig:D0sigmapPb-1}, one should be 
careful that our $pp$ parametrisation is not optimal to describe it as well 
(see \cf{fig:D0-pp}) and tend
to undershoot the $pp$ yield.

\begin{figure}[H]
\begin{center} 
\subfloat[]{\includegraphics[width=0.33\textwidth,keepaspectratio]{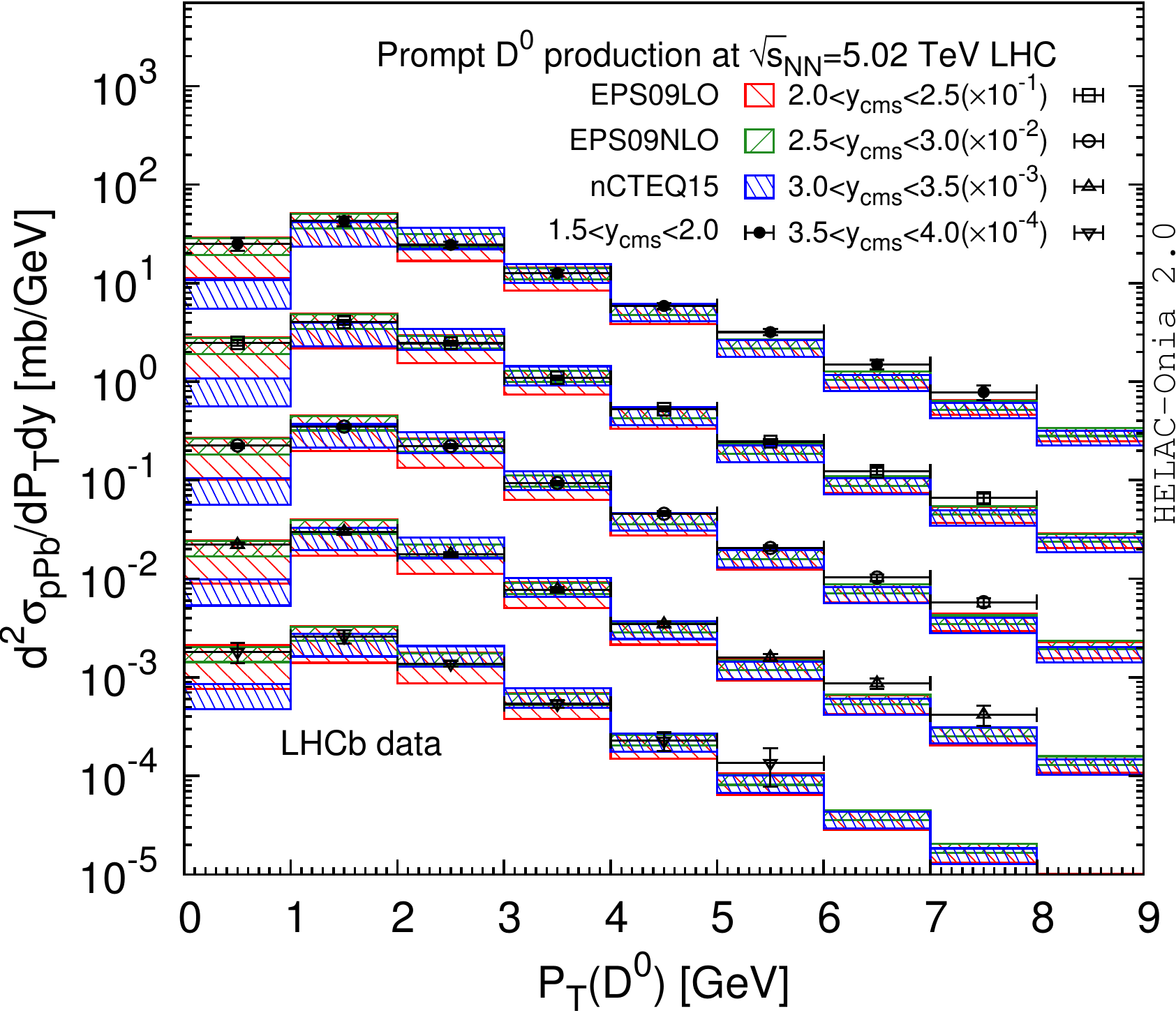}}
\subfloat[]{\includegraphics[width=0.33\textwidth,keepaspectratio]{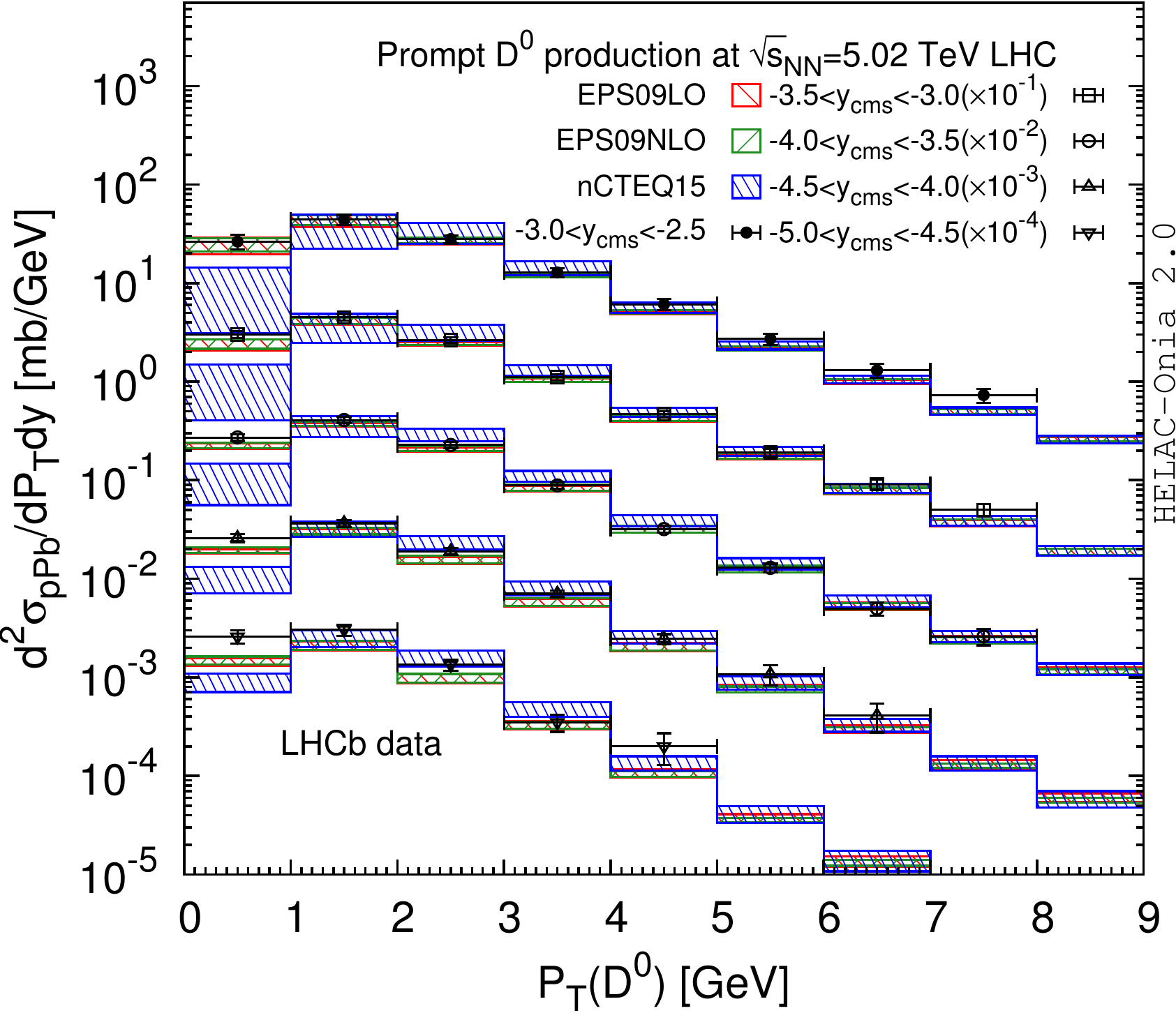}}
\subfloat[]{\includegraphics[width=0.33\textwidth,keepaspectratio]{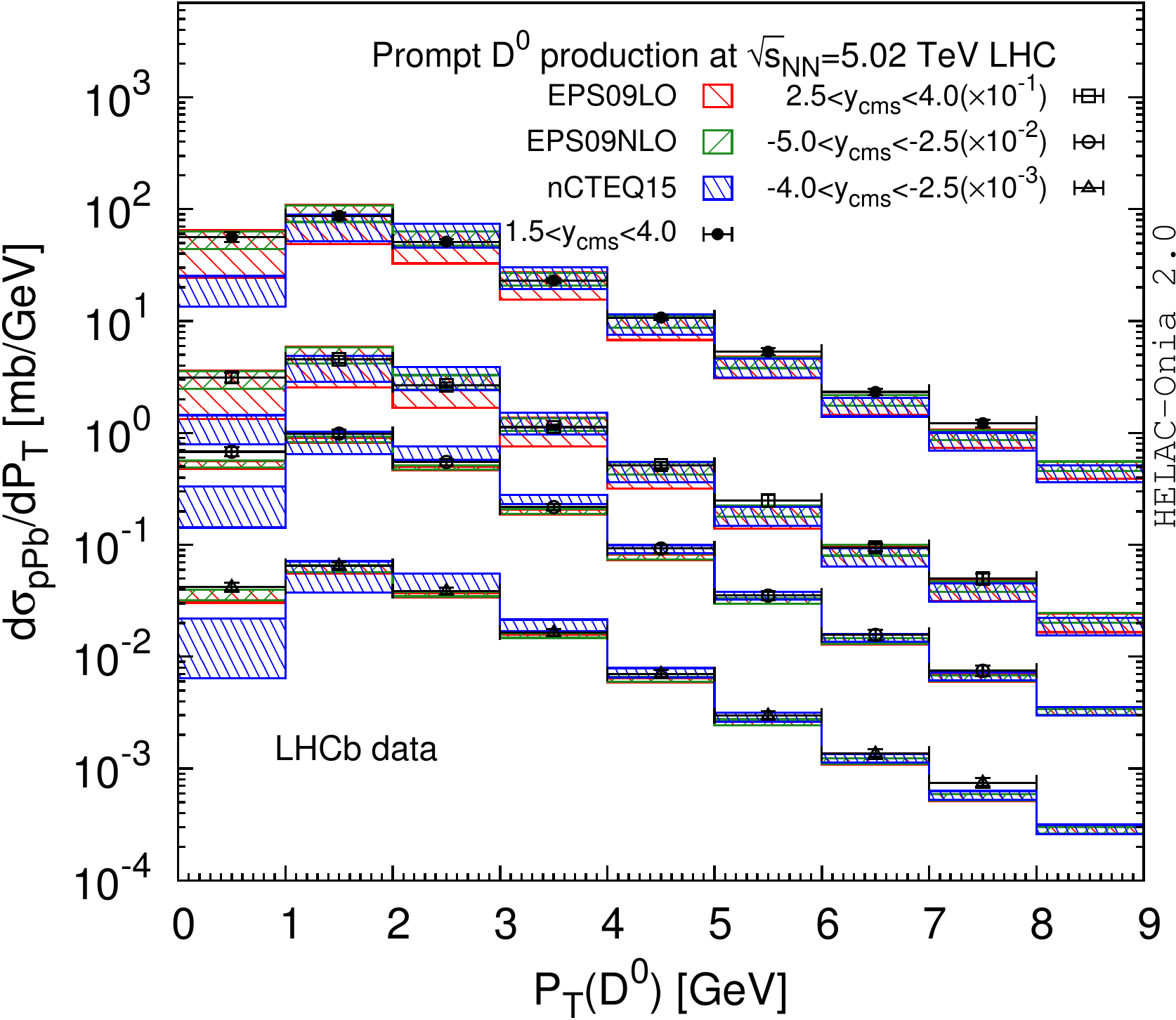}}
\caption{Transverse-momentum dependence of the production cross-section of promptly produced $D^0$ in $p$Pb collisions at $\sqrt{s_{NN}}=5.02$ TeV: comparison between our results and the measurements by LHCb~\cite{LHCb-CONF-2016-003}. 
%[The uncertainty bands represent the nuclear PDF uncertainty only.]
\label{fig:D0sigmapPb-1}}
\end{center}
\end{figure}

\begin{figure}[H]
\begin{center} 
\subfloat[LHCb~\cite{LHCb-CONF-2016-003}]{\includegraphics[width=0.33\textwidth,keepaspectratio]{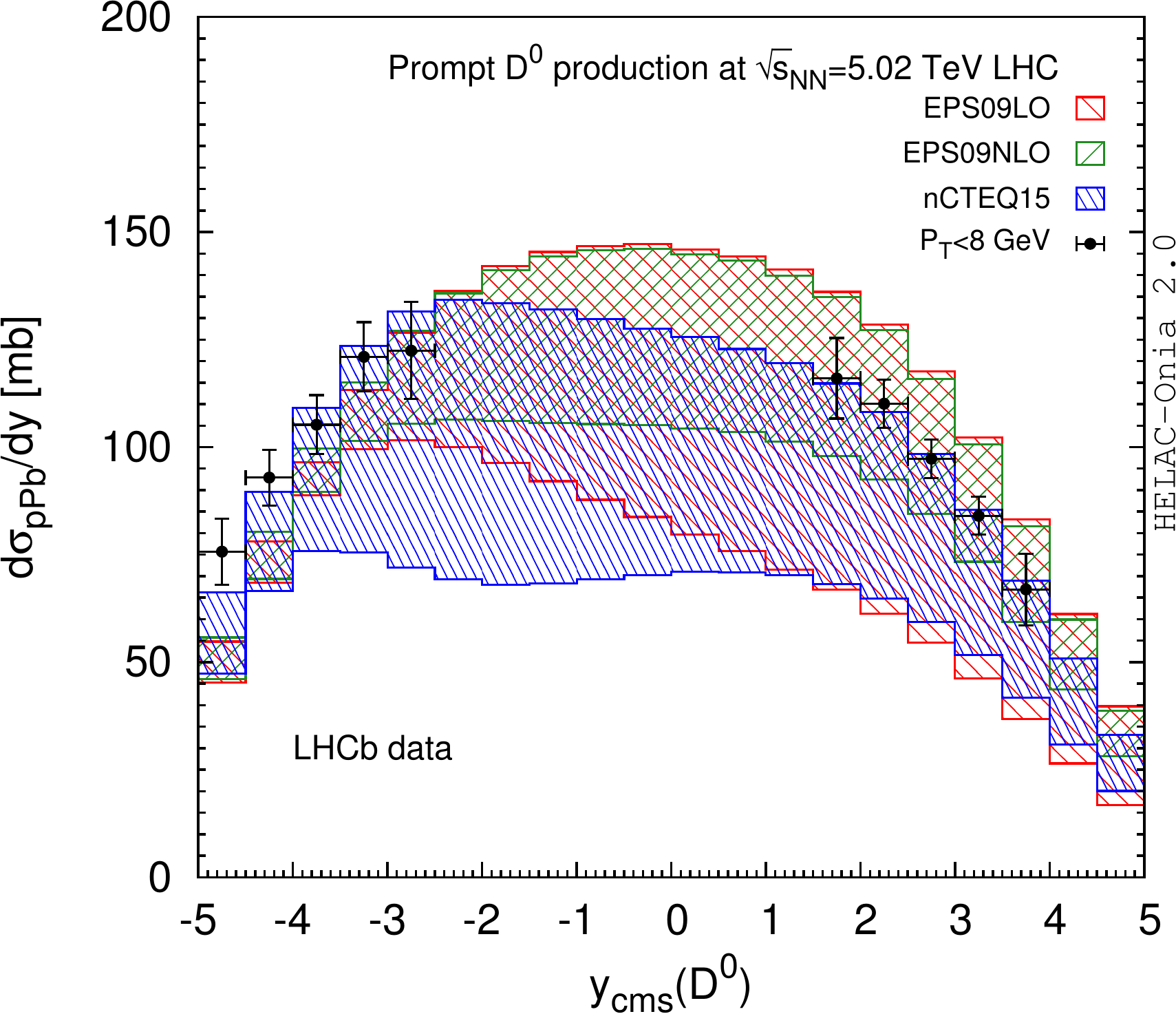}}
\subfloat[ALICE~\cite{Abelev:2014hha}]{\includegraphics[width=0.33\textwidth,keepaspectratio]{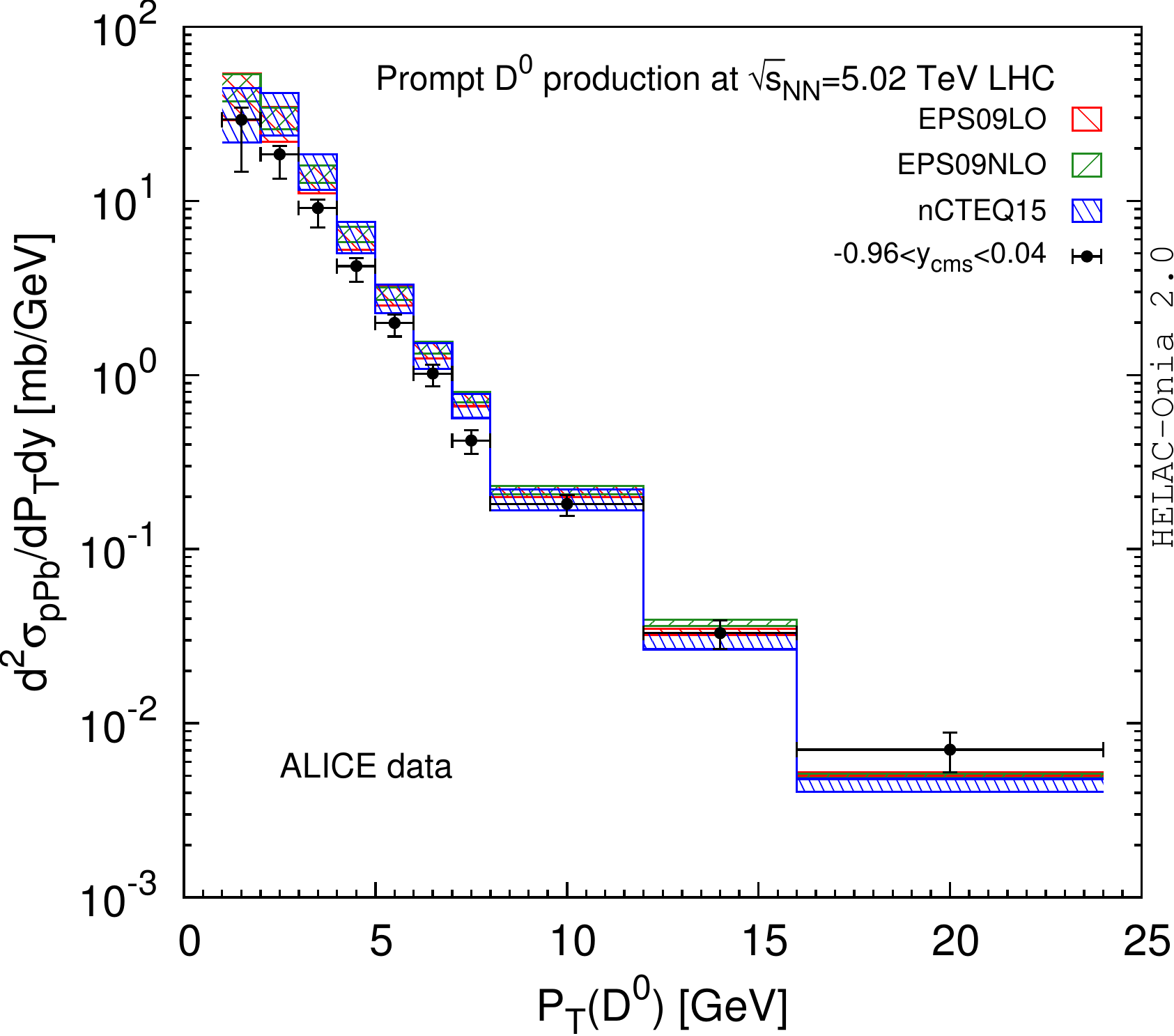}}
\caption{(a) Rapidity [(b) Transverse-momentum] dependence of the cross-section for promptly produced $D^0$ in $p$Pb at $\sqrt{s_{NN}}=5.02$ TeV:  comparison between our results and the measurements by  LHCb~\cite{LHCb-CONF-2016-003}
[ALICE~\cite{Abelev:2014hha}]. 
%[The uncertainty bands represent the nuclear PDF uncertainty only.]\label{fig:yD0sigmapPb}
}\label{fig:D0sigmapPb-2}
\end{center}
\end{figure}

Finally, \cf{fig:etacsigmapPb} show predictions --the first ever in the litterature-- for the $P_T$ and $y$ differential yield of $\eta_c$ in the LHCb acceptance.

\begin{figure}[H]
\begin{center} 
\subfloat[]{\includegraphics[width=0.33\textwidth,keepaspectratio]{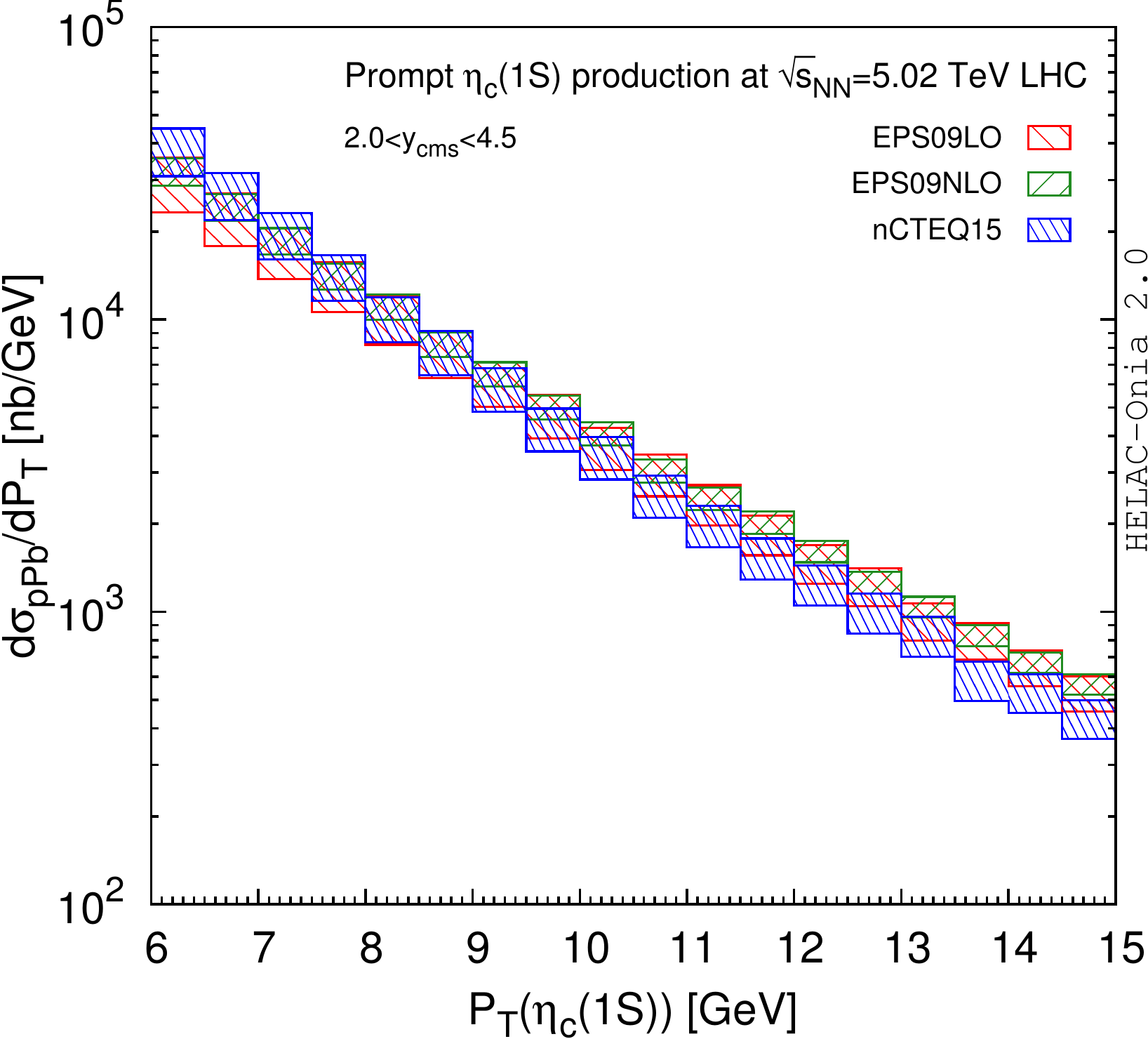}}
\subfloat[]{\includegraphics[width=0.33\textwidth,keepaspectratio]{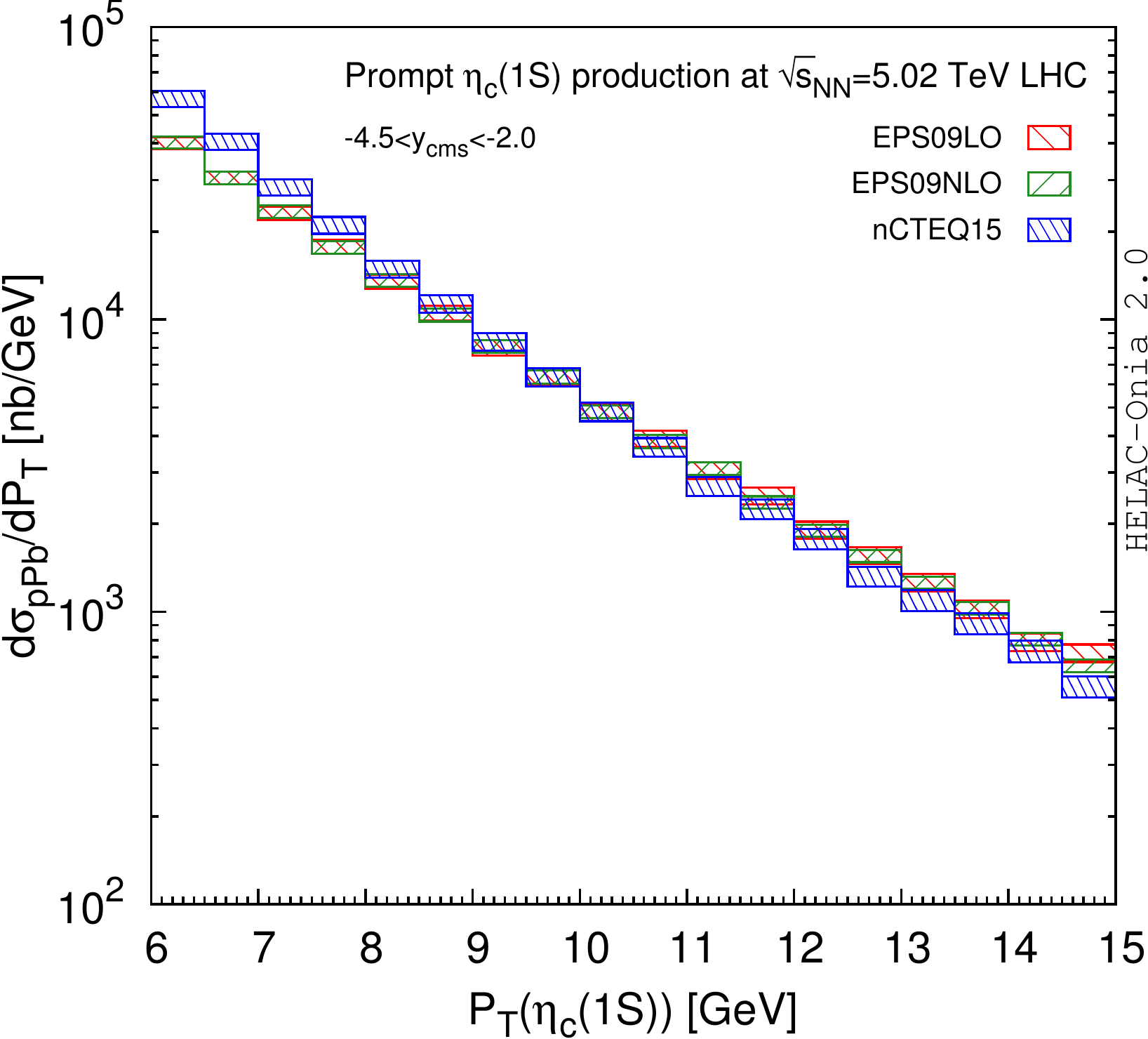}}
\subfloat[]{\includegraphics[width=0.33\textwidth,keepaspectratio]{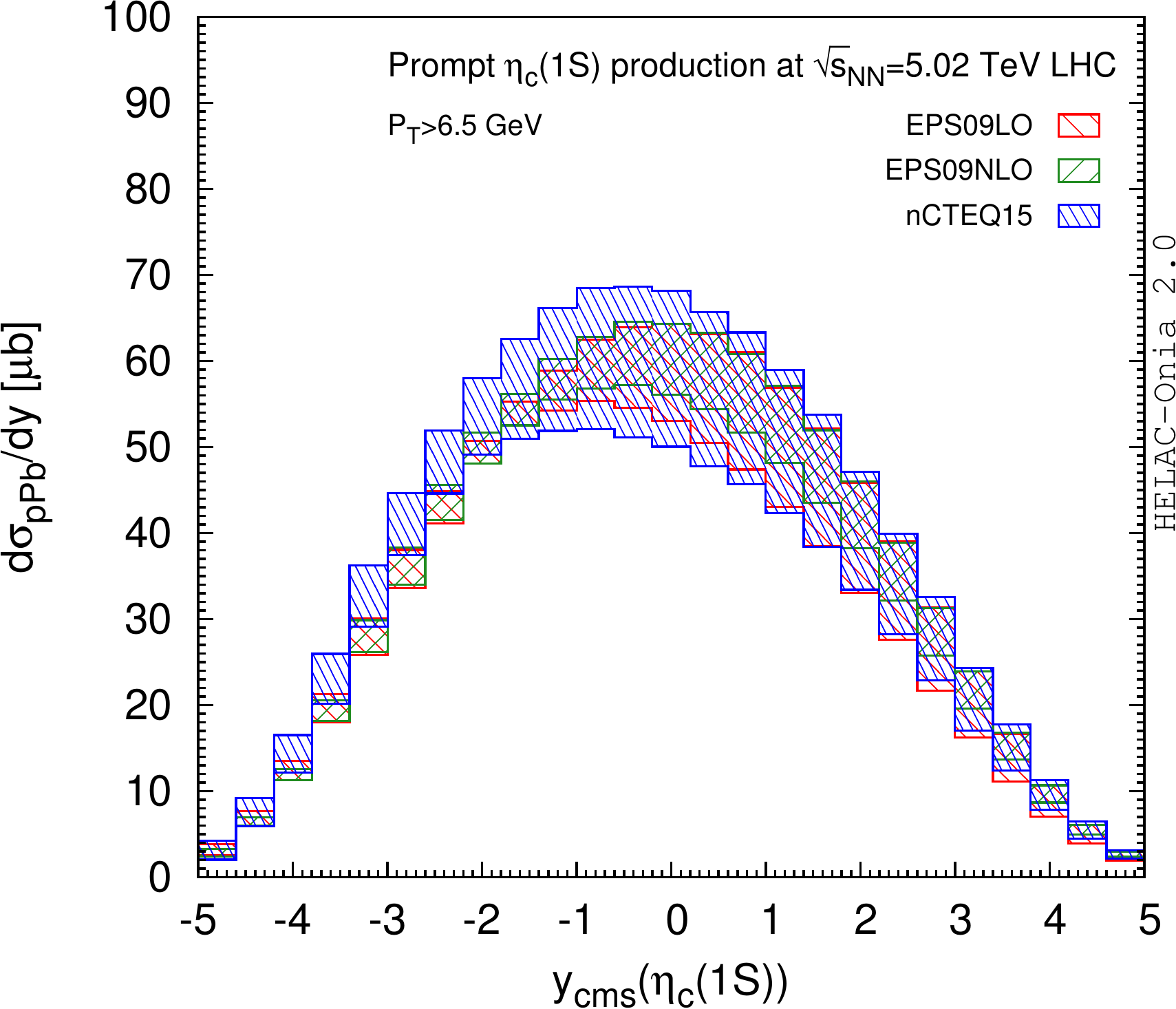}}
\caption{(a-b) Transverse-momentum ((c) rapidity) dependence of the productioncross-section of prompt $\eta_c(1S)$ in $p$Pb collisions at $\sqrt{s_{NN}}=5.02$ TeV. [The uncertainty bands represent the nuclear PDF uncertainty only.].\label{fig:etacsigmapPb}}
\end{center}
\end{figure}

\subsection{Rapidity and transverse-momentum dependence of \RpPb\ at $\sqrt{s_{NN}}=5.02$~TeV}

We now present and discuss our results for the \emph{nuclear modification factor} \RpPb\  which characterises the yield 
modification of a given probe, say ${\cal H}$, in $p$Pb collisions relative to $pp$ collisions. It is the ratio obtained by normalising the $\cal H$ yield in \pPb\ collisions to the $\cal H$ yield in $pp$ collisions in the same kinematical conditions ($y$, $P_T$, nucleon-nucleon energy, etc.) times the average number of binary 
inelastic nucleon-nucleon collisions. When mininum bias collisions are considered, that
is when all the possible geometrical configurations are summed over, it simplifies to the ratio of cross sections
corrected by the atomic number of the nucleus ($A=208$ for Pb):
\begin{equation}
\RpPb= \frac{d\sigma^{\cal H}_{p\rm Pb}}{A d\sigma^{\cal H}_{pp}}.
\end{equation}

\begin{figure}[H]
\begin{center} 
\subfloat[]{\includegraphics[width=0.33\textwidth,keepaspectratio]{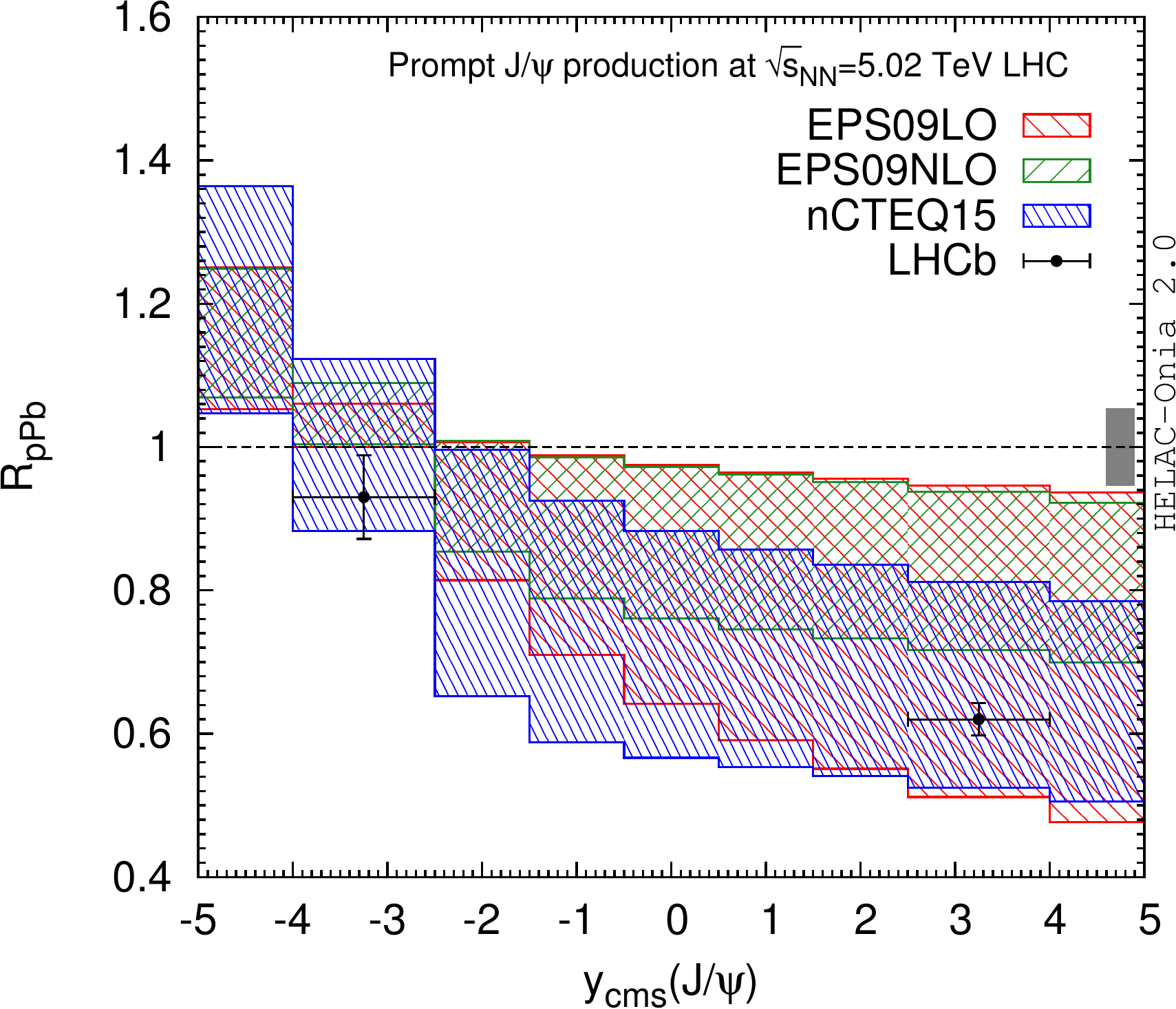}\label{fig:yjpsiRpPba}}
\subfloat[]{\includegraphics[width=0.33\textwidth,keepaspectratio]{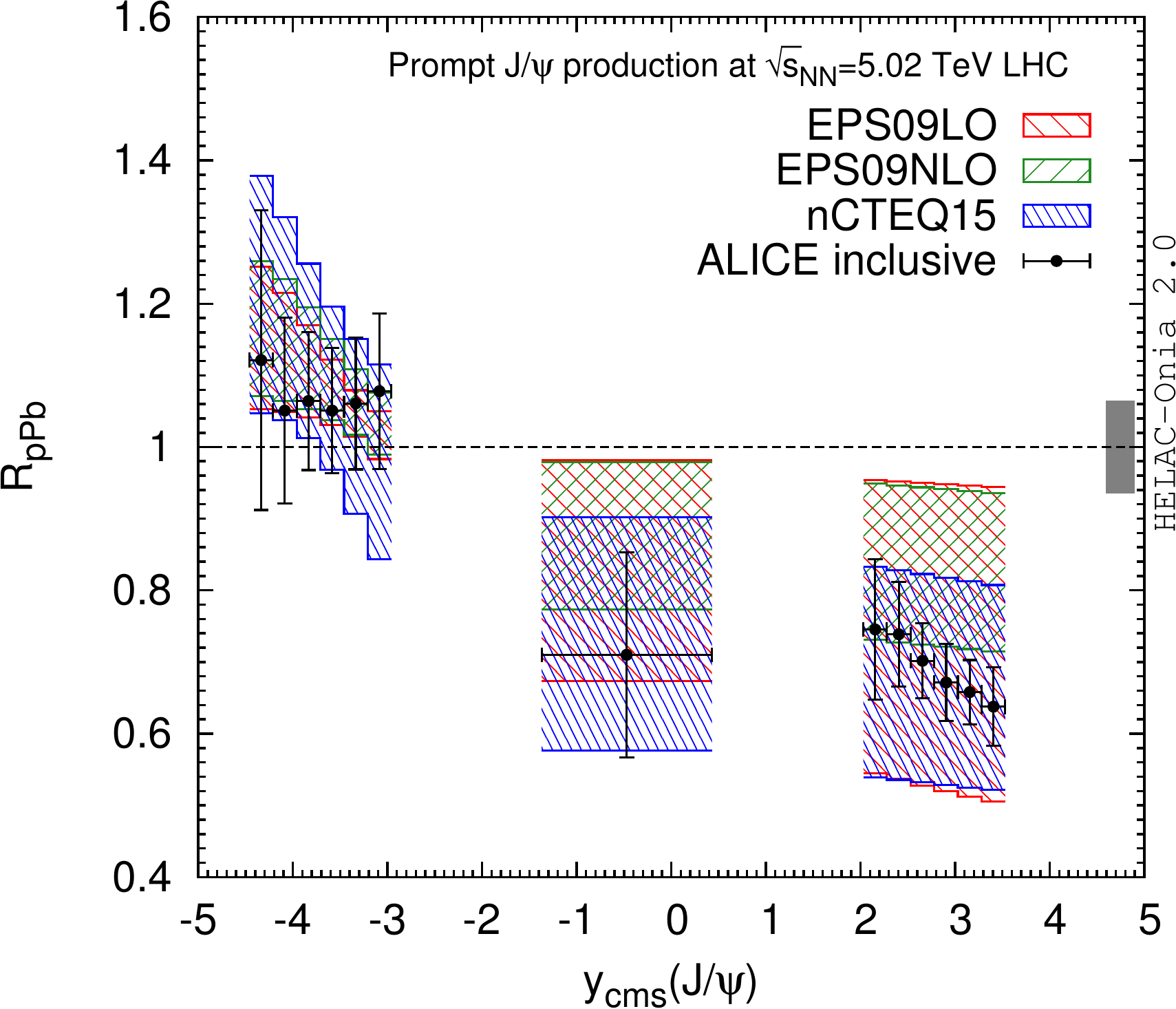}\label{fig:yjpsiRpPbb}}
\subfloat[]{\includegraphics[width=0.33\textwidth,keepaspectratio]{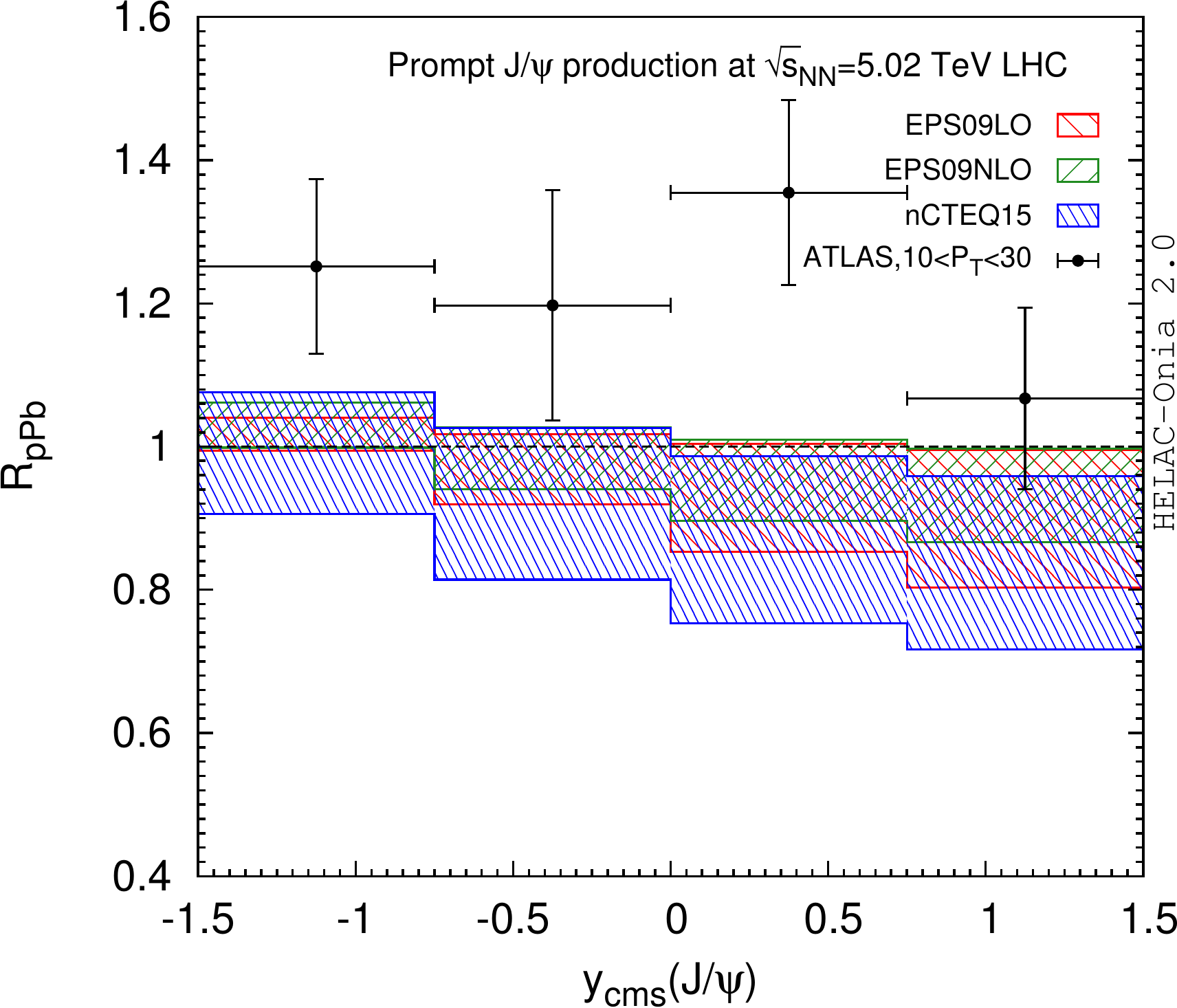}\label{fig:yjpsiRpPbc}}
\caption{Rapidity dependence of \RpPb\ of prompt $J/\psi$ in \pPb\ collisions at $\sqrt{s_{NN}}=5.02$ TeV: comparison between our results and the measurements by LHCb~\cite{Aaij:2013zxa}, ALICE~\cite{Abelev:2013yxa,Adam:2015iga} and ATLAS~\cite{ATLAS:2015pua}. 
%[The uncertainty bands represent the nuclear PDF uncertainty only.]
\label{fig:yjpsiRpPb}}
\end{center}
\end{figure}

We first discuss the rapidity dependence of \RpPb\ at the LHC with $\sqrt{s_{NN}}=5.02$ TeV for $J/\psi$ production. Our results  obtained for the three nPDFs, EPS09LO, EPS09NLO and nCTEQ15 with their associated uncertainties are compared in Fig.~\ref{fig:yjpsiRpPb} to the different experiments. Fig.~\ref{fig:yjpsiRpPba} and Fig.~\ref{fig:yjpsiRpPbb} show low $P_T$ data~\cite{Aaij:2013zxa,Abelev:2013yxa,Adam:2015iga}. It is expected that the suppression in the forward region is due to the shadowing effect, while the enhancement in the backward region is due to the anti-shadowing effect. The experimental data are compatible with these expectations. Among the three different nPDFs, the data tend to favour the result obtained with nCTEQ15.

\begin{figure}[H]
\begin{center} 
\subfloat[]{\includegraphics[width=0.33\textwidth,keepaspectratio]{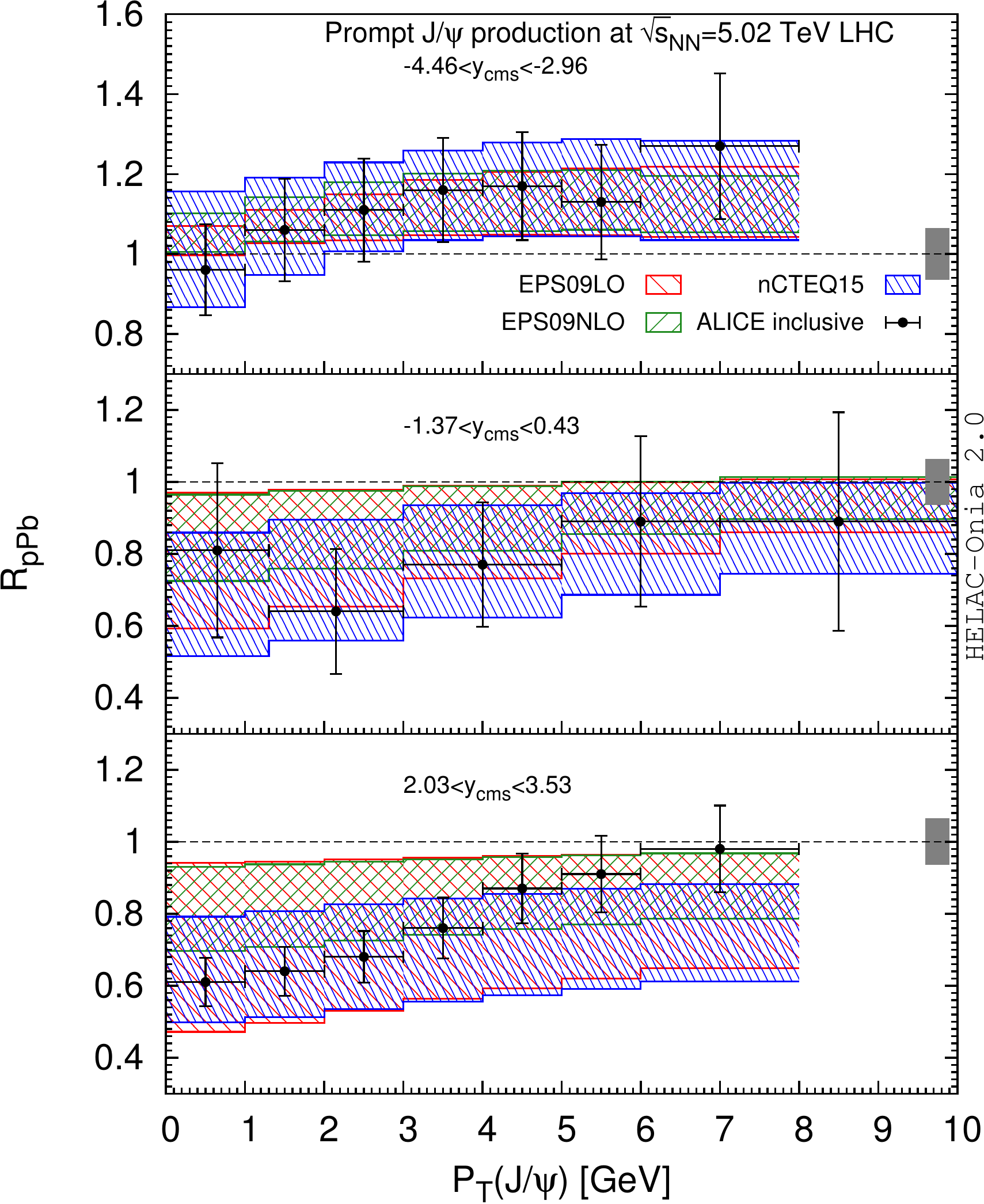}\label{fig:ptjpsiRpPba}}
\subfloat[]{\includegraphics[width=0.33\textwidth,keepaspectratio]{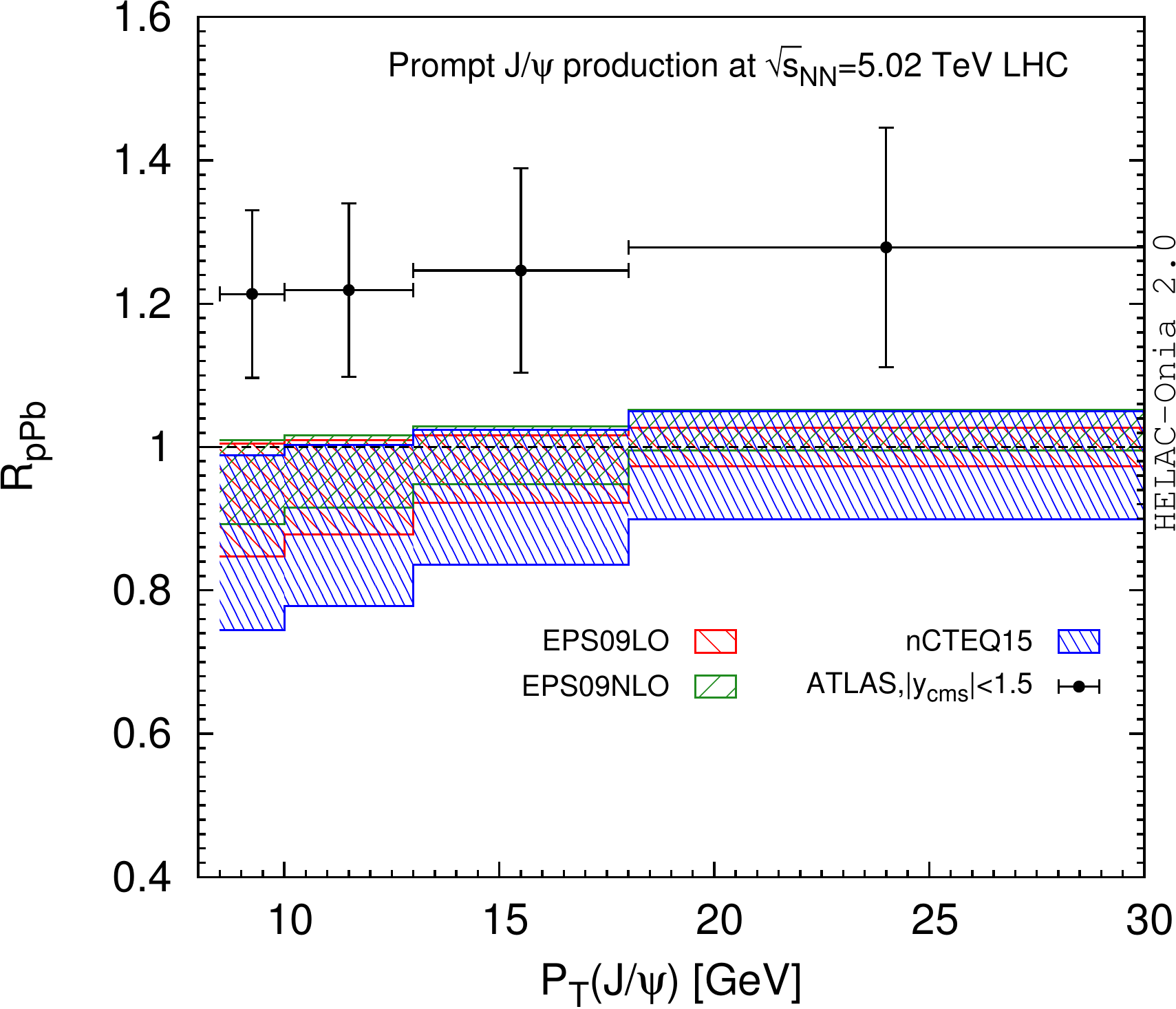}\label{fig:ptjpsiRpPbb}}
\subfloat[]{\includegraphics[width=0.33\textwidth,keepaspectratio]{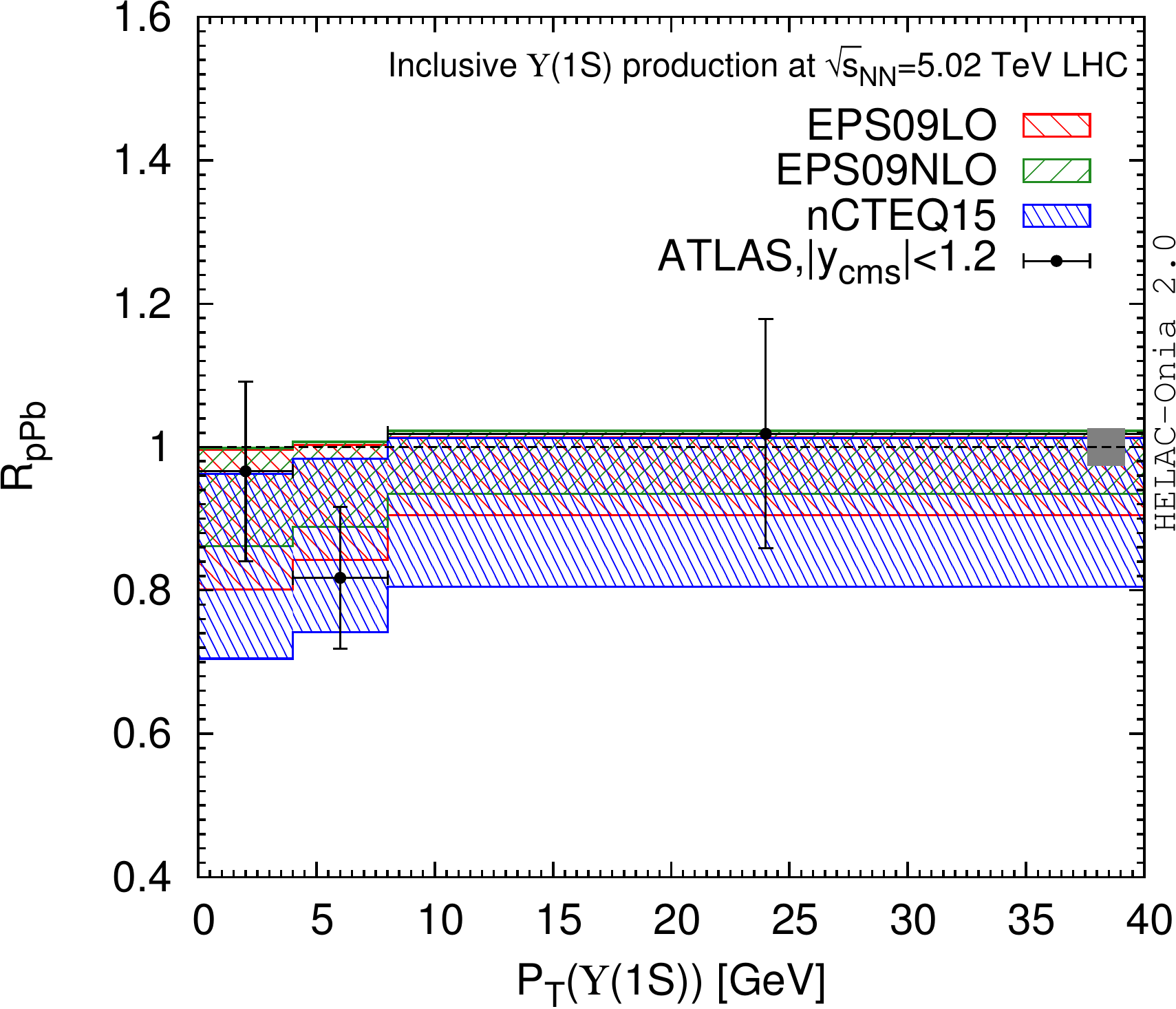}
\label{fig:pty1sRpPb}}
\caption{(a-b) [(c)] $P_T$ dependence of  \RpPb\ of prompt $J/\psi$ [inclusive $\Upsilon(1S)$] in \pPb\ collisions at $\sqrt{s_{NN}}=5.02$ TeV : comparison between our results and the measurements by ALICE~\cite{Adam:2015iga} and ATLAS~\cite{ATLAS:2015pua} [ATLAS~\cite{ATLAS-CONF-2015-050}]. %[The uncertainty bands represent the nuclear PDF uncertainty only.]
\label{fig:ptjpsiRpPb}}
\end{center}
\end{figure}

It is also interesting to note that the precision of the current data is already  better than the nPDF uncertainties, especially in the forward region. This gives some hope that these measurements could ultimately be used to constrain the gluon density in heavy ions, provided that the impact of other nuclear effects could be disentangled. We also note that the shaded boxes on the right of the first two plots refer to the global systematical uncertainty. Such an information is not available for the ATLAS data. A good agreement with the LHCb and ALICE data is obtained; a slight discrepancy with the ATLAS data is observed. It is not clear whether
it could be attributed to an offset in the data normalisation.
In Fig.~\ref{fig:ptjpsiRpPb}, we show further comparisons of \RpPb\ vs $P_T^{J/\psi}$ between our curves and the 
ALICE~\cite{Adam:2015iga} and ATLAS~\cite{ATLAS:2015pua} data. Similar to the rapidity distribution, a slight discrepancy is observed  in~\cf{fig:ptjpsiRpPbb}.

\vspace*{-1cm}

\vspace*{-1cm}

\begin{figure}[H]
\begin{center} 
\subfloat[]{\includegraphics[width=0.33\textwidth,keepaspectratio]{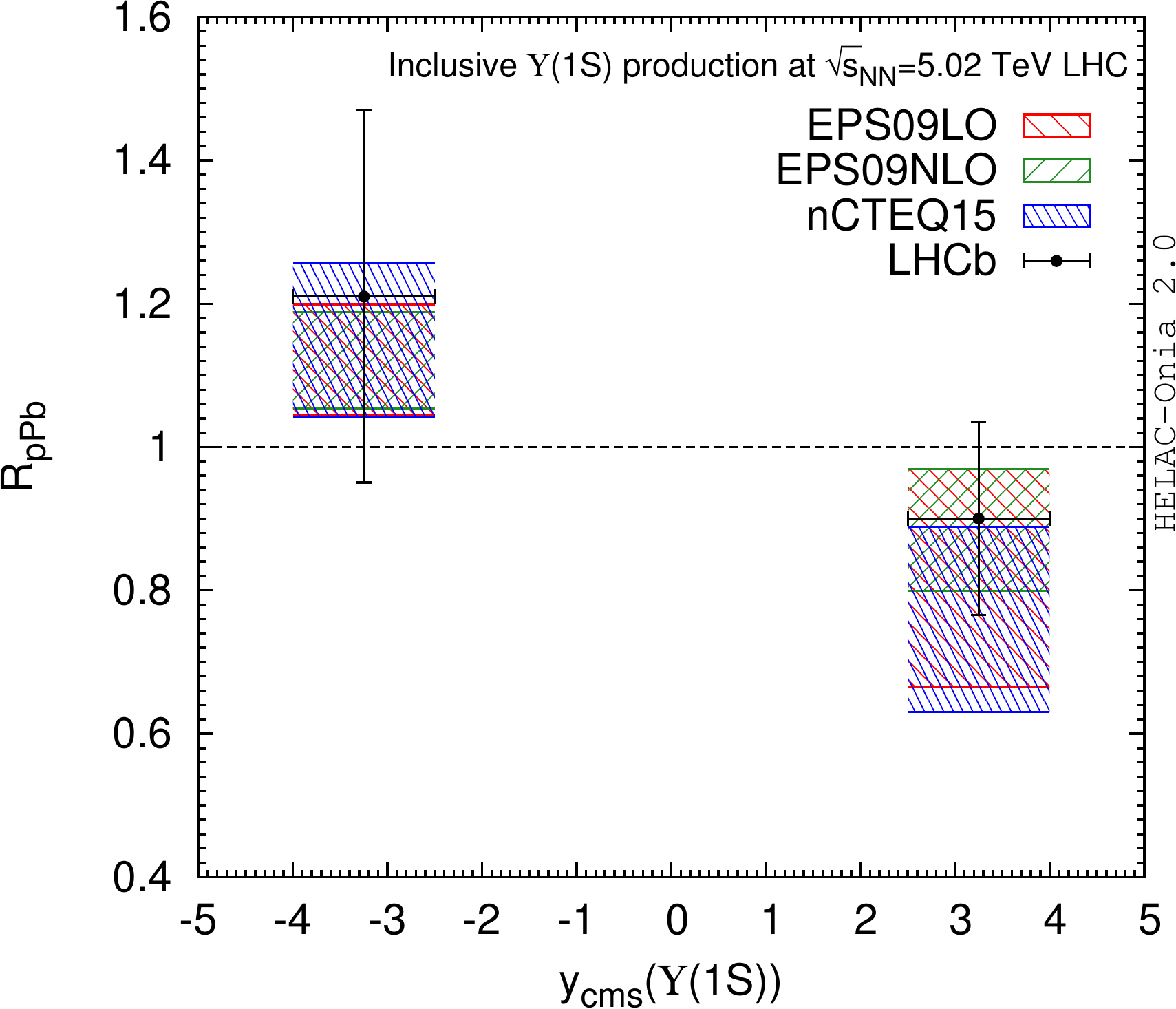}\label{fig:yy1sRpPba}}
\subfloat[]{\includegraphics[width=0.33\textwidth,keepaspectratio]{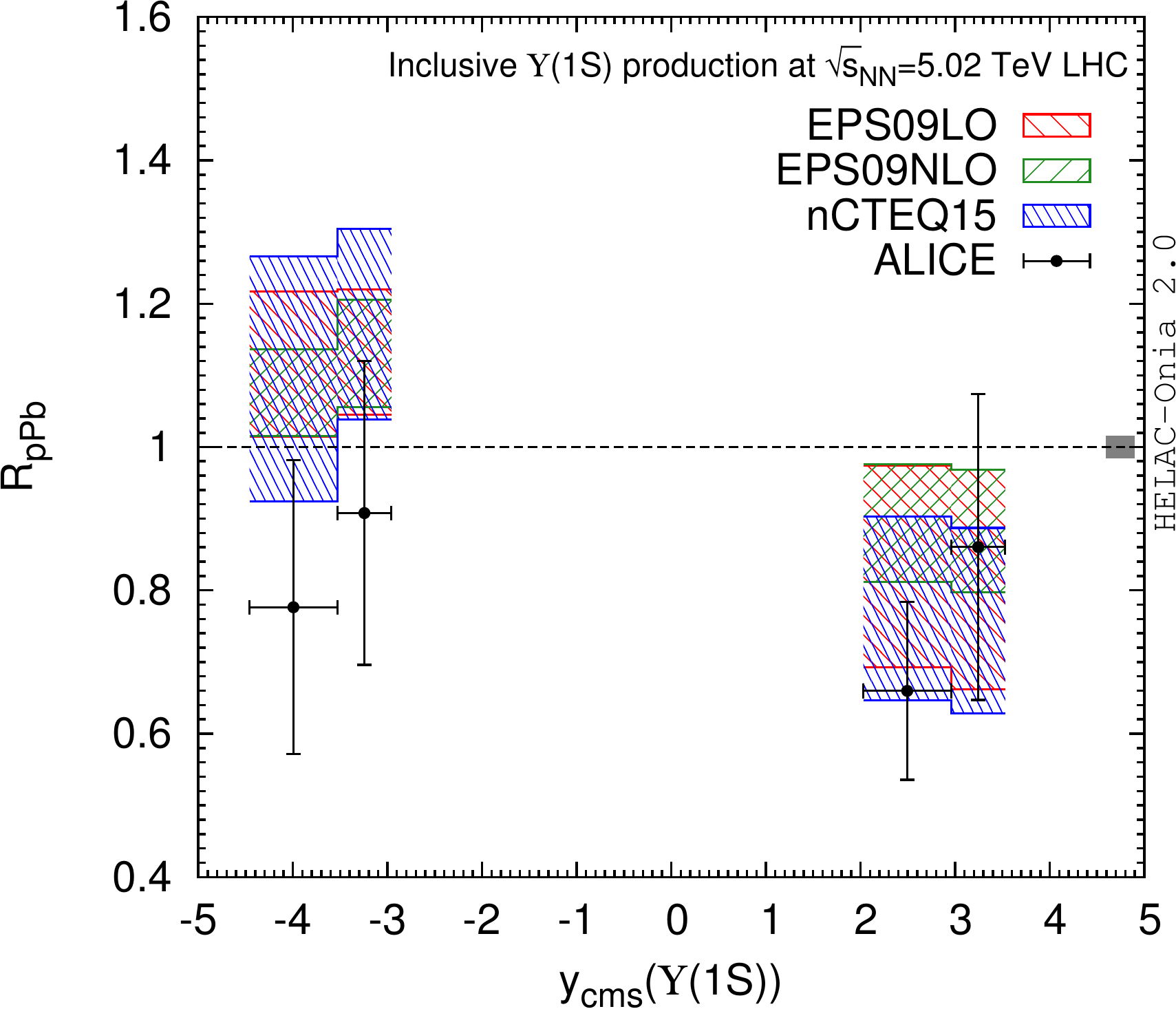}\label{fig:yy1sRpPbb}}
\subfloat[]{\includegraphics[width=0.33\textwidth,keepaspectratio]{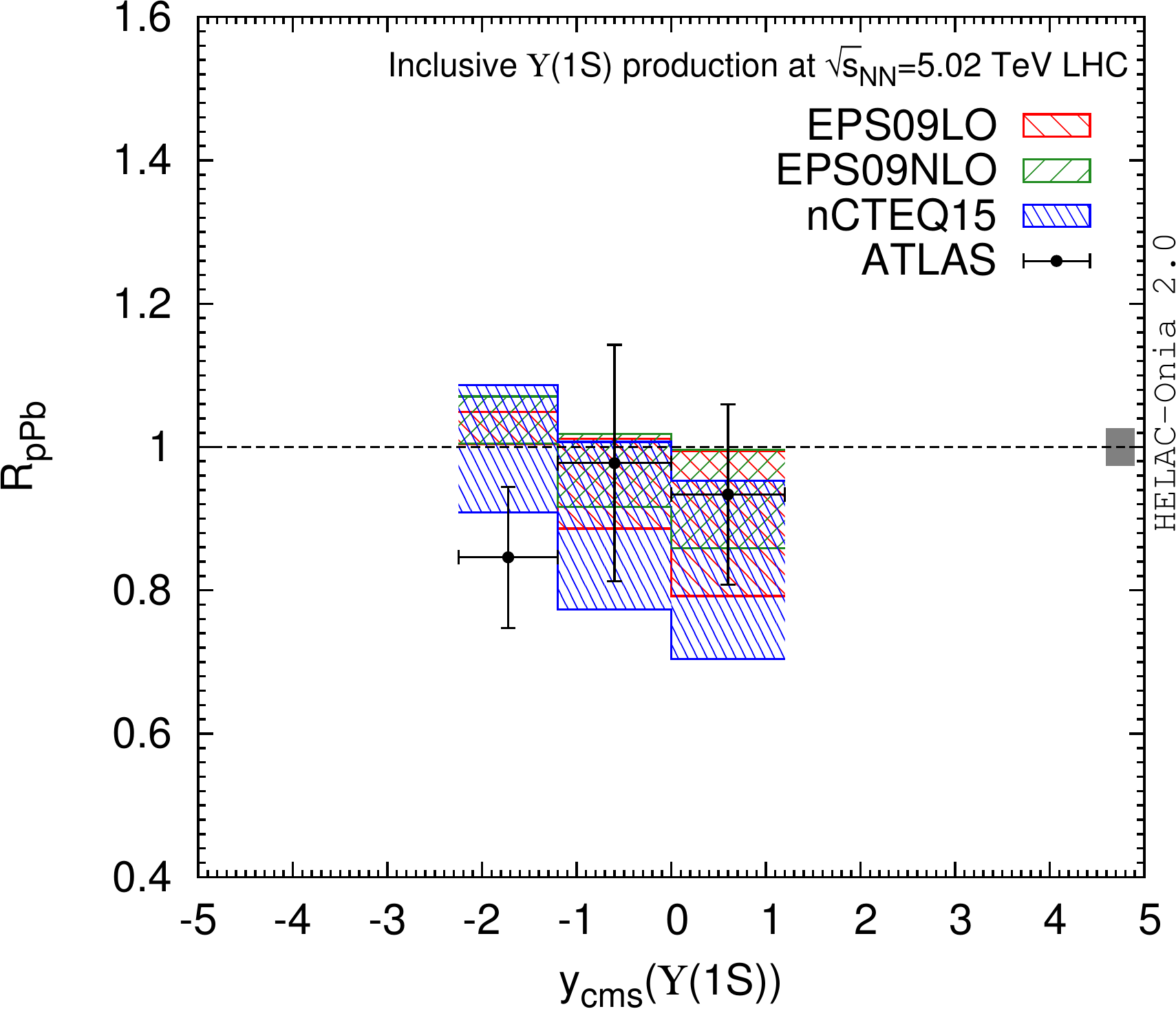}\label{fig:yy1sRpPbc}}
\caption{Rapidity dependence of \RpPb\ of inclusive $\Upsilon(1S)$ in \pPb\ collisions at $\sqrt{s_{NN}}=5.02$ TeV: comparison between our results and the measurements by LHCb~\cite{Aaij:2014mza}, ALICE~\cite{Abelev:2014oea} and ATLAS~\cite{ATLAS-CONF-2015-050}.
%[The uncertainty bands represent the nuclear PDF uncertainty only.]
\label{fig:yy1sRpPb}}
\end{center}
\end{figure}

Similar comparisons are shown for $\Upsilon(1S)$ on \cf{fig:yy1sRpPb} and \ref{fig:pty1sRpPb}. The overall agreement is acceptable given the large nPDF and experimental uncertainties.  Further comparisons with the $D^0$ results are presented on~\cf{fig:D0RpPb}. The agreement is also satisfactory and seems to indicate that EPS09 NLO is providing the best predictions. We however postpone further conclusions to the discussion of the $R_{\rm FB}$ results which however do not necessarily confirm this observation. To complete this exhaustive list of comparisons, we present our predictions for \RpPb\ of $\eta_c$ in the LHCb acceptance of its $pp$ analysis on \cf{fig:etacRpPb}. We are hopeful that it will motivate the first ever experimental studies of $\eta_c$  in \pPb\ collisions at the LHC.

\begin{figure}[H]
\begin{center} 
\subfloat[]{\includegraphics[width=0.33\textwidth,keepaspectratio]{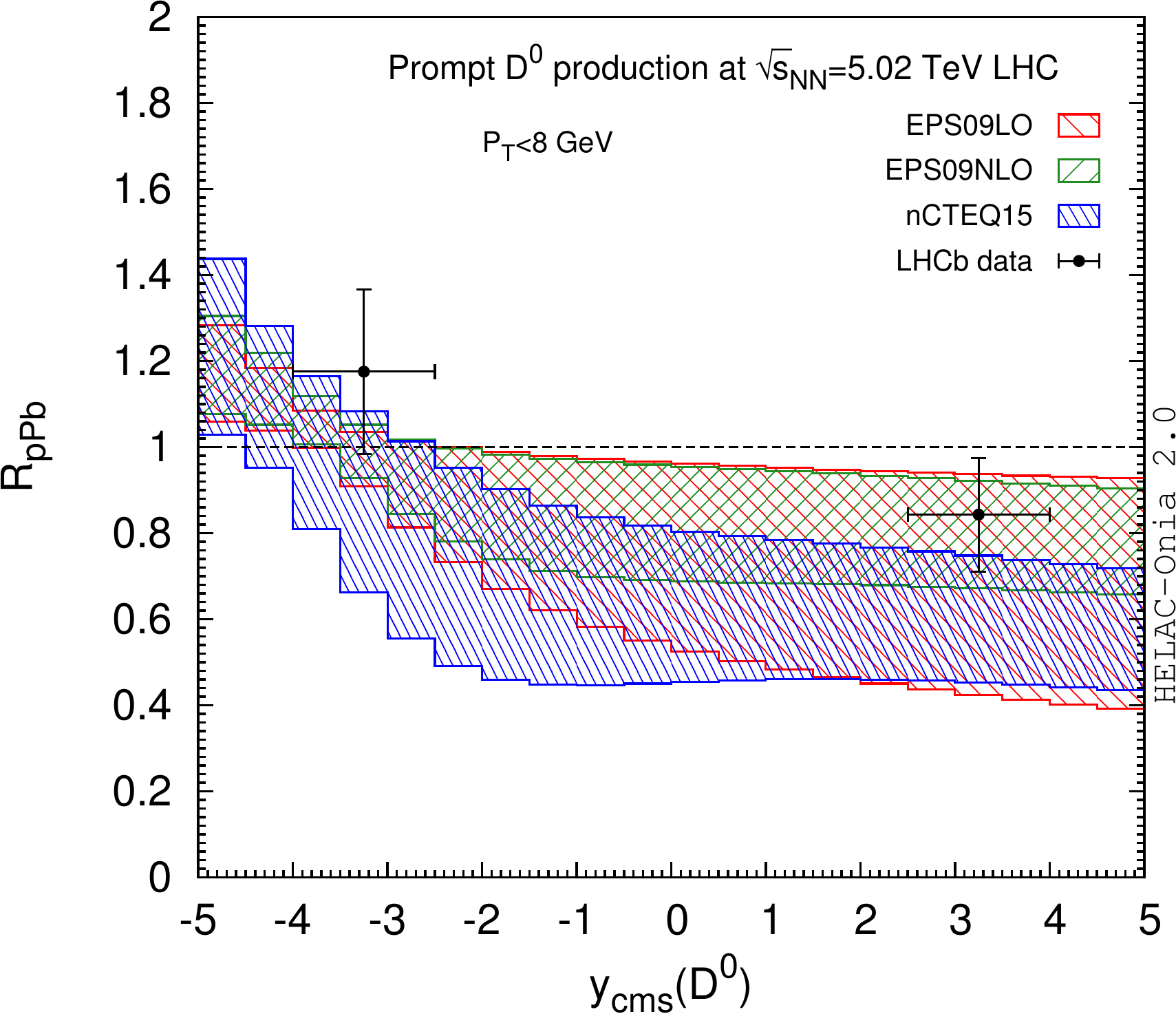}\label{fig:yD0RpPb}}
\subfloat[]{\includegraphics[width=0.25\textwidth,keepaspectratio]{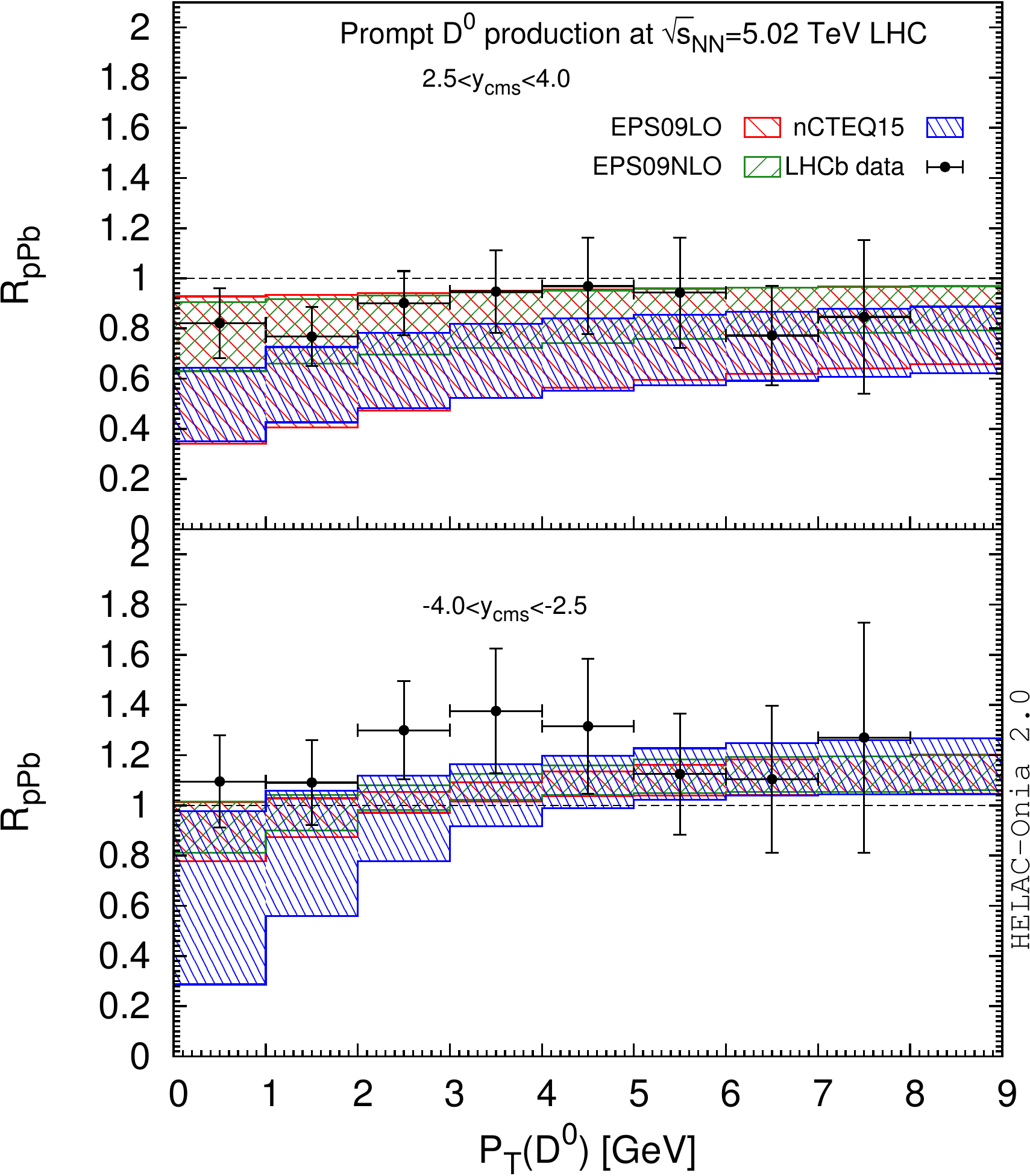}\label{fig:ptD0RpPb}}
\subfloat[]{\includegraphics[width=0.33\textwidth,keepaspectratio]{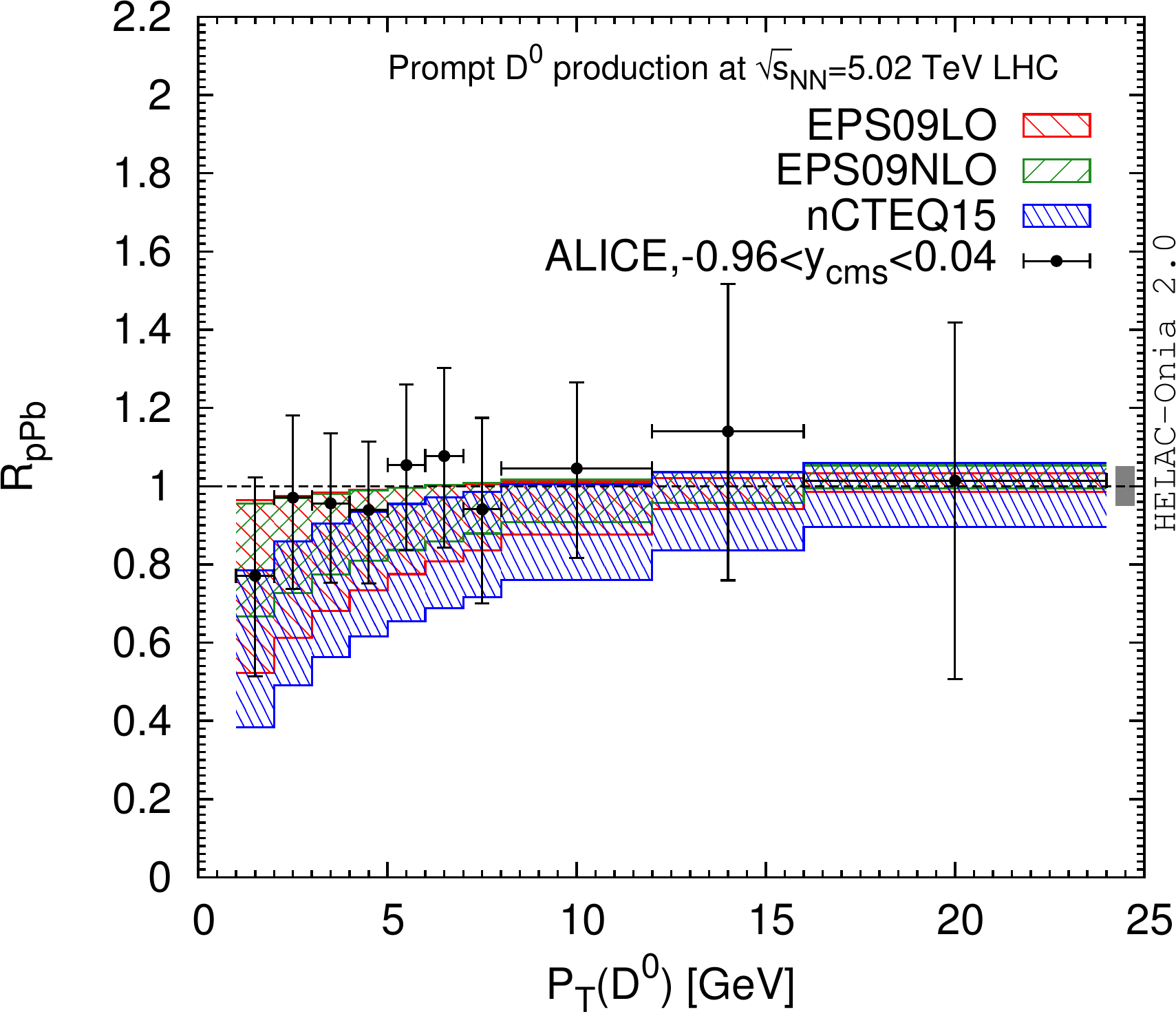}\label{fig:ptD0RpPb-2}}
\caption{Rapidity (a) and transverse-momentum (b-c) dependence of \RpPb\ of promptly produced $D^0$ in \pPb\ collisions at $\sqrt{s_{NN}}=5.02$ TeV: comparison between our results and the measurements by LHCb~\cite{LHCb-CONF-2016-003} (a-b) and ALICE~\cite{Abelev:2014hha} (c). 
%[The uncertainty bands represent the nuclear PDF uncertainty only.]
 \label{fig:D0RpPb}
}
\end{center}
\end{figure}

\vspace*{-1cm}
% eta_c

\begin{figure}[H]
\begin{center} 
\subfloat[]{\includegraphics[width=0.33\textwidth,keepaspectratio]{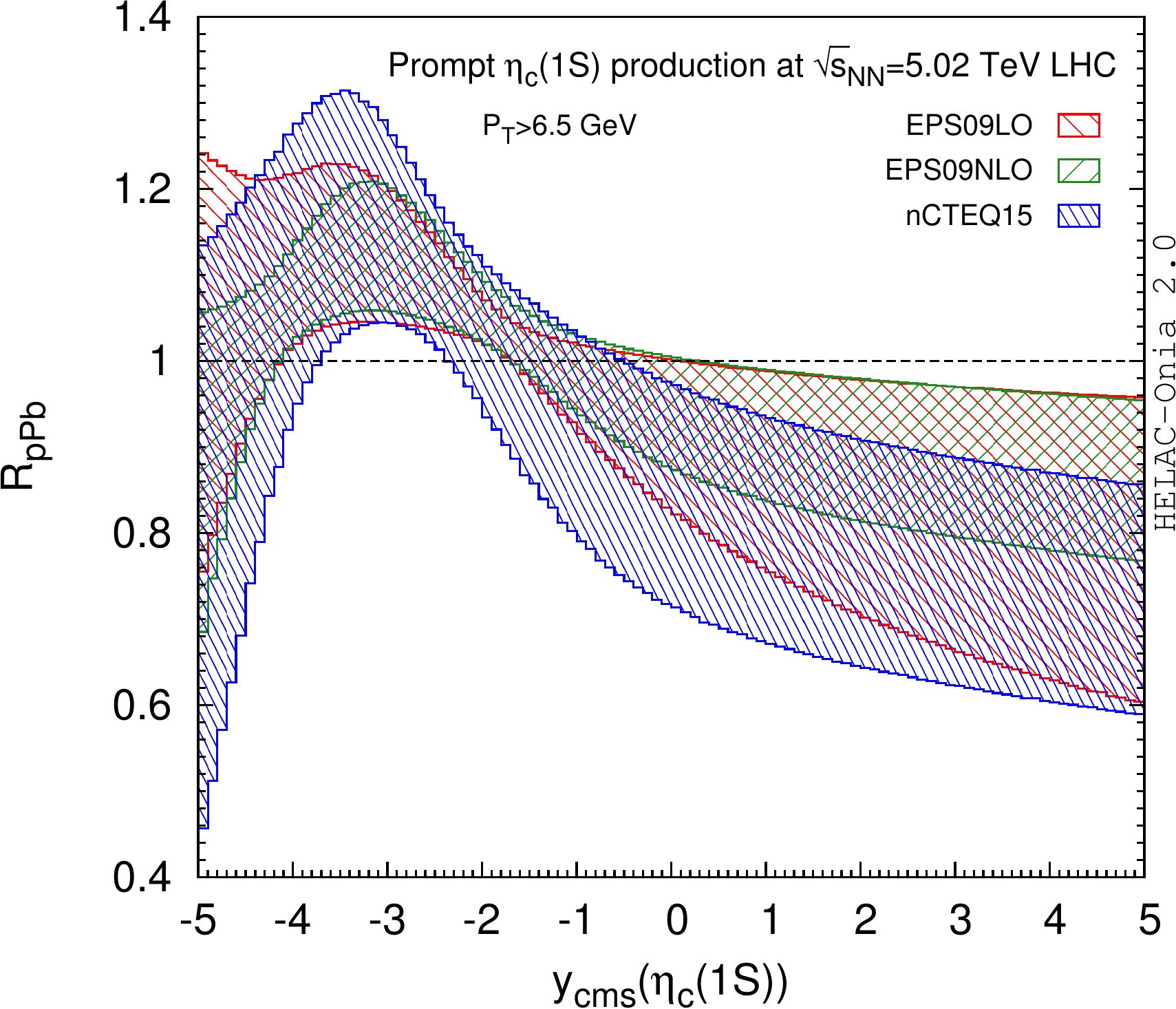}\label{fig:yetacRpPb}}
\subfloat[]{\includegraphics[width=0.33\textwidth,keepaspectratio]{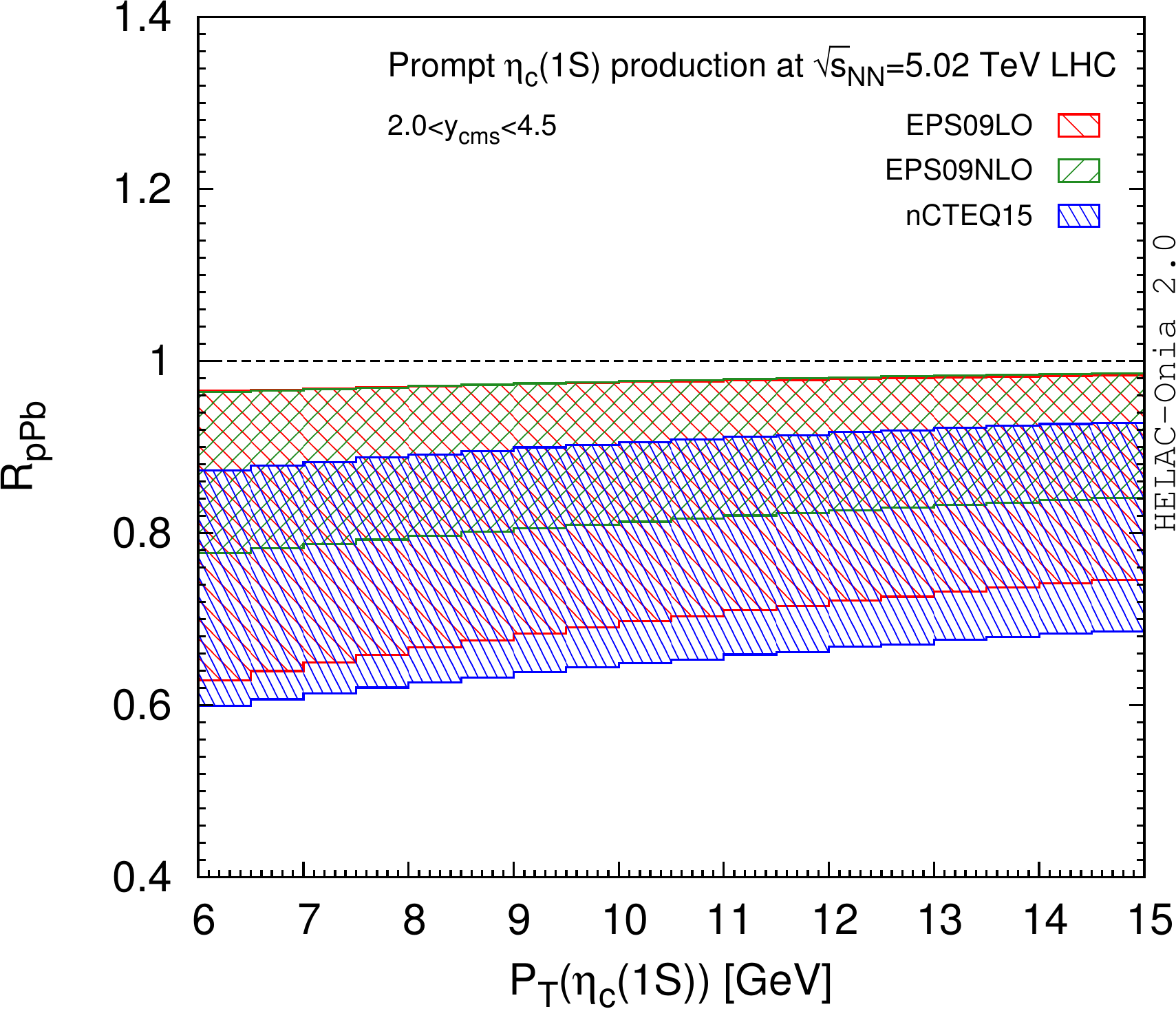}\label{fig:ptetacRpPba}}
\subfloat[]{\includegraphics[width=0.33\textwidth,keepaspectratio]{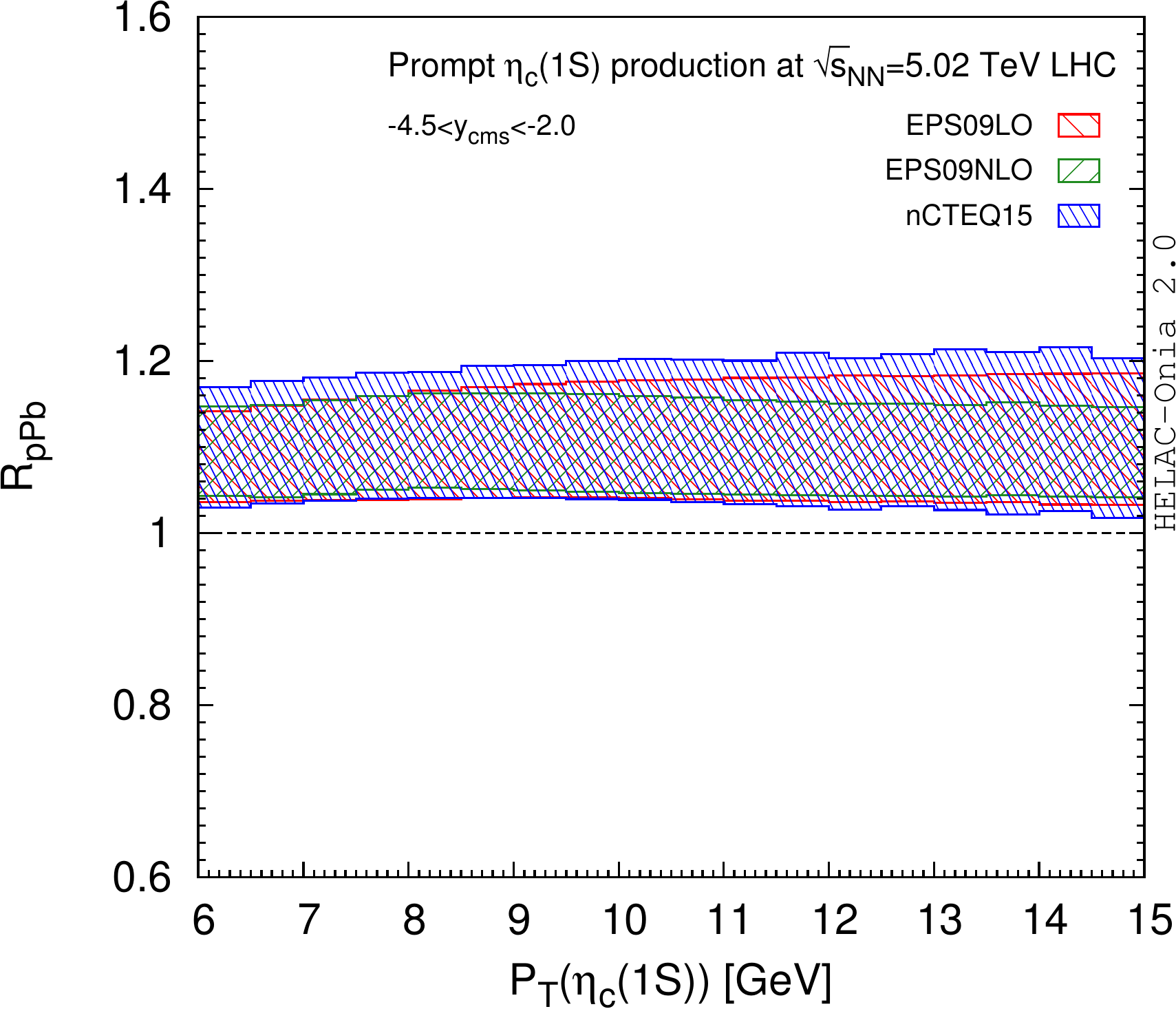}\label{fig:ptetacRpPbb}}
\caption{Rapidity (a) and transverse-momentum (b-c) dependence of \RpPb\ of prompt $\eta_c(1S)$ in \pPb\ collisions at $\sqrt{s_{NN}}=5.02$ TeV. 
%[The uncertainty bands represent the nuclear PDF uncertainty only.]
\label{fig:etacRpPb}}
\end{center}
\end{figure}

\vspace*{-1cm}
%\clearpage

\subsection{Rapidity and transverse-momentum dependence of $R_{\rm FB}$ at $\sqrt{s_{NN}}=5.02$ TeV}

In this section, we discuss the forward-to-backward production ratio $R_{\rm FB}$ 
which results from the asymmetry of the proton-nucleus collision and is thus
also sensitive to the nuclear effects. In addition, it has the advantage to be a ratio in which 
many of the systematic uncertainties of the data cancel, in particular that from the $pp$ 
yield or cross section. It is defined as 
\begin{equation}
 R_{\rm FB} = \frac{\RpPb(y_{\rm c.m.s.} >0)}{\RpPb (y_{\rm c.m.s.} <0)} =\frac{d\sigma^{\cal H}_{p\rm Pb}(y_{\rm c.m.s.} >0)}{d\sigma^{\cal H}_{p\rm Pb}(y_{\rm c.m.s.} >0)},
\end{equation}
where the "forward" direction was defined as the flight direction of the proton beam.

We stress that $R_{\rm FB}$ is identically unity at $y_{\rm c.m.s.} =0$. It tends to remain close to one
if the nuclear effects cancel between the forward and backward regions, otherwise, it tends to increase
more or less quickly for increasing $|y_{\rm c.m.s.}|$. We further note that in the current implementation of our code the nPDF uncertainties in $R_{\rm FB}$ are generally smaller than in \RpPb\ (or in the cross sections). Indeed, our current code uses the same nPDF 
eigenset to compute the forward and the backward yields used in a given ratio. This amounts to consider
that the uncertainties in the \RpPb\ are correlated. This interpretation (or rather use) of the information given by the nPDF is not unique and we could have considered that the nPDF uncertainties in \RpPb\ in the forward and backward regions are not necessarily correlated (as the widespread use of theory "bands" may suggest). Doing so, 
the uncertainties in $R_{\rm FB}$  would have been significantly larger. 
Finally, we recall that the global systematical uncertainties in the experimental data do cancel. On the experimental side, these results are usually much more reliable.

\cf{fig:yjpsiRFB} displays our results for the rapidity dependence of $R_{\rm FB}$ for the three gluon nPDFs used before (EPS09LO, EPS09NLO, nCTEQ15). For the low $P_T$ data of LHCb and ALICE, the magnitude of the asymmetry is
well compatible with that of nCTEQ15 and EPS09 LO, at the lower edge of the EPS09 NLO range.
As for the ATLAS data with a $P_T$ cut, their current uncertainties and the reduced magnitude of the effects (since 
$|y_{\rm c.m.s.}|$ is smaller) do not allow for any conclusions.

\begin{figure}[H]
\begin{center} 
\subfloat[LHCb~\cite{Aaij:2013zxa}]{\includegraphics[width=0.33\textwidth,keepaspectratio]{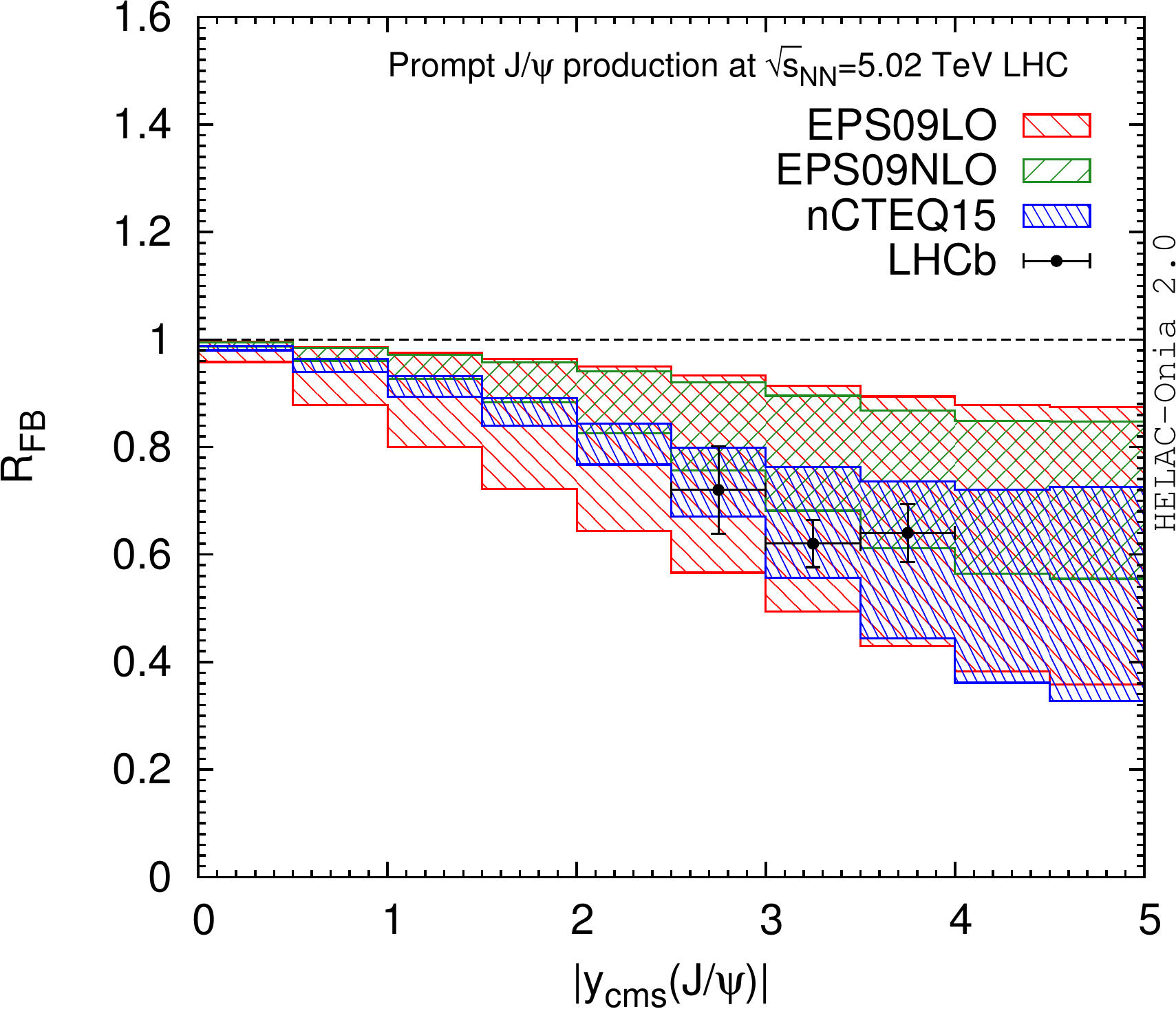}\label{fig:yjpsiRFBa}}
\subfloat[ALICE~\cite{Abelev:2013yxa}]{\includegraphics[width=0.33\textwidth,keepaspectratio]{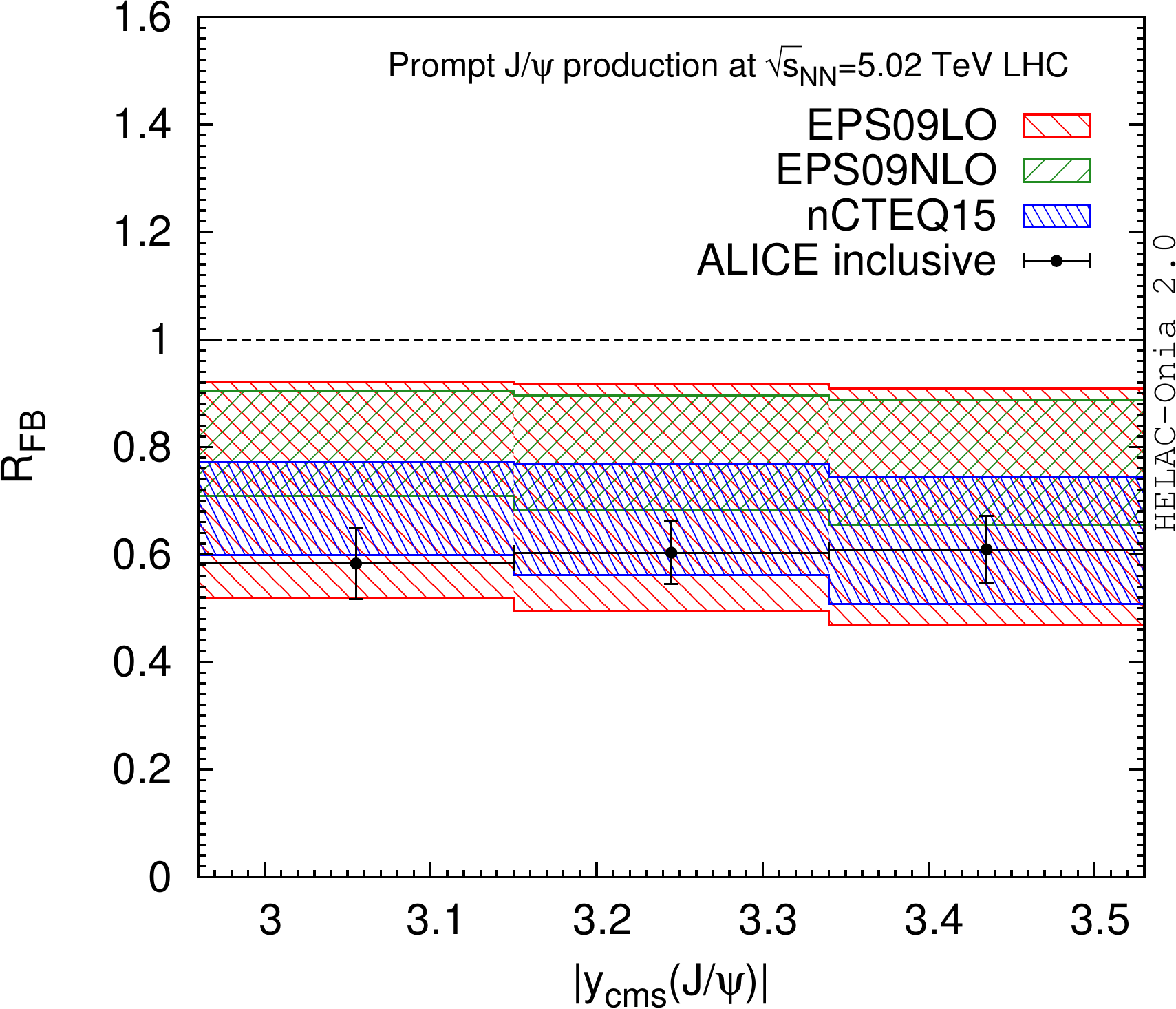}\label{fig:yjpsiRFBb}}
\subfloat[ATLAS~\cite{Aad:2015ddl}]{\includegraphics[width=0.33\textwidth,keepaspectratio]{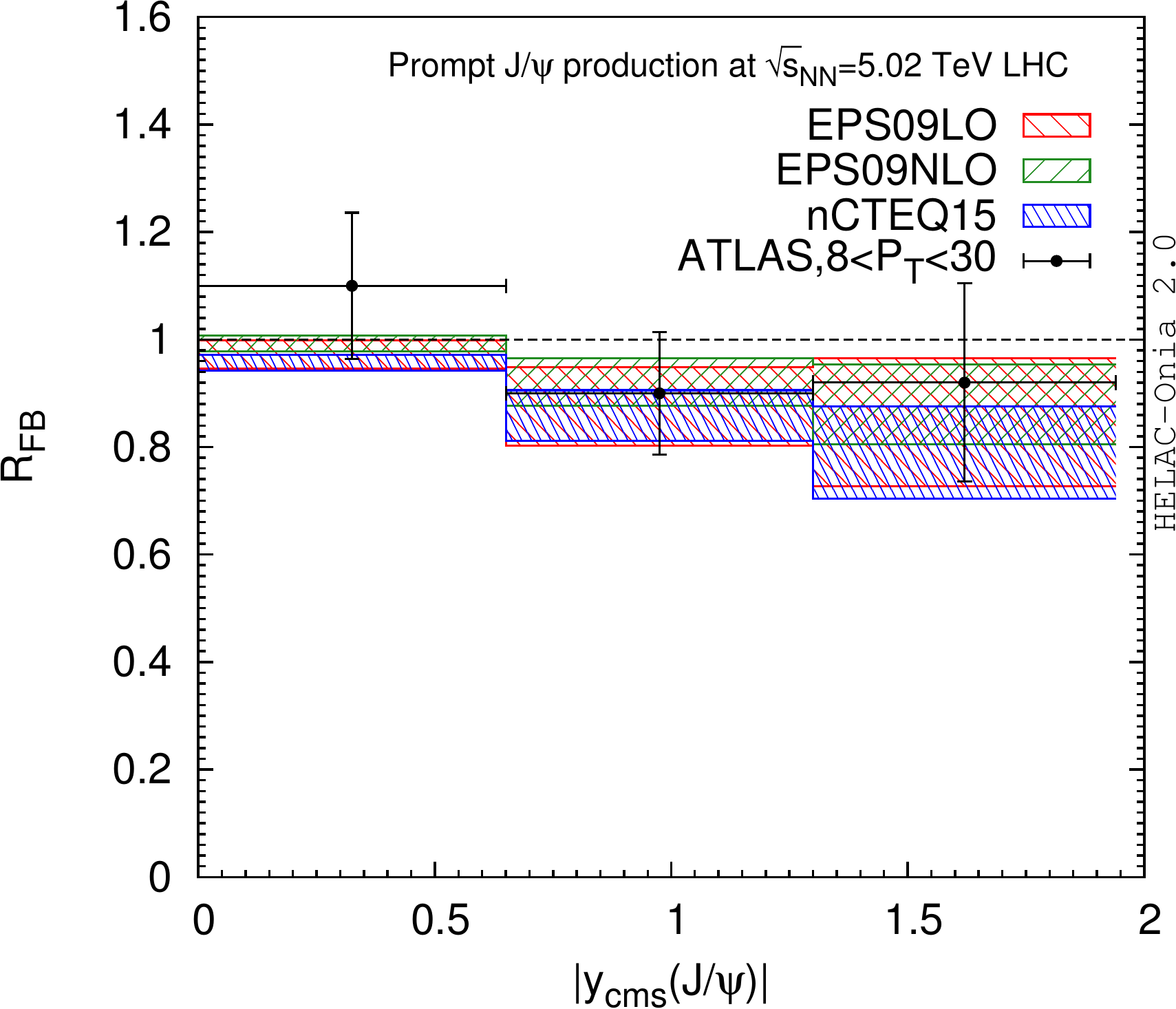}\label{fig:yjpsiRFBc}}
\caption{Rapidity dependence of $R_{\rm FB}$ of  $J/\psi$ in \pPb\ collisions at $\sqrt{s_{NN}}=5.02$ TeV: comparison between our results and the measurements by LHCb~\cite{Aaij:2013zxa}, ALICE~\cite{Abelev:2013yxa} and ATLAS~\cite{Aad:2015ddl}. [The uncertainty bands represent the nuclear PDF uncertainty only. ALICE data are for inclusive
$J/\psi$.]\label{fig:yjpsiRFB}}
\end{center}
\end{figure}

\vspace*{-1cm}

\begin{figure}[H]
\begin{center} 
\subfloat[]{\includegraphics[width=0.33\textwidth,keepaspectratio]{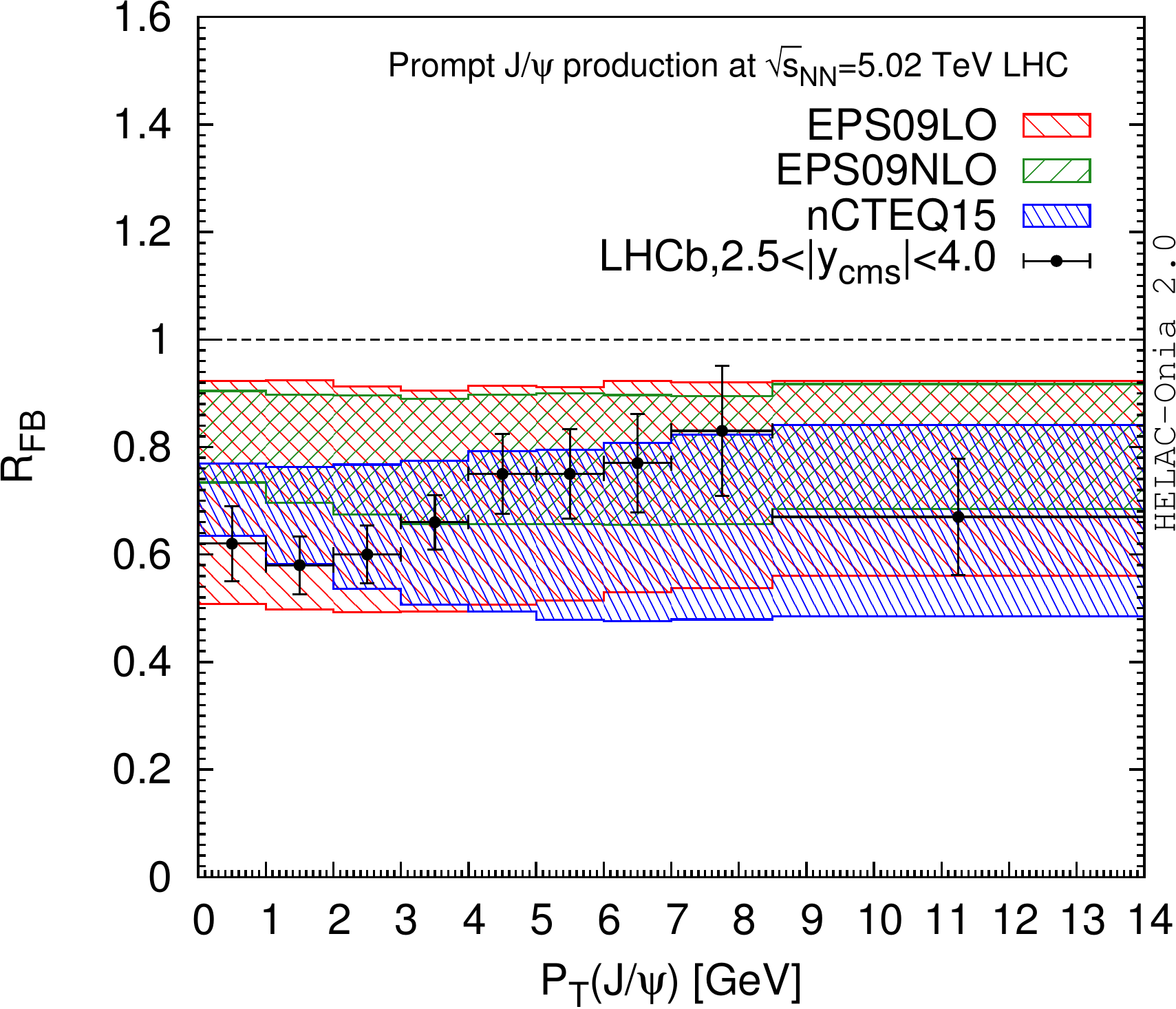}\label{fig:ptjpsiRFBa}}
\subfloat[]{\includegraphics[width=0.33\textwidth,keepaspectratio]{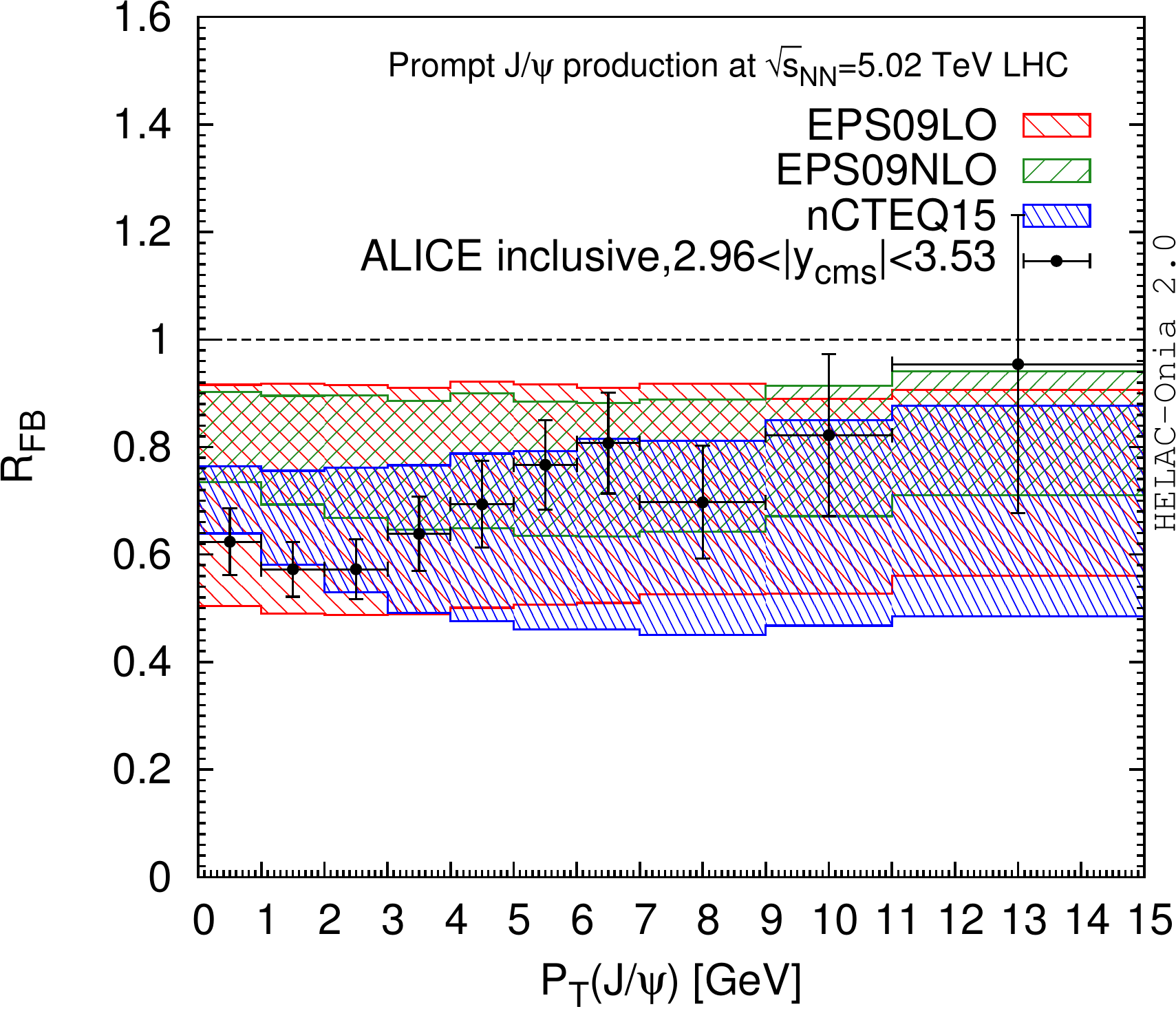}\label{fig:ptjpsiRFBb}}
\subfloat[]{\includegraphics[width=0.33\textwidth,keepaspectratio]{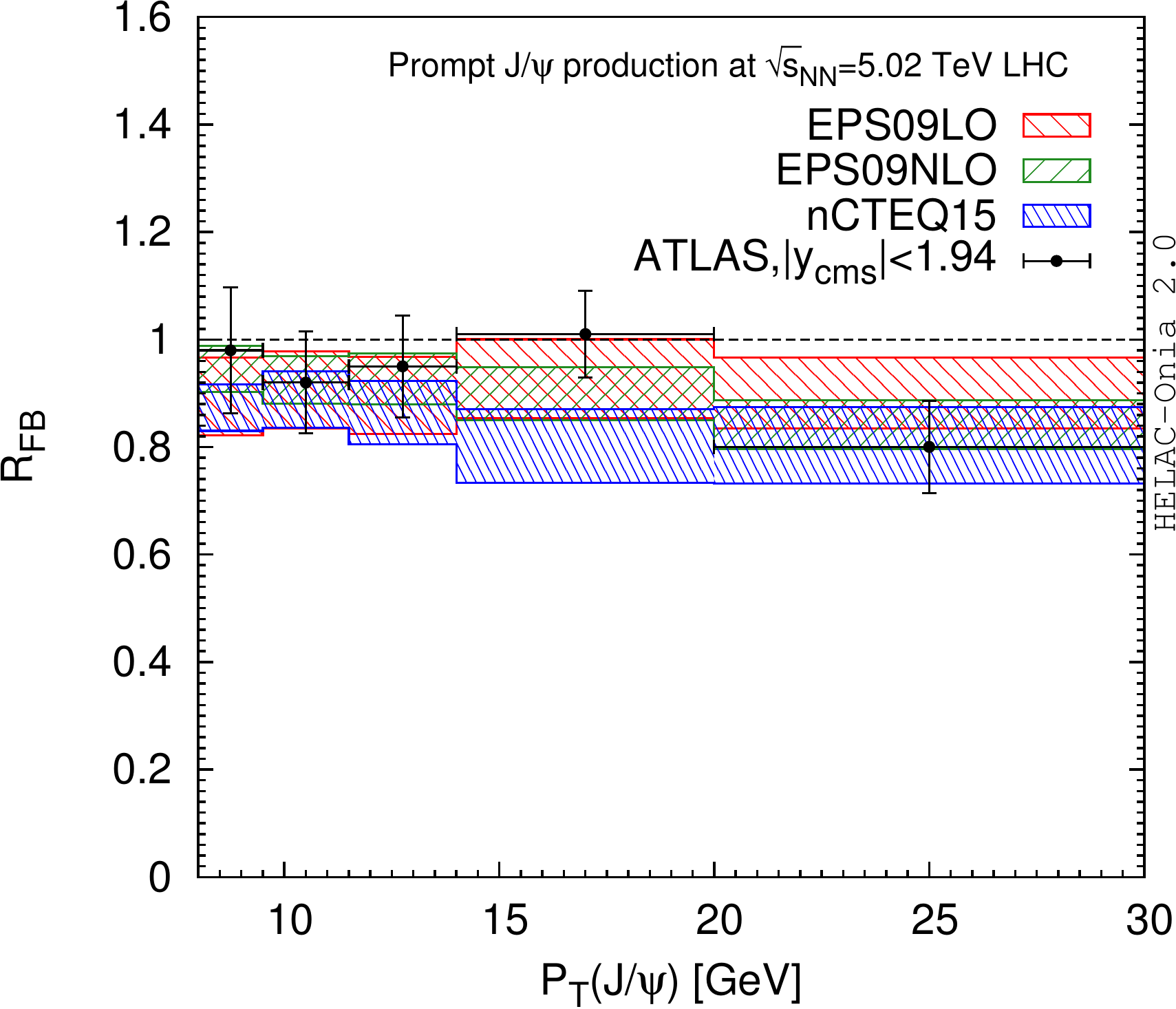}\label{fig:ptjpsiRFBc}}
\caption{Transverse-momentum dependence of the $R_{\rm FB}$ of $J/\psi$ in \pPb\ collisions at $\sqrt{s_{NN}}=5.02$~TeV: comparison between our results and the LHCb~\cite{Aaij:2013zxa}, ALICE~\cite{Abelev:2013yxa} and ATLAS~\cite{Aad:2015ddl} data.\label{fig:ptjpsiRFB}}
\end{center}
\end{figure}

Fig.~\ref{fig:ptjpsiRFB} shows our results for $R_{\rm FB}$ versus $P_T^{J/\psi}$.
A clear trend is seen in the LHCb and ALICE  results with a ratio increasing
 with $P_T$, starting at 0.6. $R_{\rm FB}$ at $P_T$ above 10 GeV are compatible with unity, 
but the larger uncertainties do not exclude values smaller than one. 
The magnitude of the ratio in the data is compatible with the 3 nPDFs. More advanced studies
are needed to go further in the interpretation of the $P_T$ dependence using specific eigensets
as opposed to bands. We also recall that a given nPDF set can be compatible with $R_{\rm FB}$
and not with \RpPb. This can happen due to specific cancellations in the magnitude of the
forward and backward nuclear modifications or to a normalisation offset. In particular, 
we note that there is no tension at all with the ATLAS data for $R_{\rm FB}$ (\cf{fig:yjpsiRFBc} and \cf{fig:ptjpsiRFBc})
unlike the case of \RpPb\ (\cf{fig:yjpsiRpPbc} and \cf{fig:ptjpsiRpPbb}). We are inclined to attribute
this to a normalisation offset from the $pp$ baseline whose effect disappears in $R_{\rm FB}$.

\begin{figure}[H]
\begin{center} 
\subfloat[]{\includegraphics[width=0.33\textwidth,keepaspectratio]{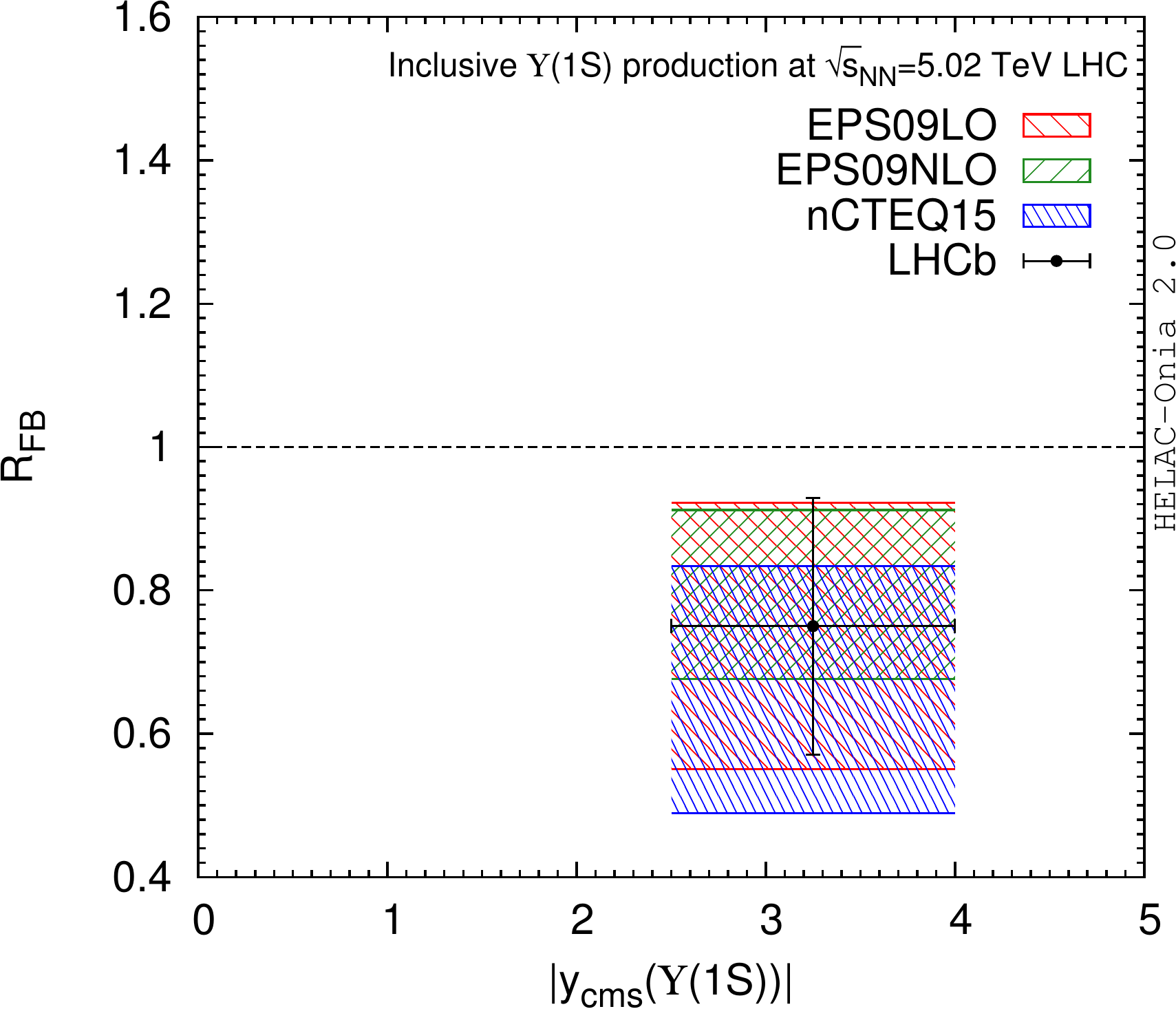}\label{fig:yy1sRFBa}}
\subfloat[]{\includegraphics[width=0.33\textwidth,keepaspectratio]{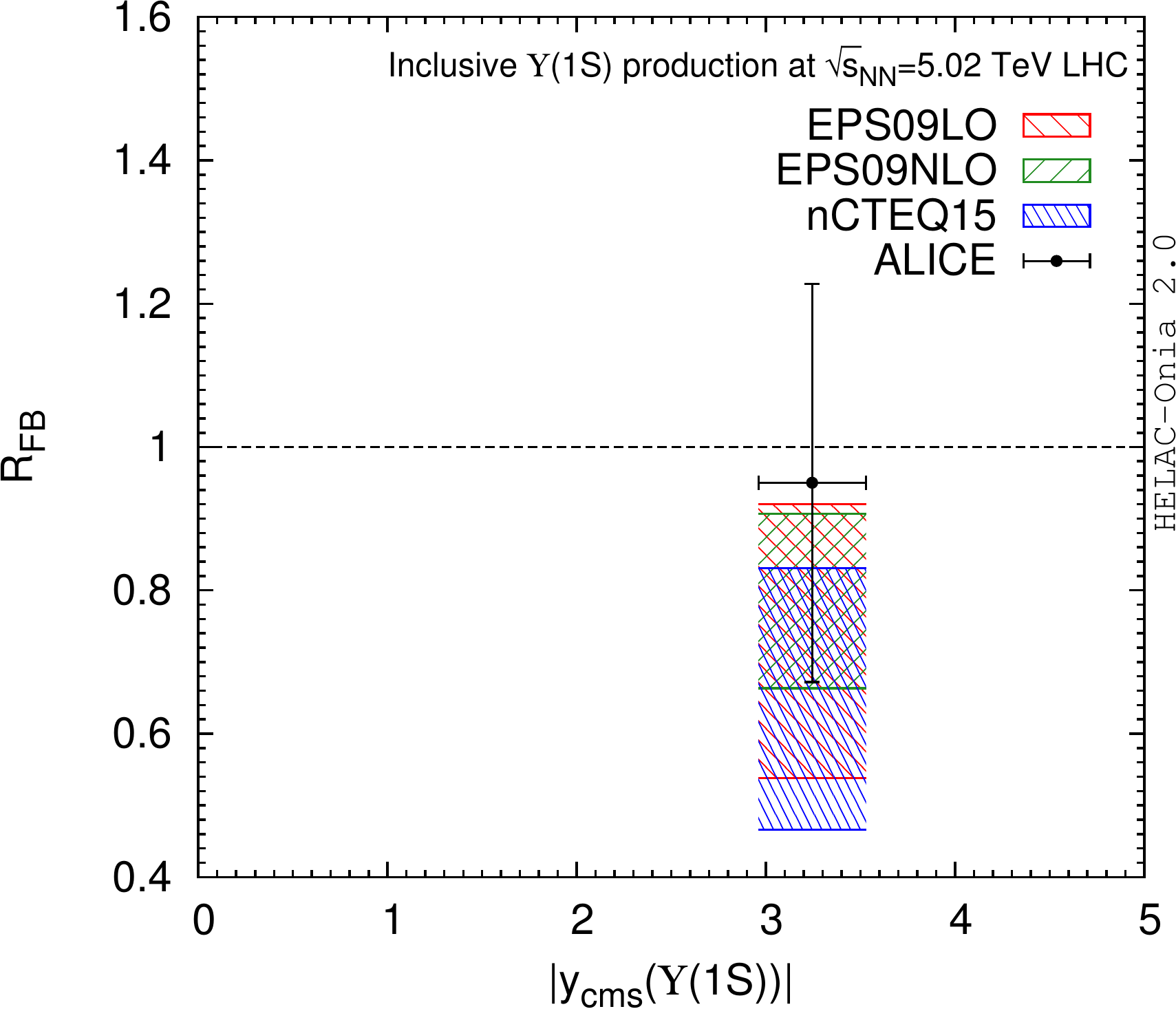}\label{fig:yy1sRFBb}}
\caption{Rapidity dependence of $R_{\rm FB}$ of inclusive $\Upsilon(1S)$ in \pPb\ collisions at $\sqrt{s_{NN}}=5.02$ TeV: comparison between our results and the measurements of LHCb~\cite{Aaij:2014mza} and ALICE~\cite{Abelev:2014oea}.\label{fig:yy1sRFB}}
\end{center}
\end{figure}

We have also computed $R_{\rm FB}$  for $\Upsilon(1S)$  (\cf{fig:yy1sRFB}) and $D^0$
(\cf{fig:D0RFB}). The same remarks as for the $J/\psi$ case apply. No tension
between the data and our computation are found. Just as for the ATLAS $J/\psi$ data, 
the good agreement with the $D^0$ LHCb data may indicate that a slight offset in the normalisation
affects \RpPb\ as plotted on \cf{fig:D0RpPb}. Whereas the \RpPb\ values point at a smaller
suppression than those encoded in the nPDFs (in particular nCTEQ15), the magnitude
of $R_{\rm FB}$ is very well accounted by nCTEQ15 and corresponds to the strongest
magnitude encoded in EPS09 NLO.
For completeness, we have also computed $R_{\rm FB}$ of prompt $\eta_c(1S)$ (see \cf{fig:etacRFB}).

\begin{figure}[H]
\begin{center} 
\subfloat[]{\includegraphics[width=0.33\textwidth,keepaspectratio]{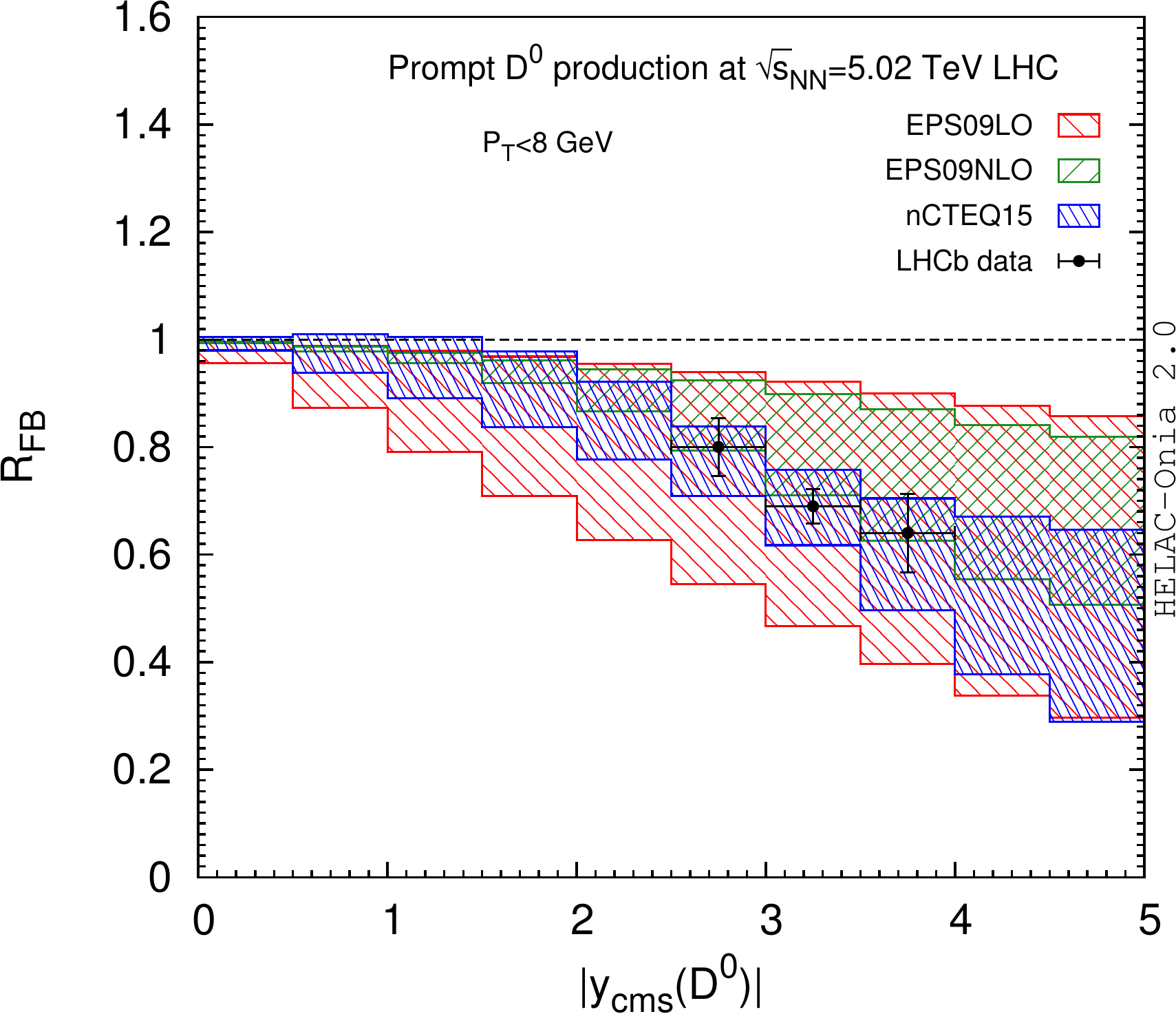}\label{fig:yD0RFB}}
\subfloat[]{\includegraphics[width=0.33\textwidth,keepaspectratio]{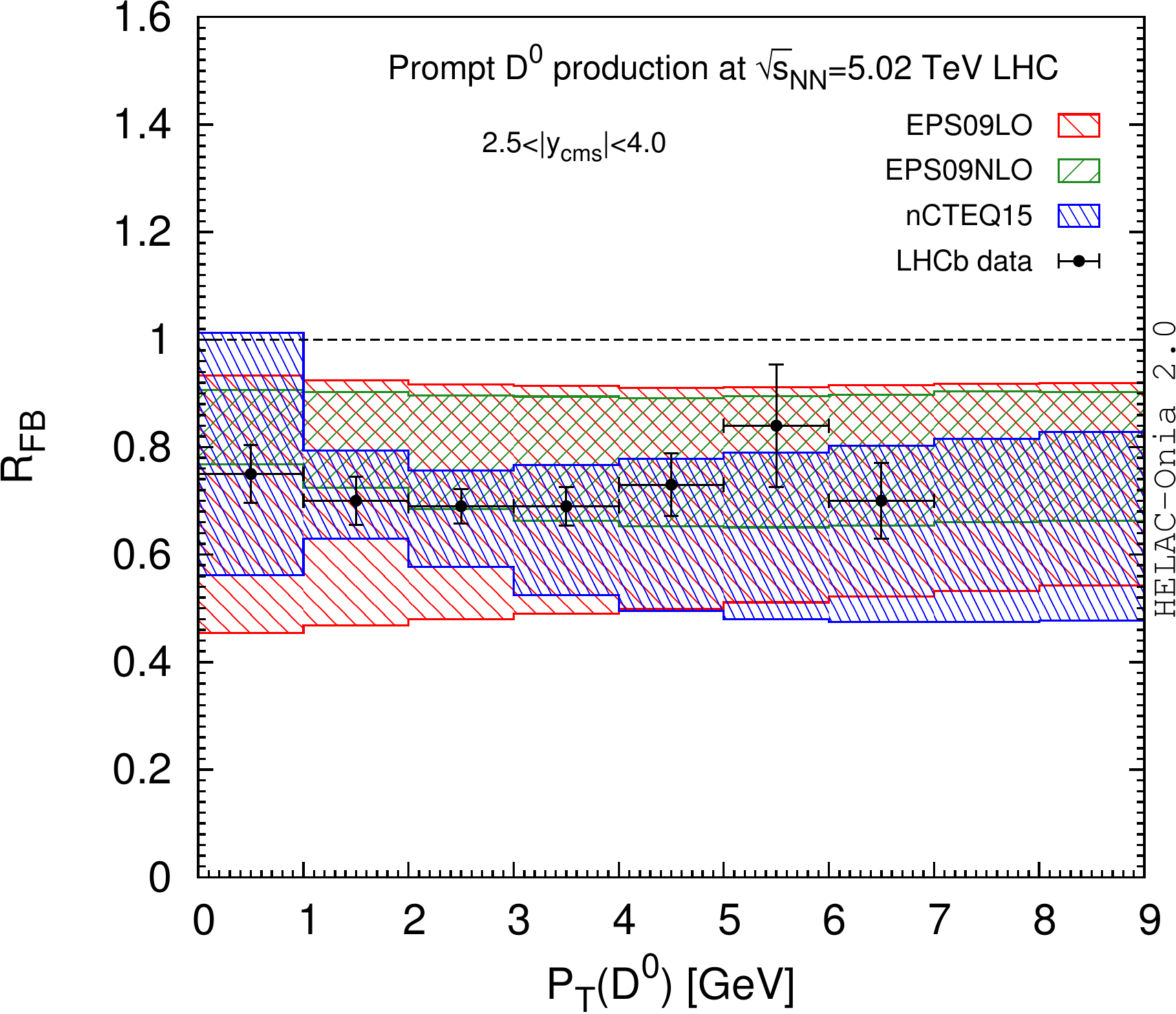}\label{fig:ptD0RFB}}
\caption{Rapidity (a) and transverse-momentum (b) dependence of $R_{\rm FB}$ of prompt $D^0$ production in \pPb\ collisions at $\sqrt{s_{NN}}=5.02$ TeV: comparison between our results and the measurements by LHCb~\cite{LHCb-CONF-2016-003}.\label{fig:D0RFB}}
\end{center}
\end{figure}

\begin{figure}[H]
\begin{center} 
\subfloat[]{\includegraphics[width=0.33\textwidth,keepaspectratio]{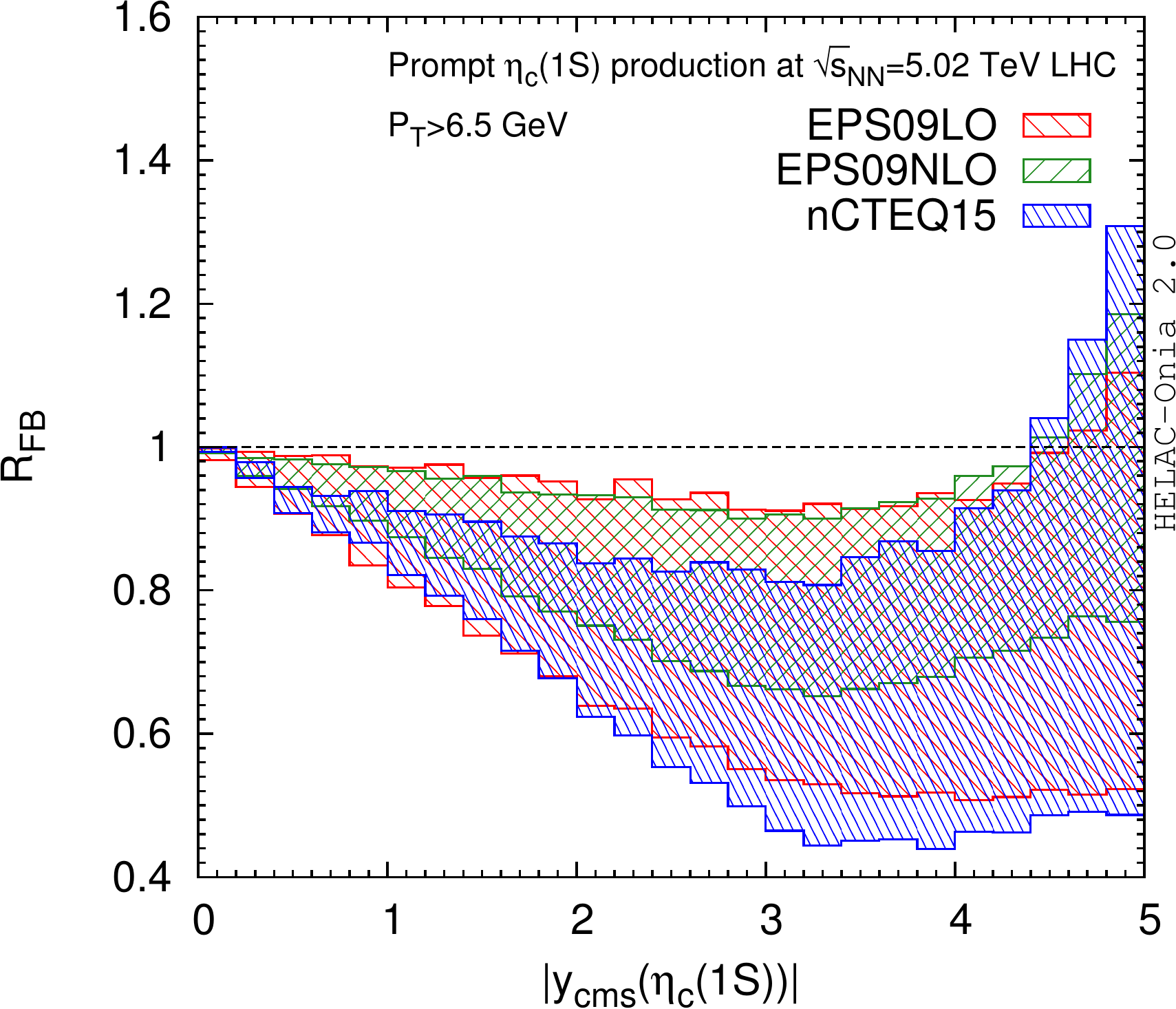}\label{fig:yetacRFB}}
\subfloat[]{\includegraphics[width=0.33\textwidth,keepaspectratio]{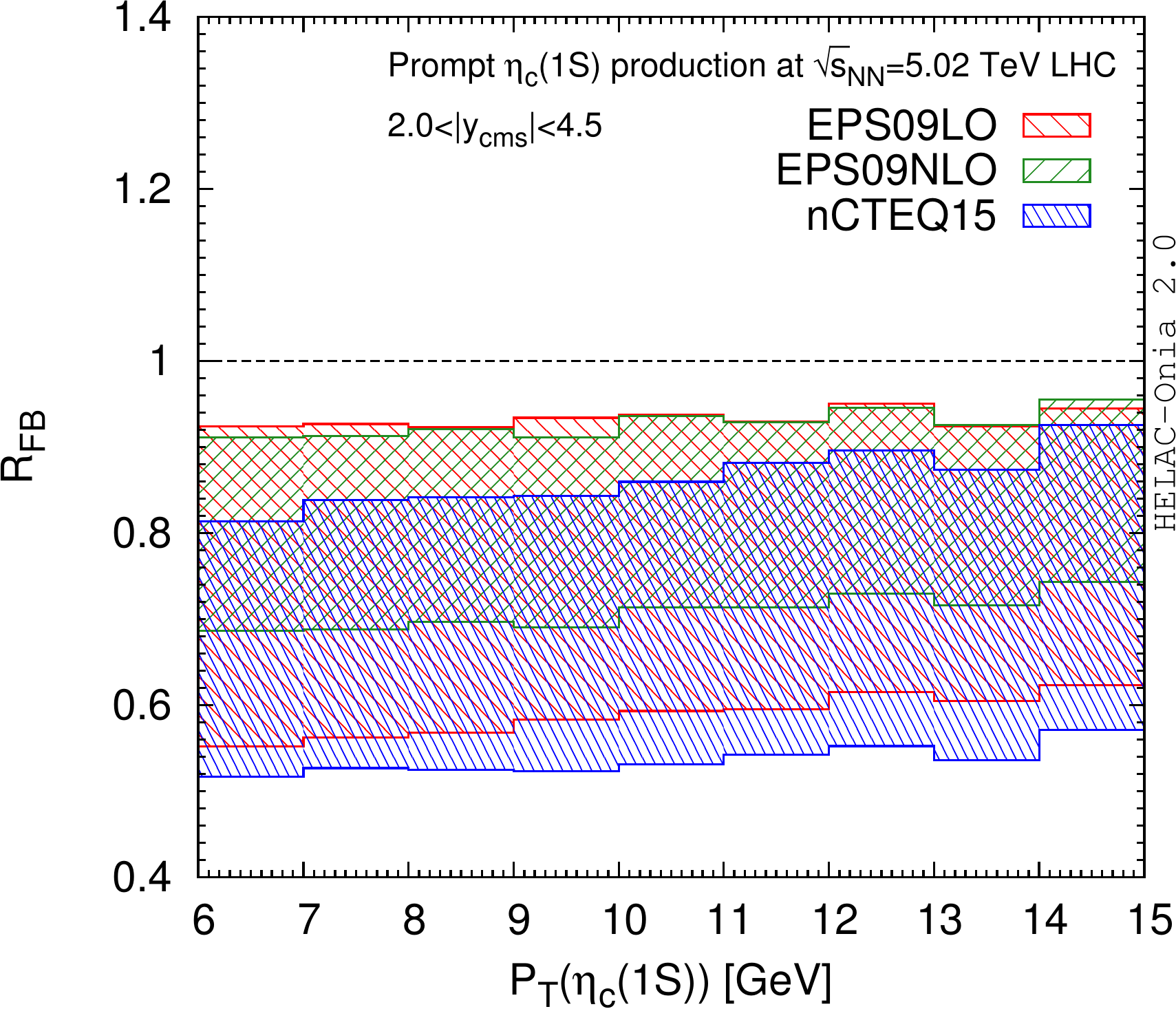}\label{fig:ptetacRFB}}
\caption{Rapidity (a) and transverse-momentum (b) dependence of $R_{\rm FB}$ of prompt $\eta_c(1S)$ in \pPb\ collisions at $\sqrt{s_{NN}}=5.02$ TeV.
}\label{fig:etacRFB}
\end{center}
\end{figure}

\section{Conclusions}

We have devised a model-independent procedure to evaluate the impact of the nuclear modification
of the gluon densities on hard probes produced in proton-nucleus collisions at colliders energies.
It is particularly tailored for two-to-two partonic scatterings, relevant for quarkonium and heavy-meson
production. The model independence of our procedure lies in the parametrisation of the
partonic amplitude squared with parameters fit to $pp$ collision data in similar kinematical
conditions as the \pPb\ data to be described.

We have illustrated the capabilities of our approach by computing the cross sections as well
as  the nuclear modifications factor for $J/\psi$, $\Upsilon$ and $D^0$ production at the LHC.
Even though our objective was not to argue that the nPDF effect is the dominant one in 
this energy range, we have not found out any significant tension between our computations
using three common nPDFs (EPS09 LO \& NLO and nCTEQ15) and the exisiting data. To further 
highlight the potentialities of the approach, we have made predictions for $\eta_c$ 
production which might be at reach for the LHCb collaboration. We have also made predictions 
for the 8 TeV \pPb\ run (see the appendix).

As outlooks for physics studies, our method can easily be transposed to $B$ hadron production. It should also
be possible to apply it for non-prompt charmonia provided that the kinematical shift
between the $b$-quark and the charmonium is correctly accounted for. On the side of the tool itself, 
we plan improve it such that it could automatically provide the user with the nuclear modification factors
starting for measured $pp$ data.

\section*{Acknowledgements}
We are very grateful to R. Arnaldi, E.G. Ferreiro, F. Fleuret,  C. Hadjidakis, D. Kikola, Y. Kim, A. Kusina, S.~Lee, A.~Rakatozafindrabe, C. Salgado, I. Schienbein, R. Vogt, Z. Yang, Y. Zhang for stimulating discussions. The work of J.P.L. is supported in part by the French CNRS via the LIA FCPPL (Quarkonium4AFTER) and the D\'efi
Inphyniti-Th\'eorie LHC France. H.S.S. is supported by the ERC grant 291377 {\it LHCtheory:
Theoretical predictions and analyses of LHC physics: advancing the precision frontier}.

\bibliographystyle{utphys}

\bibliography{pA-quarkonium_181016}

\providecommand{\href}[2]{#2}\begingroup\raggedright\begin{thebibliography}{10}

\bibitem{Andronic:2015wma}
A.~Andronic {\em et~al.}, ``{Heavy-flavour and quarkonium production in the LHC
  era: from proton–proton to heavy-ion collisions},''
  \href{http://dx.doi.org/10.1140/epjc/s10052-015-3819-5}{{\em Eur. Phys. J.}
  {\bfseries C76} no.~3, (2016) 107},
\href{http://arxiv.org/abs/1506.03981}{{\ttfamily arXiv:1506.03981 [nucl-ex]}}.
%%CITATION = ARXIV:1506.03981;%%.

\bibitem{Lansberg:2015uxa}
J.~P. Lansberg, ``{Theory status of quarkonium production in proton-nucleus
  collisions},'' \href{http://dx.doi.org/10.1088/1742-6596/668/1/012019}{{\em
  J. Phys. Conf. Ser.} {\bfseries 668} no.~1, (2016) 012019},
\href{http://arxiv.org/abs/1510.01818}{{\ttfamily arXiv:1510.01818 [nucl-th]}}.
%%CITATION = ARXIV:1510.01818;%%.

\bibitem{Brambilla:2010cs}
N.~Brambilla {\em et~al.}, ``{Heavy quarkonium: progress, puzzles, and
  opportunities},''
  \href{http://dx.doi.org/10.1140/epjc/s10052-010-1534-9}{{\em Eur. Phys. J.}
  {\bfseries C71} (2011) 1534},
\href{http://arxiv.org/abs/1010.5827}{{\ttfamily arXiv:1010.5827 [hep-ph]}}.
%%CITATION = ARXIV:1010.5827;%%.

\bibitem{Rapp:2008tf}
R.~Rapp, D.~Blaschke, and P.~Crochet, ``{Charmonium and bottomonium production
  in heavy-ion collisions},''
  \href{http://dx.doi.org/10.1016/j.ppnp.2010.07.002}{{\em Prog. Part. Nucl.
  Phys.} {\bfseries 65} (2010) 209--266},
\href{http://arxiv.org/abs/0807.2470}{{\ttfamily arXiv:0807.2470 [hep-ph]}}.
%%CITATION = ARXIV:0807.2470;%%.

\bibitem{Frawley:2008kk}
A.~D. Frawley, T.~Ullrich, and R.~Vogt, ``{Heavy flavor in heavy-ion collisions
  at RHIC and RHIC II},''
  \href{http://dx.doi.org/10.1016/j.physrep.2008.04.002}{{\em Phys. Rept.}
  {\bfseries 462} (2008) 125--175},
\href{http://arxiv.org/abs/0806.1013}{{\ttfamily arXiv:0806.1013 [nucl-ex]}}.
%%CITATION = ARXIV:0806.1013;%%.

\bibitem{Lansberg:2006dh}
J.~P. Lansberg, ``{$J/\psi$, $\psi$ ' and $\Upsilon$ production at hadron
  colliders: A Review},''
  \href{http://dx.doi.org/10.1142/S0217751X06033180}{{\em Int. J. Mod. Phys.}
  {\bfseries A21} (2006) 3857--3916},
\href{http://arxiv.org/abs/hep-ph/0602091}{{\ttfamily arXiv:hep-ph/0602091
  [hep-ph]}}.
%%CITATION = HEP-PH/0602091;%%.

\bibitem{Gerschel:1988wn}
C.~Gerschel and J.~Hufner, ``{A Contribution to the Suppression of the J/psi
  Meson Produced in High-Energy Nucleus Nucleus Collisions},''
\href{http://dx.doi.org/10.1016/0370-2693(88)90570-9}{{\em Phys. Lett.}
  {\bfseries B207} (1988) 253--256}.
%%CITATION = PHLTA,B207,253;%%.

\bibitem{Vogt:1999cu}
R.~Vogt, ``{$J/\psi$ production and suppression},''
\href{http://dx.doi.org/10.1016/S0370-1573(98)00074-X}{{\em Phys. Rept.}
  {\bfseries 310} (1999) 197--260}.
%%CITATION = PRPLC,310,197;%%.

\bibitem{Ferreiro:2014bia}
E.~G. Ferreiro, ``{Excited charmonium suppression in proton–nucleus
  collisions as a consequence of comovers},''
  \href{http://dx.doi.org/10.1016/j.physletb.2015.07.066}{{\em Phys. Lett.}
  {\bfseries B749} (2015) 98--103},
\href{http://arxiv.org/abs/1411.0549}{{\ttfamily arXiv:1411.0549 [hep-ph]}}.
%%CITATION = ARXIV:1411.0549;%%.

\bibitem{Capella:2005cn}
A.~Capella and E.~G. Ferreiro, ``{J/ psi suppression at s**(1/2) = 200-GeV in
  the comovers interaction model},''
  \href{http://dx.doi.org/10.1140/epjc/s2005-02348-0}{{\em Eur. Phys. J.}
  {\bfseries C42} (2005) 419--424},
\href{http://arxiv.org/abs/hep-ph/0505032}{{\ttfamily arXiv:hep-ph/0505032
  [hep-ph]}}.
%%CITATION = HEP-PH/0505032;%%.

\bibitem{Capella:2000zp}
A.~Capella, E.~G. Ferreiro, and A.~B. Kaidalov, ``{Nonsaturation of the J / psi
  suppression at large transverse energy in the comovers approach},''
  \href{http://dx.doi.org/10.1103/PhysRevLett.85.2080}{{\em Phys. Rev. Lett.}
  {\bfseries 85} (2000) 2080--2083},
\href{http://arxiv.org/abs/hep-ph/0002300}{{\ttfamily arXiv:hep-ph/0002300
  [hep-ph]}}.
%%CITATION = HEP-PH/0002300;%%.

\bibitem{Gavin:1990gm}
S.~Gavin and R.~Vogt, ``{$J/\psi$ Suppression From Hadron - Nucleus to
  Nucleus-nucleus Collisions},''
\href{http://dx.doi.org/10.1016/0550-3213(90)90610-P}{{\em Nucl. Phys.}
  {\bfseries B345} (1990) 104--124}.
%%CITATION = NUPHA,B345,104;%%.

\bibitem{Arleo:2012hn}
F.~Arleo and S.~Peigne, ``{J/$\psi$ suppression in p-A collisions from parton
  energy loss in cold QCD matter},''
  \href{http://dx.doi.org/10.1103/PhysRevLett.109.122301}{{\em Phys. Rev.
  Lett.} {\bfseries 109} (2012) 122301},
\href{http://arxiv.org/abs/1204.4609}{{\ttfamily arXiv:1204.4609 [hep-ph]}}.
%%CITATION = ARXIV:1204.4609;%%.

\bibitem{Sharma:2012dy}
R.~Sharma and I.~Vitev, ``{High transverse momentum quarkonium production and
  dissociation in heavy ion collisions},''
  \href{http://dx.doi.org/10.1103/PhysRevC.87.044905}{{\em Phys. Rev.}
  {\bfseries C87} no.~4, (2013) 044905},
\href{http://arxiv.org/abs/1203.0329}{{\ttfamily arXiv:1203.0329 [hep-ph]}}.
%%CITATION = ARXIV:1203.0329;%%.

\bibitem{Arleo:2010rb}
F.~Arleo, S.~Peigne, and T.~Sami, ``{Revisiting scaling properties of
  medium-induced gluon radiation},''
  \href{http://dx.doi.org/10.1103/PhysRevD.83.114036}{{\em Phys. Rev.}
  {\bfseries D83} (2011) 114036},
\href{http://arxiv.org/abs/1006.0818}{{\ttfamily arXiv:1006.0818 [hep-ph]}}.
%%CITATION = ARXIV:1006.0818;%%.

\bibitem{Brodsky:1992nq}
S.~J. Brodsky and P.~Hoyer, ``{A Bound on the energy loss of partons in
  nuclei},'' \href{http://dx.doi.org/10.1016/0370-2693(93)91724-2}{{\em Phys.
  Lett.} {\bfseries B298} (1993) 165--170},
\href{http://arxiv.org/abs/hep-ph/9210262}{{\ttfamily arXiv:hep-ph/9210262
  [hep-ph]}}.
%%CITATION = HEP-PH/9210262;%%.

\bibitem{Gavin:1991qk}
S.~Gavin and J.~Milana, ``{Energy loss at large x(F) in nuclear collisions},''
\href{http://dx.doi.org/10.1103/PhysRevLett.68.1834}{{\em Phys. Rev. Lett.}
  {\bfseries 68} (1992) 1834--1837}.
%%CITATION = PRLTA,68,1834;%%.

\bibitem{Brodsky:1989ex}
S.~J. Brodsky and P.~Hoyer, ``{The Nucleus as a Color Filter in {QCD} Decays:
  Hadroproduction in Nuclei},''
\href{http://dx.doi.org/10.1103/PhysRevLett.63.1566}{{\em Phys. Rev. Lett.}
  {\bfseries 63} (1989) 1566}.
%%CITATION = PRLTA,63,1566;%%.

\bibitem{Ducloue:2015gfa}
B.~Duclou\'e, T.~Lappi, and H.~Mantysaari, ``{Forward $J/\psi$ production in
  proton-nucleus collisions at high energy},''
  \href{http://dx.doi.org/10.1103/PhysRevD.91.114005}{{\em Phys. Rev.}
  {\bfseries D91} no.~11, (2015) 114005},
\href{http://arxiv.org/abs/1503.02789}{{\ttfamily arXiv:1503.02789 [hep-ph]}}.
%%CITATION = ARXIV:1503.02789;%%.

\bibitem{Ma:2015sia}
Y.-Q. Ma, R.~Venugopalan, and H.-F. Zhang, ``{$J/\psi$ production and
  suppression in high energy proton-nucleus collisions},''
  \href{http://dx.doi.org/10.1103/PhysRevD.92.071901}{{\em Phys. Rev.}
  {\bfseries D92} (2015) 071901},
\href{http://arxiv.org/abs/1503.07772}{{\ttfamily arXiv:1503.07772 [hep-ph]}}.
%%CITATION = ARXIV:1503.07772;%%.

\bibitem{Fujii:2013gxa}
H.~Fujii and K.~Watanabe, ``{Heavy quark pair production in high energy pA
  collisions: Quarkonium},''
  \href{http://dx.doi.org/10.1016/j.nuclphysa.2013.06.011}{{\em Nucl. Phys.}
  {\bfseries A915} (2013) 1--23},
\href{http://arxiv.org/abs/1304.2221}{{\ttfamily arXiv:1304.2221 [hep-ph]}}.
%%CITATION = ARXIV:1304.2221;%%.

\bibitem{Qiu:2013qka}
J.-W. Qiu, P.~Sun, B.-W. Xiao, and F.~Yuan, ``{Universal Suppression of Heavy
  Quarkonium Production in pA Collisions at Low Transverse Momentum},''
  \href{http://dx.doi.org/10.1103/PhysRevD.89.034007}{{\em Phys. Rev.}
  {\bfseries D89} no.~3, (2014) 034007},
\href{http://arxiv.org/abs/1310.2230}{{\ttfamily arXiv:1310.2230 [hep-ph]}}.
%%CITATION = ARXIV:1310.2230;%%.

\bibitem{Kopeliovich:2001ee}
B.~Kopeliovich, A.~Tarasov, and J.~Hufner, ``{Coherence phenomena in charmonium
  production off nuclei at the energies of RHIC and LHC},''
  \href{http://dx.doi.org/10.1016/S0375-9474(01)01220-9}{{\em Nucl. Phys.}
  {\bfseries A696} (2001) 669--714},
\href{http://arxiv.org/abs/hep-ph/0104256}{{\ttfamily arXiv:hep-ph/0104256
  [hep-ph]}}.
%%CITATION = HEP-PH/0104256;%%.

\bibitem{Kovarik:2015cma}
K.~Kovarik {\em et~al.}, ``{nCTEQ15 - Global analysis of nuclear parton
  distributions with uncertainties in the CTEQ framework},''
  \href{http://dx.doi.org/10.1103/PhysRevD.93.085037}{{\em Phys. Rev.}
  {\bfseries D93} no.~8, (2016) 085037},
\href{http://arxiv.org/abs/1509.00792}{{\ttfamily arXiv:1509.00792 [hep-ph]}}.
%%CITATION = ARXIV:1509.00792;%%.

\bibitem{Owens:2012bv}
J.~F. Owens, A.~Accardi, and W.~Melnitchouk, ``{Global parton distributions
  with nuclear and finite-$Q^2$ corrections},''
  \href{http://dx.doi.org/10.1103/PhysRevD.87.094012}{{\em Phys. Rev.}
  {\bfseries D87} no.~9, (2013) 094012},
\href{http://arxiv.org/abs/1212.1702}{{\ttfamily arXiv:1212.1702 [hep-ph]}}.
%%CITATION = ARXIV:1212.1702;%%.

\bibitem{deFlorian:2011fp}
D.~de~Florian, R.~Sassot, P.~Zurita, and M.~Stratmann, ``{Global Analysis of
  Nuclear Parton Distributions},''
  \href{http://dx.doi.org/10.1103/PhysRevD.85.074028}{{\em Phys. Rev.}
  {\bfseries D85} (2012) 074028},
\href{http://arxiv.org/abs/1112.6324}{{\ttfamily arXiv:1112.6324 [hep-ph]}}.
%%CITATION = ARXIV:1112.6324;%%.

\bibitem{Eskola:2009uj}
K.~J. Eskola, H.~Paukkunen, and C.~A. Salgado, ``{EPS09: A New Generation of
  NLO and LO Nuclear Parton Distribution Functions},''
  \href{http://dx.doi.org/10.1088/1126-6708/2009/04/065}{{\em JHEP} {\bfseries
  04} (2009) 065},
\href{http://arxiv.org/abs/0902.4154}{{\ttfamily arXiv:0902.4154 [hep-ph]}}.
%%CITATION = ARXIV:0902.4154;%%.

\bibitem{Hirai:2007sx}
M.~Hirai, S.~Kumano, and T.~H. Nagai, ``{Determination of nuclear parton
  distribution functions and their uncertainties in next-to-leading order},''
  \href{http://dx.doi.org/10.1103/PhysRevC.76.065207}{{\em Phys. Rev.}
  {\bfseries C76} (2007) 065207},
\href{http://arxiv.org/abs/0709.3038}{{\ttfamily arXiv:0709.3038 [hep-ph]}}.
%%CITATION = ARXIV:0709.3038;%%.

\bibitem{Albacete:2016veq}
J.~L. Albacete {\em et~al.}, ``{Predictions for $p+$Pb Collisions at
  $\sqrt{s_{NN}} = 5$ TeV: Comparison with Data},''
  \href{http://dx.doi.org/10.1142/S0218301316300058}{{\em Int. J. Mod. Phys.}
  {\bfseries E25} no.~09, (2016) 1630005},
\href{http://arxiv.org/abs/1605.09479}{{\ttfamily arXiv:1605.09479 [hep-ph]}}.
%%CITATION = ARXIV:1605.09479;%%.

\bibitem{Ferreiro:2013pua}
E.~G. Ferreiro, F.~Fleuret, J.~P. Lansberg, and A.~Rakotozafindrabe, ``{Impact
  of the Nuclear Modification of the Gluon Densities on $J/\psi$ production in
  $p$Pb collisions at $\sqrt{s_{NN}} =$ 5 TeV},''
  \href{http://dx.doi.org/10.1103/PhysRevC.88.047901}{{\em Phys. Rev.}
  {\bfseries C88} no.~4, (2013) 047901},
\href{http://arxiv.org/abs/1305.4569}{{\ttfamily arXiv:1305.4569 [hep-ph]}}.
%%CITATION = ARXIV:1305.4569;%%.

\bibitem{Vogt:2004dh}
R.~Vogt, ``{Shadowing and absorption effects on J/psi production in dA
  collisions},'' \href{http://dx.doi.org/10.1103/PhysRevC.71.054902}{{\em Phys.
  Rev.} {\bfseries C71} (2005) 054902},
\href{http://arxiv.org/abs/hep-ph/0411378}{{\ttfamily arXiv:hep-ph/0411378
  [hep-ph]}}.
%%CITATION = HEP-PH/0411378;%%.

\bibitem{Arleo:2006qk}
F.~Arleo and V.-N. Tram, ``{A Systematic study of J/psi suppression in cold
  nuclear matter},''
  \href{http://dx.doi.org/10.1140/epjc/s10052-008-0604-8}{{\em Eur. Phys. J.}
  {\bfseries C55} (2008) 449--461},
\href{http://arxiv.org/abs/hep-ph/0612043}{{\ttfamily arXiv:hep-ph/0612043
  [hep-ph]}}.
%%CITATION = HEP-PH/0612043;%%.

\bibitem{Arleo:2008zc}
F.~Arleo, ``{Constraints on nuclear gluon densities from J/psi data},''
  \href{http://dx.doi.org/10.1016/j.physletb.2008.06.074}{{\em Phys. Lett.}
  {\bfseries B666} (2008) 31--33},
\href{http://arxiv.org/abs/0804.2802}{{\ttfamily arXiv:0804.2802 [hep-ph]}}.
%%CITATION = ARXIV:0804.2802;%%.

\bibitem{Lourenco:2008sk}
C.~Lourenco, R.~Vogt, and H.~K. Woehri, ``{Energy dependence of J/psi
  absorption in proton-nucleus collisions},''
  \href{http://dx.doi.org/10.1088/1126-6708/2009/02/014}{{\em JHEP} {\bfseries
  02} (2009) 014},
\href{http://arxiv.org/abs/0901.3054}{{\ttfamily arXiv:0901.3054 [hep-ph]}}.
%%CITATION = ARXIV:0901.3054;%%.

\bibitem{Vogt:2010aa}
R.~Vogt, ``{Cold Nuclear Matter Effects on $J/\psi$ and $\Upsilon$ Production
  at the LHC},'' \href{http://dx.doi.org/10.1103/PhysRevC.81.044903}{{\em Phys.
  Rev.} {\bfseries C81} (2010) 044903},
\href{http://arxiv.org/abs/1003.3497}{{\ttfamily arXiv:1003.3497 [hep-ph]}}.
%%CITATION = ARXIV:1003.3497;%%.

\bibitem{Ferreiro:2008qj}
E.~G. Ferreiro, F.~Fleuret, and A.~Rakotozafindrabe, ``{Transverse momentum
  dependence of J/psi shadowing effects},''
  \href{http://dx.doi.org/10.1140/epjc/s10052-008-0843-8}{{\em Eur. Phys. J.}
  {\bfseries C61} (2009) 859--864},
\href{http://arxiv.org/abs/0801.4949}{{\ttfamily arXiv:0801.4949 [hep-ph]}}.
%%CITATION = ARXIV:0801.4949;%%.

\bibitem{Ferreiro:2008wc}
E.~G. Ferreiro, F.~Fleuret, J.~P. Lansberg, and A.~Rakotozafindrabe, ``{Cold
  nuclear matter effects on J/psi production: Intrinsic and extrinsic
  transverse momentum effects},''
  \href{http://dx.doi.org/10.1016/j.physletb.2009.07.076}{{\em Phys. Lett.}
  {\bfseries B680} (2009) 50--55},
\href{http://arxiv.org/abs/0809.4684}{{\ttfamily arXiv:0809.4684 [hep-ph]}}.
%%CITATION = ARXIV:0809.4684;%%.

\bibitem{Ferreiro:2009ur}
E.~G. Ferreiro, F.~Fleuret, J.~P. Lansberg, and A.~Rakotozafindrabe,
  ``{Centrality, Rapidity and Transverse-Momentum Dependence of Cold Nuclear
  Matter Effects on J/Psi Production in d Au, Cu Cu and Au Au Collisions at
  s(NN)**(1/2) = 200-GeV},''
  \href{http://dx.doi.org/10.1103/PhysRevC.81.064911}{{\em Phys. Rev.}
  {\bfseries C81} (2010) 064911},
\href{http://arxiv.org/abs/0912.4498}{{\ttfamily arXiv:0912.4498 [hep-ph]}}.
%%CITATION = ARXIV:0912.4498;%%.

\bibitem{delValle:2014wha}
Z.~Conesa~del Valle, E.~G. Ferreiro, F.~Fleuret, J.~P. Lansberg, and
  A.~Rakotozafindrabe, ``{Open-beauty production in $p$Pb collisions at
  $\sqrt{s_{NN}}$=5 TeV: effect of the gluon nuclear densities},''
  \href{http://arxiv.org/abs/1402.1747}{{\ttfamily arXiv:1402.1747 [hep-ph]}}.
[Nucl. Phys.A926,236(2014)].
%%CITATION = ARXIV:1402.1747;%%.

\bibitem{Shao:2012iz}
H.-S. Shao, ``{HELAC-Onia: An automatic matrix element generator for heavy
  quarkonium physics},''
  \href{http://dx.doi.org/10.1016/j.cpc.2013.05.023}{{\em Comput. Phys.
  Commun.} {\bfseries 184} (2013) 2562--2570},
\href{http://arxiv.org/abs/1212.5293}{{\ttfamily arXiv:1212.5293 [hep-ph]}}.
%%CITATION = ARXIV:1212.5293;%%.

\bibitem{Shao:2015vga}
H.-S. Shao, ``{HELAC-Onia 2.0: an upgraded matrix-element and event generator
  for heavy quarkonium physics},''
  \href{http://dx.doi.org/10.1016/j.cpc.2015.09.011}{{\em Comput. Phys.
  Commun.} {\bfseries 198} (2016) 238--259},
\href{http://arxiv.org/abs/1507.03435}{{\ttfamily arXiv:1507.03435 [hep-ph]}}.
%%CITATION = ARXIV:1507.03435;%%.

\bibitem{Whalley:2005nh}
M.~R. Whalley, D.~Bourilkov, and R.~C. Group, ``{The Les Houches accord PDFs
  (LHAPDF) and LHAGLUE},'' in {\em {HERA and the LHC: A Workshop on the
  implications of HERA for LHC physics. Proceedings, Part B}}.
\newblock 2005.
\newblock
\href{http://arxiv.org/abs/hep-ph/0508110}{{\ttfamily arXiv:hep-ph/0508110
  [hep-ph]}}.
\newblock
%%CITATION = HEP-PH/0508110;%%.

\bibitem{Bourilkov:2006cj}
D.~Bourilkov, R.~C. Group, and M.~R. Whalley, ``{LHAPDF: PDF use from the
  Tevatron to the LHC},'' in {\em {TeV4LHC Workshop - 4th meeting Batavia,
  Illinois, October 20-22, 2005}}.
\newblock 2006.
\newblock
\href{http://arxiv.org/abs/hep-ph/0605240}{{\ttfamily arXiv:hep-ph/0605240
  [hep-ph]}}.
\newblock
%%CITATION = HEP-PH/0605240;%%.

\bibitem{Buckley:2014ana}
A.~Buckley, J.~Ferrando, S.~Lloyd, K.~Nordström, B.~Page, M.~Rüfenacht,
  M.~Schönherr, and G.~Watt, ``{LHAPDF6: parton density access in the LHC
  precision era},''
  \href{http://dx.doi.org/10.1140/epjc/s10052-015-3318-8}{{\em Eur. Phys. J.}
  {\bfseries C75} (2015) 132},
\href{http://arxiv.org/abs/1412.7420}{{\ttfamily arXiv:1412.7420 [hep-ph]}}.
%%CITATION = ARXIV:1412.7420;%%.

\bibitem{Kom:2011bd}
C.~H. Kom, A.~Kulesza, and W.~J. Stirling, ``{Pair Production of J/psi as a
  Probe of Double Parton Scattering at LHCb},''
  \href{http://dx.doi.org/10.1103/PhysRevLett.107.082002}{{\em Phys. Rev.
  Lett.} {\bfseries 107} (2011) 082002},
\href{http://arxiv.org/abs/1105.4186}{{\ttfamily arXiv:1105.4186 [hep-ph]}}.
%%CITATION = ARXIV:1105.4186;%%.

\bibitem{Lansberg:2014swa}
J.-P. Lansberg and H.-S. Shao, ``{$J/\psi$ -pair production at large momenta:
  Indications for double parton scatterings and large $\alpha_s^5$
  contributions},''
  \href{http://dx.doi.org/10.1016/j.physletb.2015.10.083}{{\em Phys. Lett.}
  {\bfseries B751} (2015) 479--486},
\href{http://arxiv.org/abs/1410.8822}{{\ttfamily arXiv:1410.8822 [hep-ph]}}.
%%CITATION = ARXIV:1410.8822;%%.

\bibitem{Lansberg:2015lva}
J.-P. Lansberg and H.-S. Shao, ``{Double-quarkonium production at a
  fixed-target experiment at the LHC (AFTER@LHC)},''
  \href{http://dx.doi.org/10.1016/j.nuclphysb.2015.09.005}{{\em Nucl. Phys.}
  {\bfseries B900} (2015) 273--294},
\href{http://arxiv.org/abs/1504.06531}{{\ttfamily arXiv:1504.06531 [hep-ph]}}.
%%CITATION = ARXIV:1504.06531;%%.

\bibitem{Shao:2016wor}
H.-S. Shao and Y.-J. Zhang, ``{Complete study of hadroproduction of a
  $\Upsilon$ meson associated with a prompt $J/\psi$},''
  \href{http://dx.doi.org/10.1103/PhysRevLett.117.062001}{{\em Phys. Rev.
  Lett.} {\bfseries 117} no.~6, (2016) 062001},
\href{http://arxiv.org/abs/1605.03061}{{\ttfamily arXiv:1605.03061 [hep-ph]}}.
%%CITATION = ARXIV:1605.03061;%%.

\bibitem{Borschensky:2016nkv}
C.~Borschensky and A.~Kulesza, ``{Double parton scattering in pair-production
  of $J/\psi$ mesons at the LHC revisited},''
\href{http://arxiv.org/abs/1610.00666}{{\ttfamily arXiv:1610.00666 [hep-ph]}}.
%%CITATION = ARXIV:1610.00666;%%.

\bibitem{Dulat:2015mca}
S.~Dulat, T.-J. Hou, J.~Gao, M.~Guzzi, J.~Huston, P.~Nadolsky, J.~Pumplin,
  C.~Schmidt, D.~Stump, and C.~P. Yuan, ``{New parton distribution functions
  from a global analysis of quantum chromodynamics},''
  \href{http://dx.doi.org/10.1103/PhysRevD.93.033006}{{\em Phys. Rev.}
  {\bfseries D93} no.~3, (2016) 033006},
\href{http://arxiv.org/abs/1506.07443}{{\ttfamily arXiv:1506.07443 [hep-ph]}}.
%%CITATION = ARXIV:1506.07443;%%.

\bibitem{Lai:2010vv}
H.-L. Lai, M.~Guzzi, J.~Huston, Z.~Li, P.~M. Nadolsky, J.~Pumplin, and C.~P.
  Yuan, ``{New parton distributions for collider physics},''
  \href{http://dx.doi.org/10.1103/PhysRevD.82.074024}{{\em Phys. Rev.}
  {\bfseries D82} (2010) 074024},
\href{http://arxiv.org/abs/1007.2241}{{\ttfamily arXiv:1007.2241 [hep-ph]}}.
%%CITATION = ARXIV:1007.2241;%%.

\bibitem{Aaij:2011jh}
{\bfseries LHCb} Collaboration, R.~Aaij {\em et~al.}, ``{Measurement of
  $J/\psi$ production in $pp$ collisions at $\sqrt{s}=7~\rm{TeV}$},''
  \href{http://dx.doi.org/10.1140/epjc/s10052-011-1645-y}{{\em Eur. Phys. J.}
  {\bfseries C71} (2011) 1645},
\href{http://arxiv.org/abs/1103.0423}{{\ttfamily arXiv:1103.0423 [hep-ex]}}.
%%CITATION = ARXIV:1103.0423;%%.

\bibitem{Aaij:2013yaa}
{\bfseries LHCb} Collaboration, R.~Aaij {\em et~al.}, ``{Production of J/psi
  and Upsilon mesons in pp collisions at sqrt(s) = 8 TeV},''
  \href{http://dx.doi.org/10.1007/JHEP06(2013)064}{{\em JHEP} {\bfseries 06}
  (2013) 064},
\href{http://arxiv.org/abs/1304.6977}{{\ttfamily arXiv:1304.6977 [hep-ex]}}.
%%CITATION = ARXIV:1304.6977;%%.

\bibitem{Aad:2015duc}
{\bfseries ATLAS} Collaboration, G.~Aad {\em et~al.}, ``{Measurement of the
  differential cross-sections of prompt and non-prompt production of $J/\psi $
  and $\psi (2\mathrm {S})$ in $pp$ collisions at $\sqrt{s} = 7$ and 8 TeV
  with the ATLAS detector},''
  \href{http://dx.doi.org/10.1140/epjc/s10052-016-4050-8}{{\em Eur. Phys. J.}
  {\bfseries C76} no.~5, (2016) 283},
\href{http://arxiv.org/abs/1512.03657}{{\ttfamily arXiv:1512.03657 [hep-ex]}}.
%%CITATION = ARXIV:1512.03657;%%.

\bibitem{Khachatryan:2015rra}
{\bfseries CMS} Collaboration, V.~Khachatryan {\em et~al.}, ``{Measurement of
  $J/\psi$ and ψ(2S) Prompt Double-Differential Cross Sections in pp
  Collisions at $\sqrt{s}$=7  TeV},''
  \href{http://dx.doi.org/10.1103/PhysRevLett.114.191802}{{\em Phys. Rev.
  Lett.} {\bfseries 114} no.~19, (2015) 191802},
\href{http://arxiv.org/abs/1502.04155}{{\ttfamily arXiv:1502.04155 [hep-ex]}}.
%%CITATION = ARXIV:1502.04155;%%.

\bibitem{Abelev:2014qha}
{\bfseries ALICE} Collaboration, B.~B. Abelev {\em et~al.}, ``{Measurement of
  quarkonium production at forward rapidity in $pp$ collisions at $\sqrt{s} =
  7$ TeV},'' \href{http://dx.doi.org/10.1140/epjc/s10052-014-2974-4}{{\em Eur.
  Phys. J.} {\bfseries C74} no.~8, (2014) 2974},
\href{http://arxiv.org/abs/1403.3648}{{\ttfamily arXiv:1403.3648 [nucl-ex]}}.
%%CITATION = ARXIV:1403.3648;%%.

\bibitem{LHCb:2012aa}
{\bfseries LHCb} Collaboration, R.~Aaij {\em et~al.}, ``{Measurement of Upsilon
  production in pp collisions at $\sqrt{s}$ = 7 TeV},''
  \href{http://dx.doi.org/10.1140/epjc/s10052-012-2025-y}{{\em Eur. Phys. J.}
  {\bfseries C72} (2012) 2025},
\href{http://arxiv.org/abs/1202.6579}{{\ttfamily arXiv:1202.6579 [hep-ex]}}.
%%CITATION = ARXIV:1202.6579;%%.

\bibitem{Aaij:2015awa}
{\bfseries LHCb} Collaboration, R.~Aaij {\em et~al.}, ``{Forward production of
  $\Upsilon$ mesons in $pp$ collisions at $\sqrt{s}=7$ and 8TeV},''
  \href{http://dx.doi.org/10.1007/JHEP11(2015)103}{{\em JHEP} {\bfseries 11}
  (2015) 103},
\href{http://arxiv.org/abs/1509.02372}{{\ttfamily arXiv:1509.02372 [hep-ex]}}.
%%CITATION = ARXIV:1509.02372;%%.

\bibitem{Aad:2012dlq}
{\bfseries ATLAS} Collaboration, G.~Aad {\em et~al.}, ``{Measurement of Upsilon
  production in 7 TeV pp collisions at ATLAS},''
  \href{http://dx.doi.org/10.1103/PhysRevD.87.052004}{{\em Phys. Rev.}
  {\bfseries D87} no.~5, (2013) 052004},
\href{http://arxiv.org/abs/1211.7255}{{\ttfamily arXiv:1211.7255 [hep-ex]}}.
%%CITATION = ARXIV:1211.7255;%%.

\bibitem{Chatrchyan:2013yna}
{\bfseries CMS} Collaboration, S.~Chatrchyan {\em et~al.}, ``{Measurement of
  the $\Upsilon(1S), \Upsilon(2S)$, and $\Upsilon(3S)$ cross sections in pp
  collisions at $\sqrt{s}$ = 7 TeV},''
  \href{http://dx.doi.org/10.1016/j.physletb.2013.10.033}{{\em Phys. Lett.}
  {\bfseries B727} (2013) 101--125},
\href{http://arxiv.org/abs/1303.5900}{{\ttfamily arXiv:1303.5900 [hep-ex]}}.
%%CITATION = ARXIV:1303.5900;%%.

\bibitem{Aaij:2014bga}
{\bfseries LHCb} Collaboration, R.~Aaij {\em et~al.}, ``{Measurement of the
  $\eta_c (1S)$ production cross-section in proton-proton collisions via the
  decay $\eta_c (1S) \rightarrow p \bar{p}$},''
  \href{http://dx.doi.org/10.1140/epjc/s10052-015-3502-x}{{\em Eur. Phys. J.}
  {\bfseries C75} no.~7, (2015) 311},
\href{http://arxiv.org/abs/1409.3612}{{\ttfamily arXiv:1409.3612 [hep-ex]}}.
%%CITATION = ARXIV:1409.3612;%%.

\bibitem{Aaij:2013mga}
{\bfseries LHCb} Collaboration, R.~Aaij {\em et~al.}, ``{Prompt charm
  production in pp collisions at sqrt(s)=7 TeV},''
  \href{http://dx.doi.org/10.1016/j.nuclphysb.2013.02.010}{{\em Nucl. Phys.}
  {\bfseries B871} (2013) 1--20},
\href{http://arxiv.org/abs/1302.2864}{{\ttfamily arXiv:1302.2864 [hep-ex]}}.
%%CITATION = ARXIV:1302.2864;%%.

\bibitem{Aaij:2013zxa}
{\bfseries LHCb} Collaboration, R.~Aaij {\em et~al.}, ``{Study of $J/\psi$
  production and cold nuclear matter effects in $pPb$ collisions at
  $\sqrt{s_{NN}} = 5$ TeV},''
  \href{http://dx.doi.org/10.1007/JHEP02(2014)072}{{\em JHEP} {\bfseries 02}
  (2014) 072},
\href{http://arxiv.org/abs/1308.6729}{{\ttfamily arXiv:1308.6729 [nucl-ex]}}.
%%CITATION = ARXIV:1308.6729;%%.

\bibitem{Adam:2015iga}
{\bfseries ALICE} Collaboration, J.~Adam {\em et~al.}, ``{Rapidity and
  transverse-momentum dependence of the inclusive J/$\psi$ nuclear modification
  factor in p-Pb collisions at $ \sqrt{s_{N\ N}} =$ 5.02 TeV},''
  \href{http://dx.doi.org/10.1007/JHEP06(2015)055}{{\em JHEP} {\bfseries 06}
  (2015) 055},
\href{http://arxiv.org/abs/1503.07179}{{\ttfamily arXiv:1503.07179 [nucl-ex]}}.
%%CITATION = ARXIV:1503.07179;%%.

\bibitem{Aad:2015ddl}
{\bfseries ATLAS} Collaboration, G.~Aad {\em et~al.}, ``{Measurement of
  differential $J/\psi$ production cross sections and forward-backward ratios
  in p + Pb collisions with the ATLAS detector},''
  \href{http://dx.doi.org/10.1103/PhysRevC.92.034904}{{\em Phys. Rev.}
  {\bfseries C92} no.~3, (2015) 034904},
\href{http://arxiv.org/abs/1505.08141}{{\ttfamily arXiv:1505.08141 [hep-ex]}}.
%%CITATION = ARXIV:1505.08141;%%.

\bibitem{Abelev:2013yxa}
{\bfseries ALICE} Collaboration, B.~B. Abelev {\em et~al.}, ``{$J/\psi$
  production and nuclear effects in p-Pb collisions at $\sqrt{S_{NN}}$ = 5.02
  TeV},'' \href{http://dx.doi.org/10.1007/JHEP02(2014)073}{{\em JHEP}
  {\bfseries 02} (2014) 073},
\href{http://arxiv.org/abs/1308.6726}{{\ttfamily arXiv:1308.6726 [nucl-ex]}}.
%%CITATION = ARXIV:1308.6726;%%.

\bibitem{Aaij:2014mza}
{\bfseries LHCb} Collaboration, R.~Aaij {\em et~al.}, ``{Study of $\Upsilon$
  production and cold nuclear matter effects in $p$Pb collisions at
  $\sqrt{s_{NN}}$=5 TeV},''
  \href{http://dx.doi.org/10.1007/JHEP07(2014)094}{{\em JHEP} {\bfseries 07}
  (2014) 094},
\href{http://arxiv.org/abs/1405.5152}{{\ttfamily arXiv:1405.5152 [nucl-ex]}}.
%%CITATION = ARXIV:1405.5152;%%.

\bibitem{Abelev:2014oea}
{\bfseries ALICE} Collaboration, B.~B. Abelev {\em et~al.}, ``{Production of
  inclusive $\Upsilon$(1S) and $\Upsilon$(2S) in p-Pb collisions at
  $\mathbf{\sqrt{s_{{\rm NN}}} = 5.02}$ TeV},''
  \href{http://dx.doi.org/10.1016/j.physletb.2014.11.041}{{\em Phys. Lett.}
  {\bfseries B740} (2015) 105--117},
\href{http://arxiv.org/abs/1410.2234}{{\ttfamily arXiv:1410.2234 [nucl-ex]}}.
%%CITATION = ARXIV:1410.2234;%%.

\bibitem{ATLAS-CONF-2015-050}
{\bfseries ATLAS} Collaboration, ``{Measurement of $\Upsilon(\mathrm{nS})$
  production with $p$+Pb collisions at $\sqrt{s_{\rm NN}} = 5.02~\mathrm {TeV}$
  and $pp$ collisions at $\sqrt{s} = 2.76~\mathrm {TeV}$},'' Tech. Rep.
  ATLAS-CONF-2015-050, CERN, Geneva, Sep, 2015.
\newblock \url{https://cds.cern.ch/record/2055266}.

\bibitem{LHCb-CONF-2016-003}
{\bfseries LHCb} Collaboration, ``{Study of cold nuclear matter effects using
  prompt $D^0$ meson production in $p\mathrm{Pb}$ collisions at LHCb},'' Tech.
  Rep. LHCb-CONF-2016-003, CERN, Geneva, Mar, 2016.
\newblock \url{http://cds.cern.ch/record/2138946}.

\bibitem{Abelev:2014hha}
{\bfseries ALICE} Collaboration, B.~B. Abelev {\em et~al.}, ``{Measurement of
  prompt $D$-meson production in $p-Pb$ collisions at $\sqrt{s_{NN}}$ = 5.02
  TeV},'' \href{http://dx.doi.org/10.1103/PhysRevLett.113.232301}{{\em Phys.
  Rev. Lett.} {\bfseries 113} no.~23, (2014) 232301},
\href{http://arxiv.org/abs/1405.3452}{{\ttfamily arXiv:1405.3452 [nucl-ex]}}.
%%CITATION = ARXIV:1405.3452;%%.

\bibitem{ATLAS:2015pua}
{\bfseries ATLAS} Collaboration, T.~A. collaboration, ``{Study of $J/\psi$ and
  $\psi(\mathrm{2S})$ production in $\sqrt{s_{\rm NN}} = 5.02\mathrm{~TeV}$
  $p+\rm{Pb}$ and $\sqrt{s} = 2.76\mathrm{~TeV}$ $pp$ collisions with the ATLAS
  detector},'' Tech. Rep. ATLAS-CONF-2015-023, 2015.

\end{thebibliography}\endgroup

\appendix

\section{Appendix: Predictions for the \pPb\ run at 8 TeV.}
\begin{figure}[H]
\begin{center} 
\subfloat[]{\includegraphics[width=0.33\textwidth,keepaspectratio,draft=false]{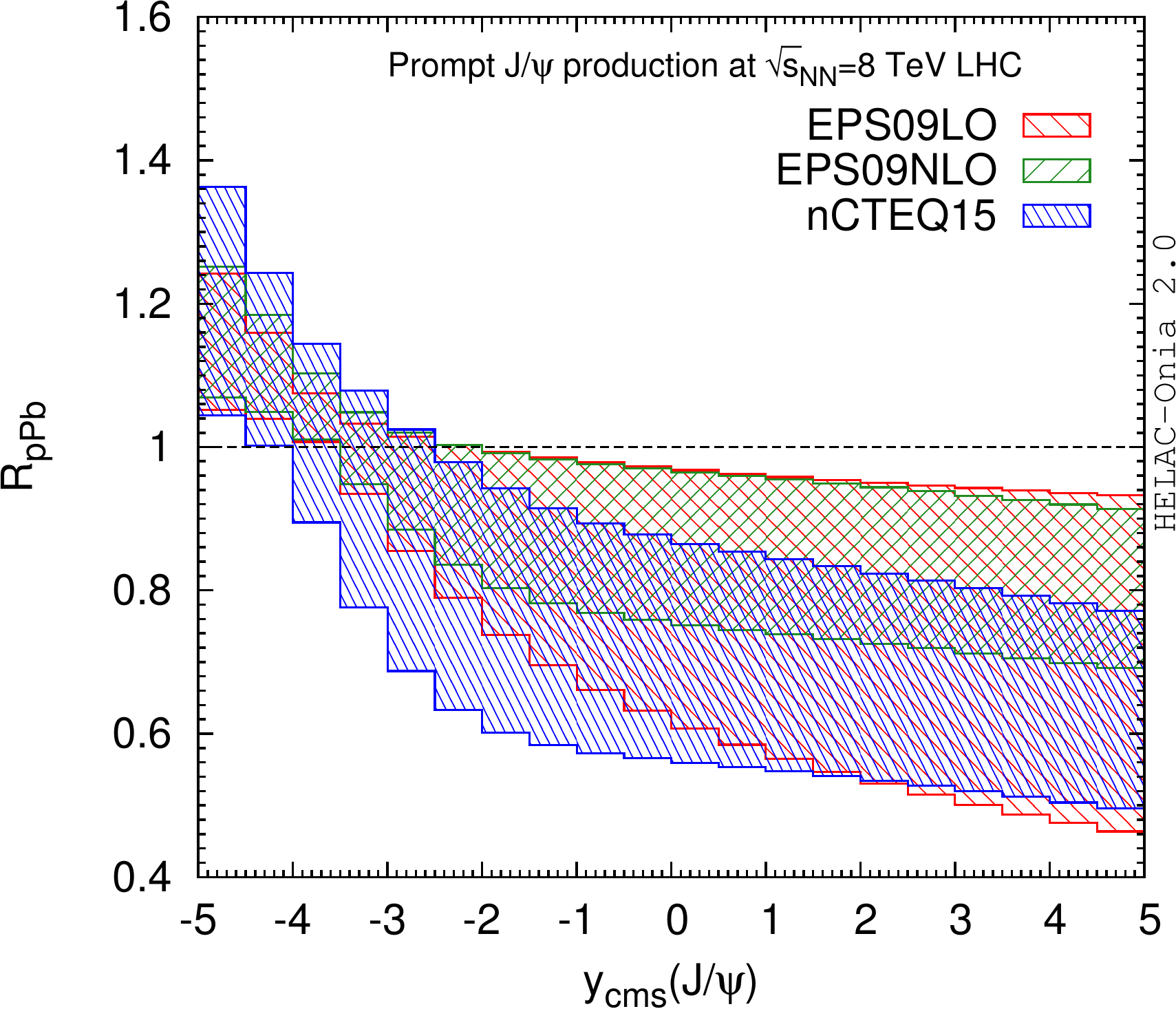}}
\subfloat[]{\includegraphics[width=0.33\textwidth,keepaspectratio,draft=false]{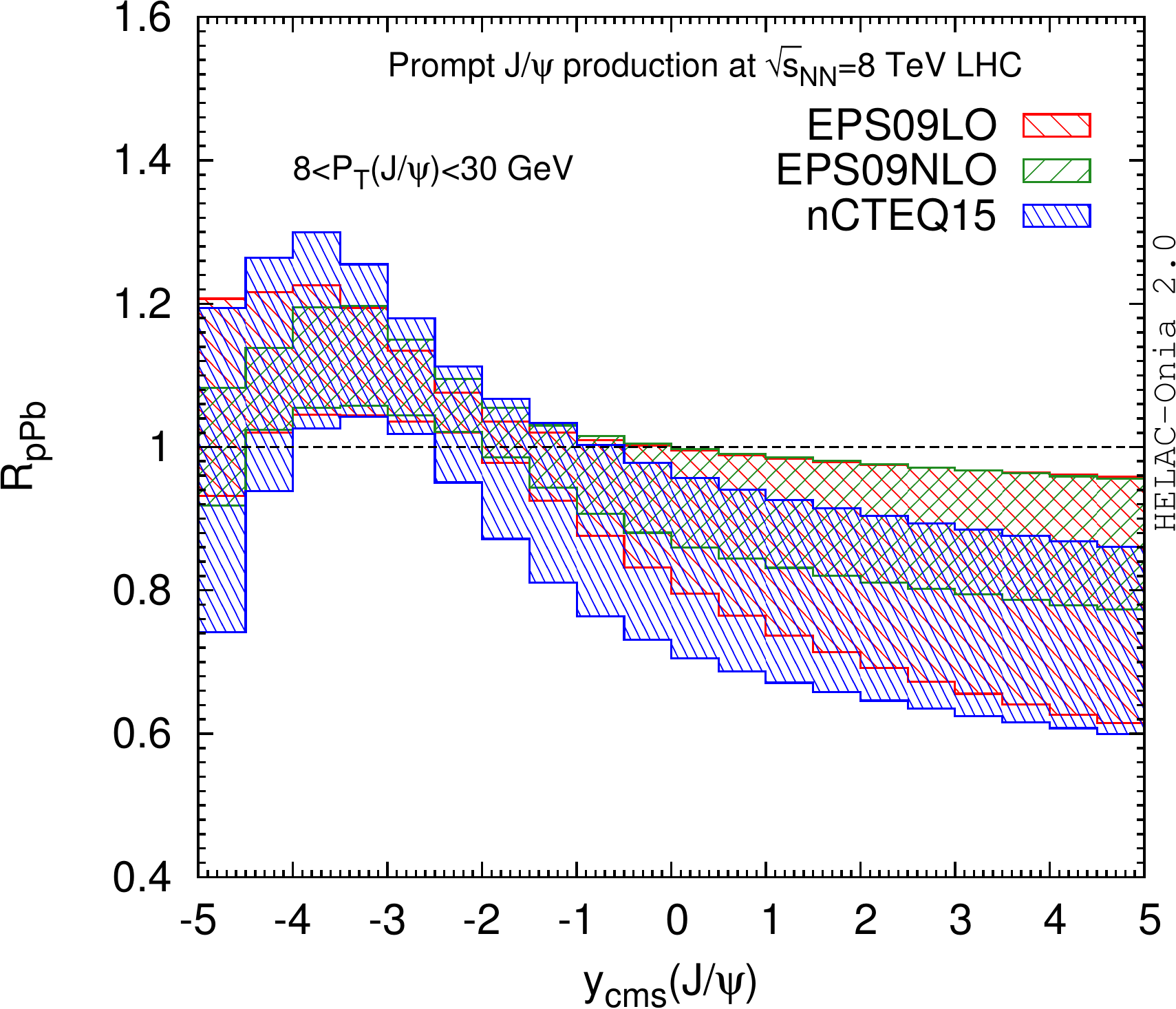}}
\subfloat[]{\includegraphics[width=0.3\textwidth,keepaspectratio,draft=false]{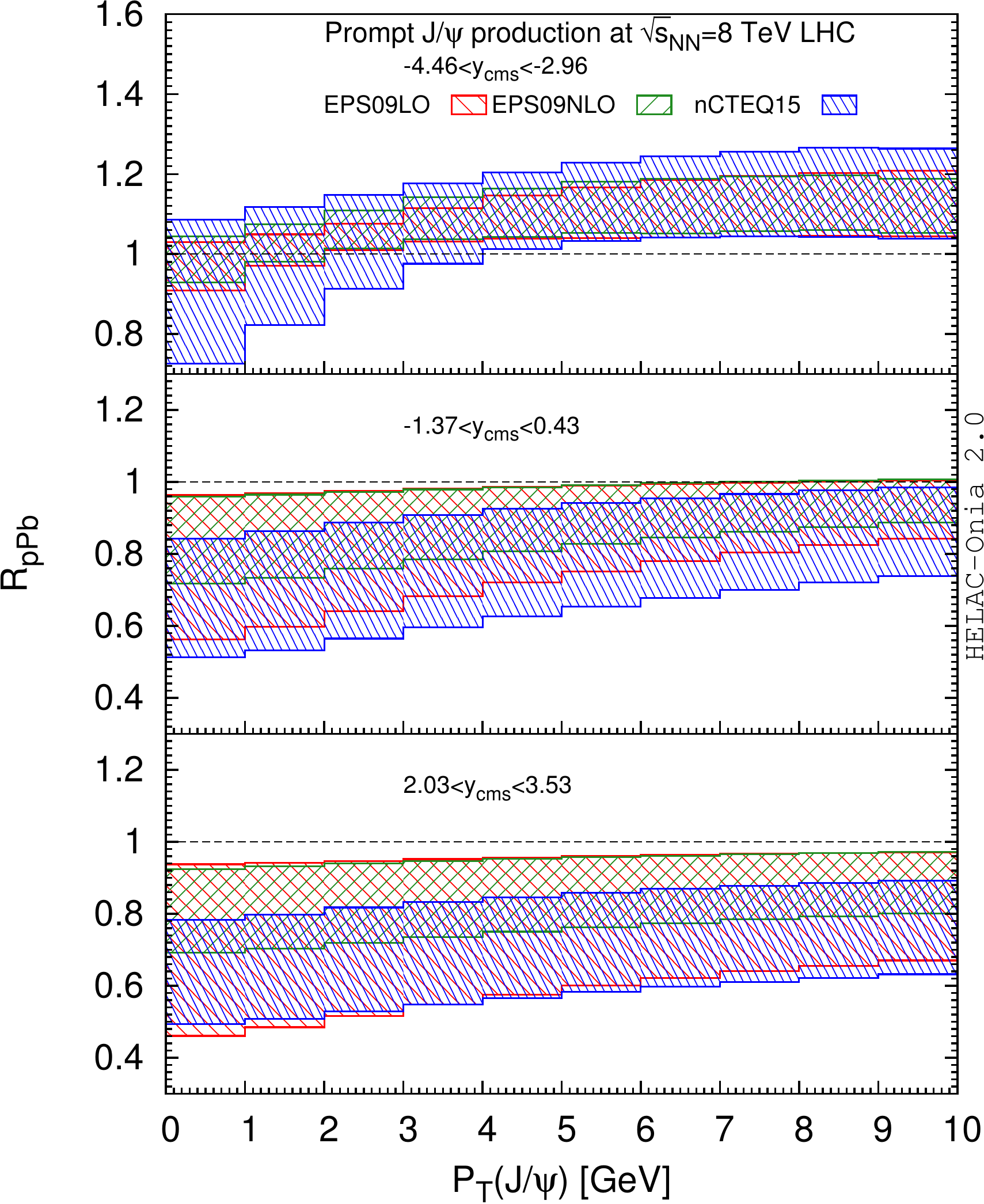}}
\\
\subfloat[]{\includegraphics[width=0.3\textwidth,keepaspectratio,draft=false]{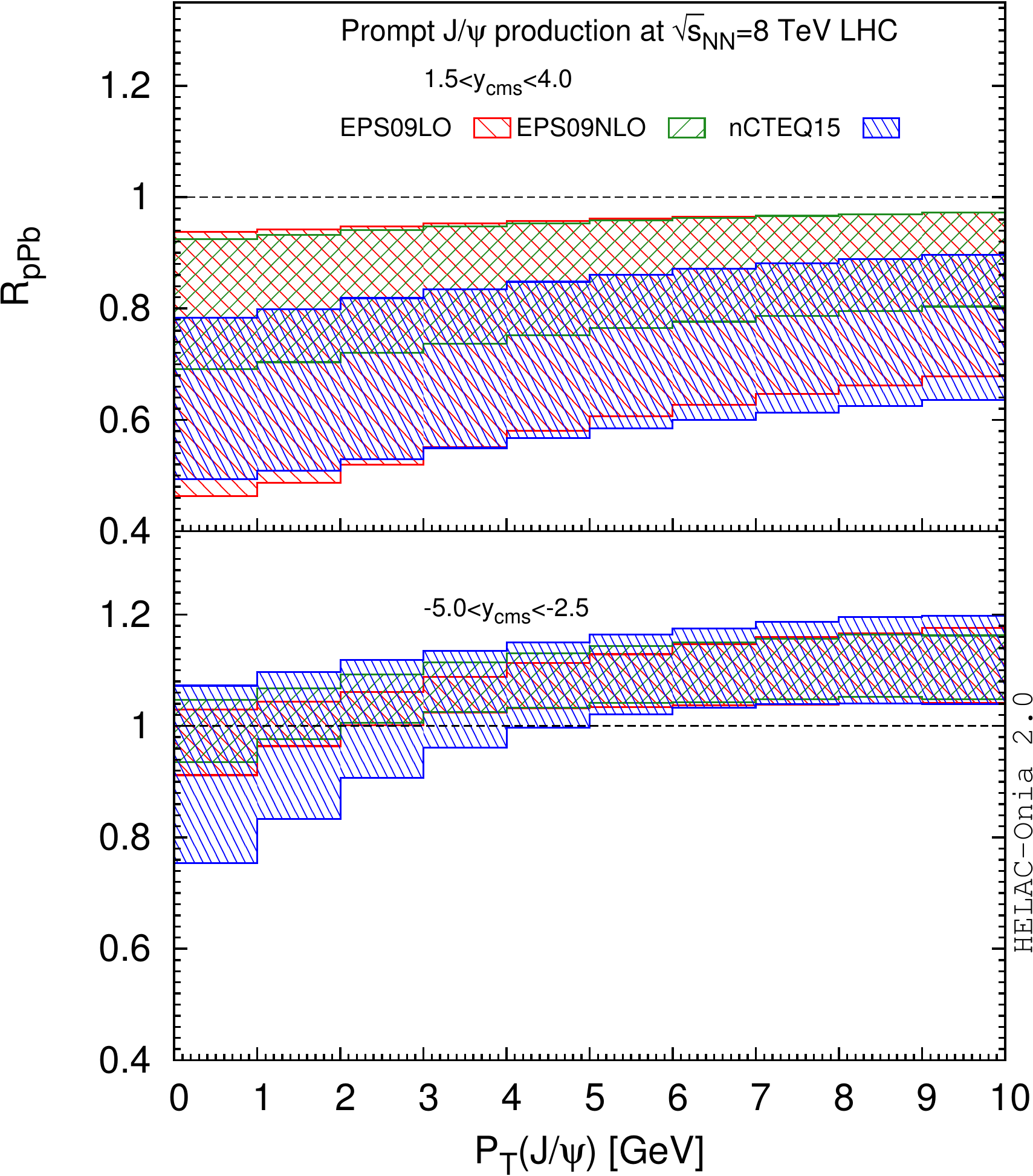}}
\subfloat[]{\includegraphics[width=0.33\textwidth,keepaspectratio,draft=false]{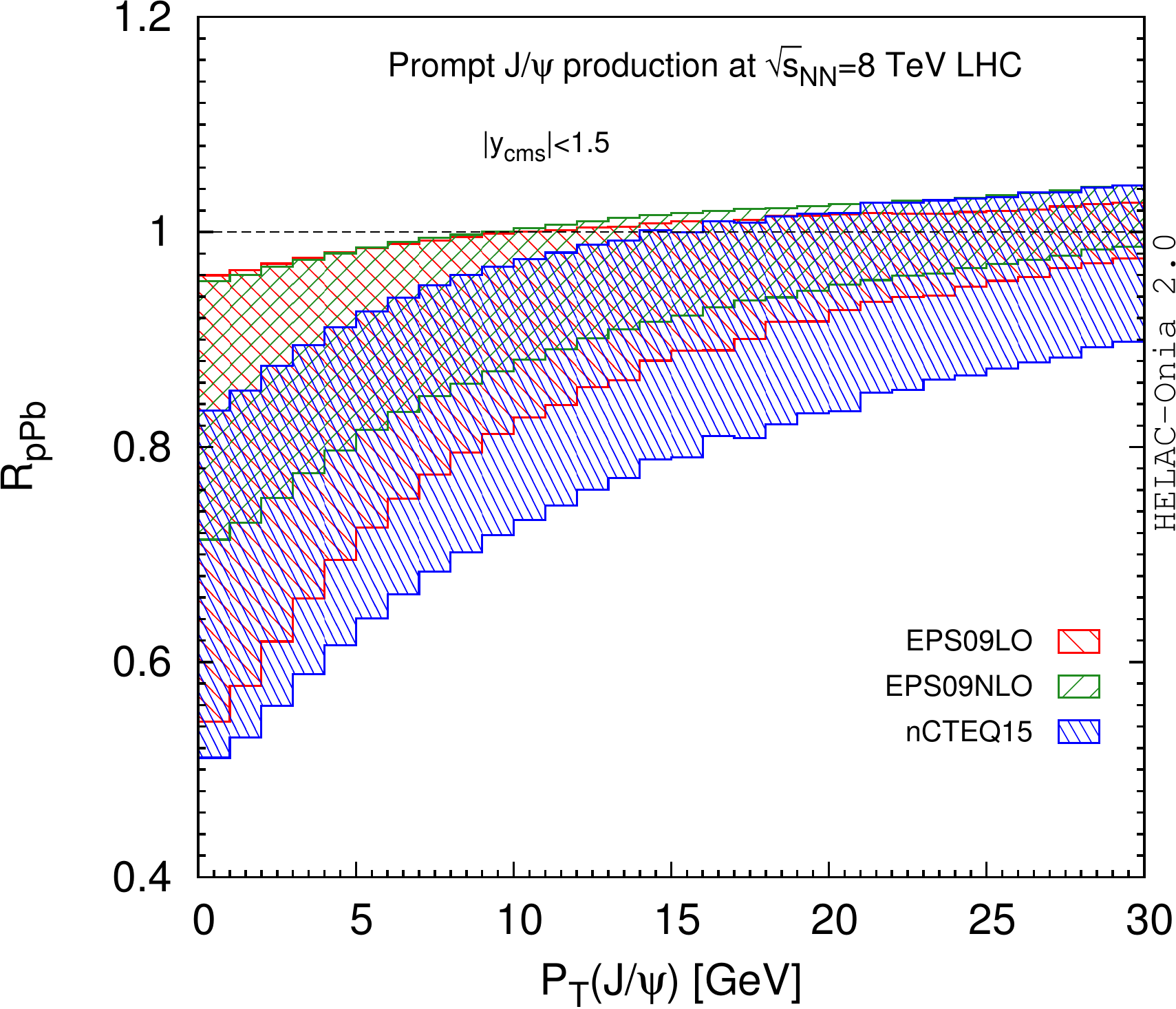}}
\subfloat[]{\includegraphics[width=0.33\textwidth,keepaspectratio,draft=false]{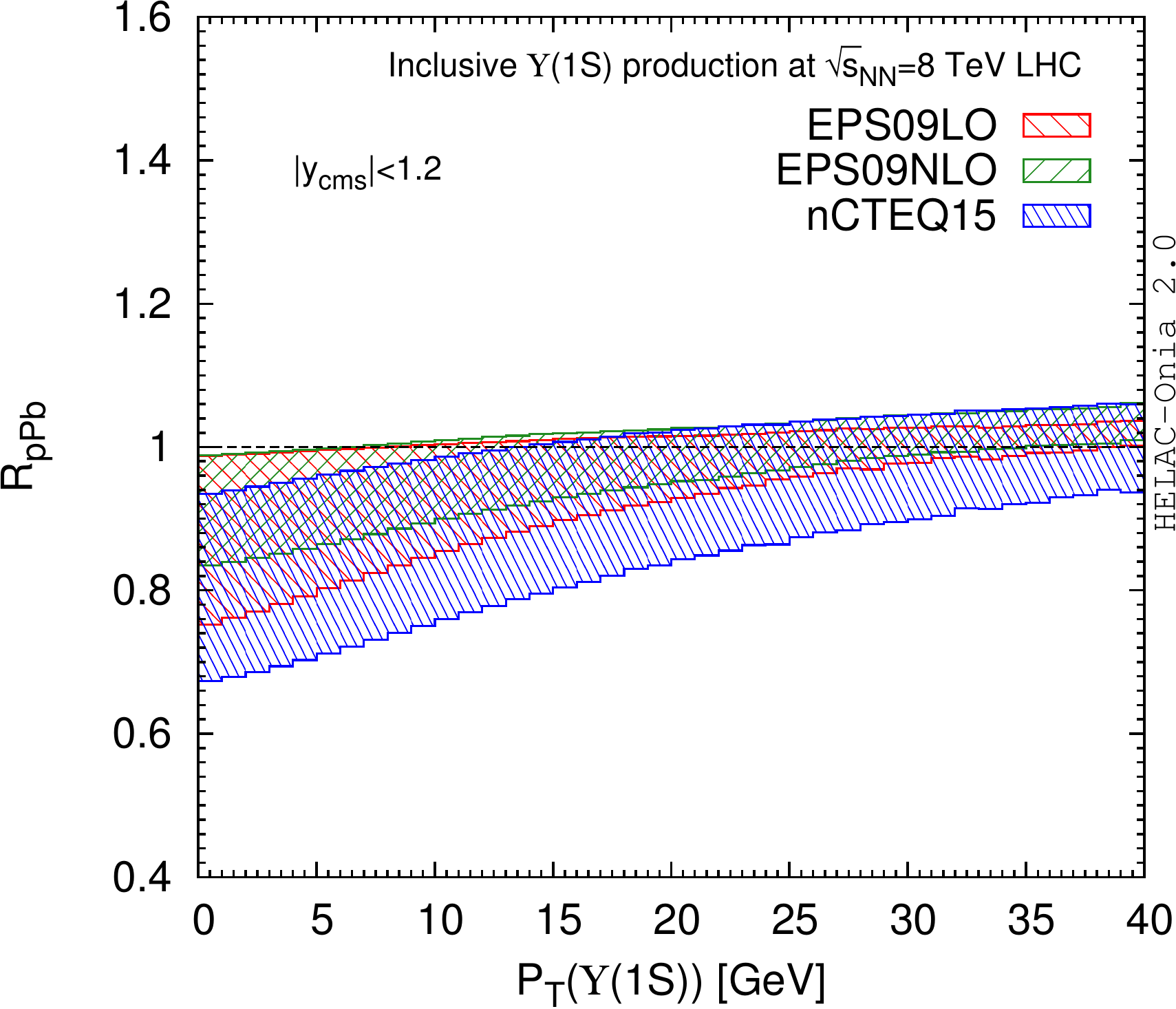}}
\\
\subfloat[]{\includegraphics[width=0.33\textwidth,keepaspectratio,draft=false]{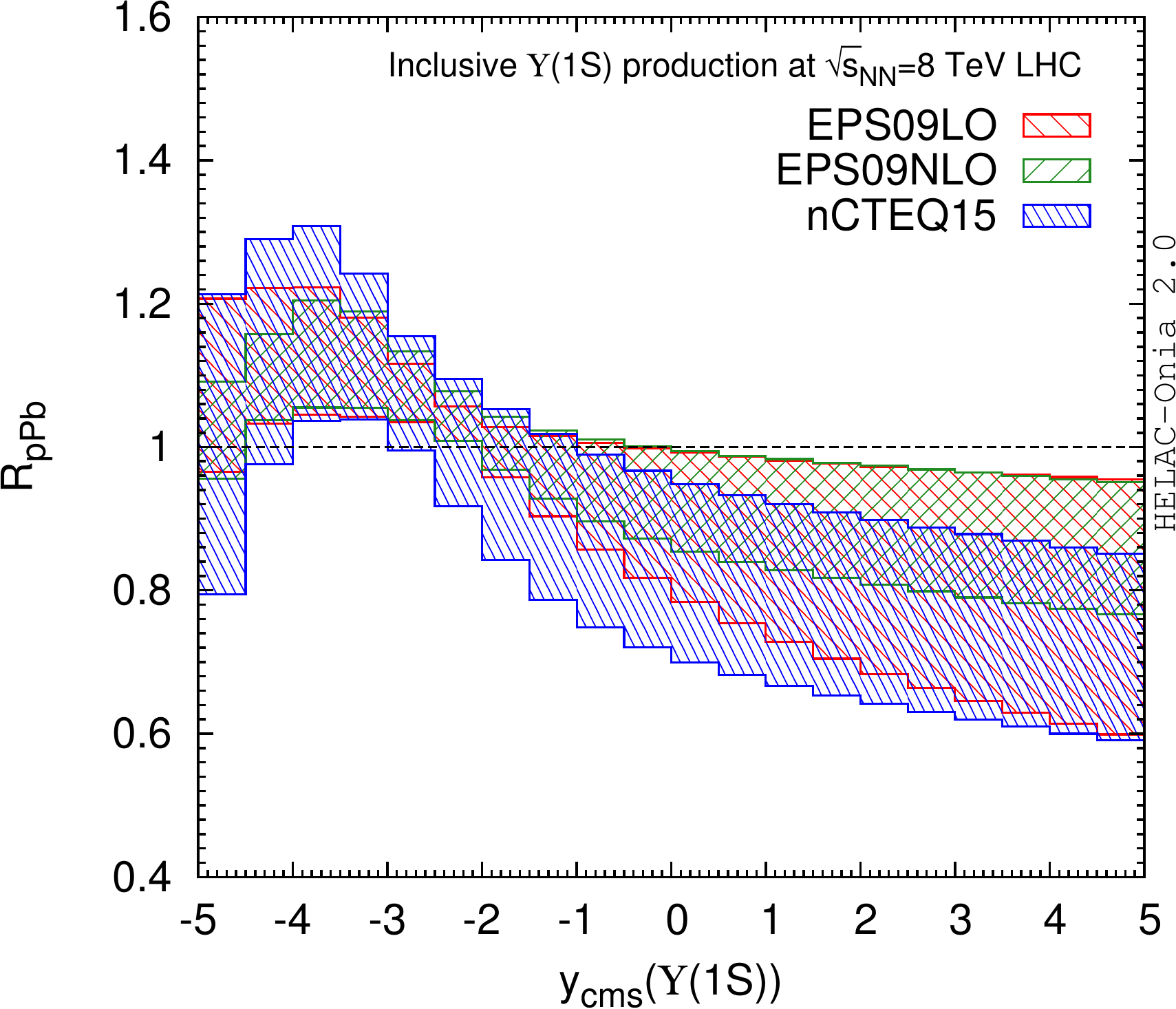}}
\subfloat[]{\includegraphics[width=0.33\textwidth,keepaspectratio,draft=false]{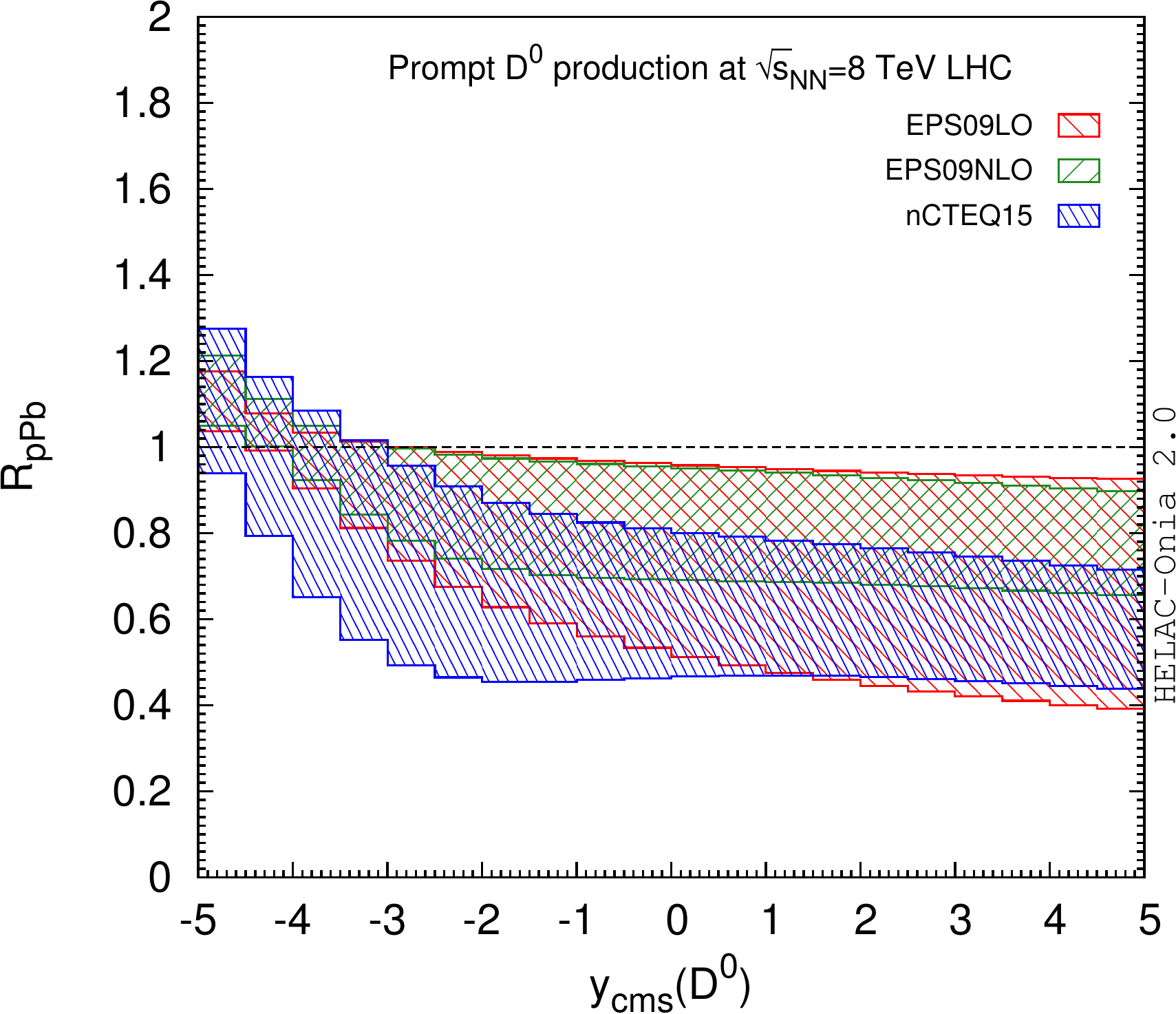}}
\subfloat[]{\includegraphics[width=0.33\textwidth,keepaspectratio,draft=false]{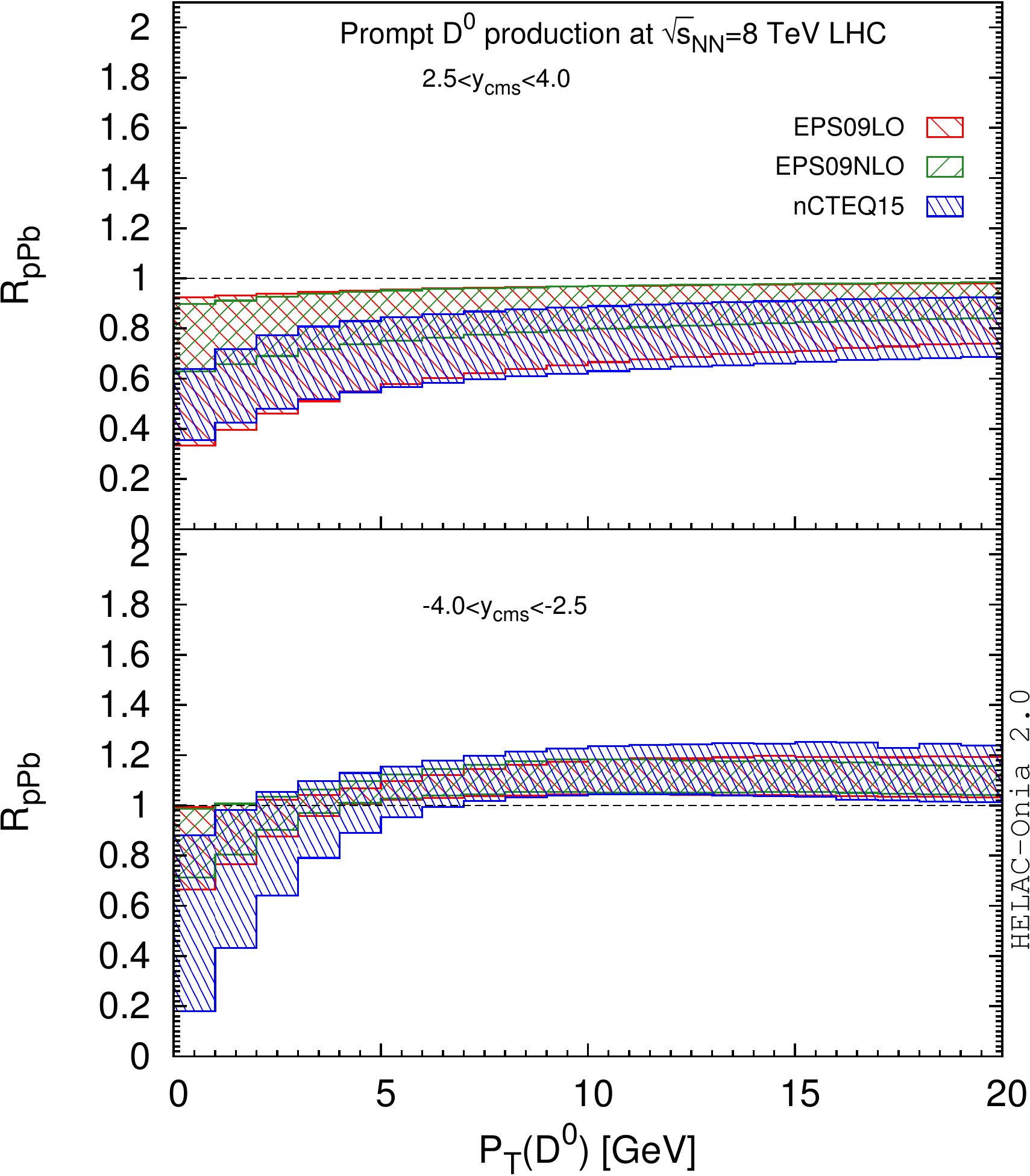}}
\caption{Predictions for 8 TeV in  different rapidity and transverse momentum regions.
}\label{fig:8TeV}
\end{center}
\end{figure}

\end{document}